\documentclass[11pt,a4paper]{article}
\usepackage{jcappub}
\usepackage{subfigure}

\title{Axion cosmology with long-lived domain walls}

\author[a]{Takashi Hiramatsu,}
\author[b,c]{Masahiro Kawasaki,}
\author[b]{Ken'ichi Saikawa}
\author[d]{and Toyokazu Sekiguchi}

\affiliation[a]{Yukawa Institute for Theoretical Physics, Kyoto University,\\
Kitashirakawa Oiwake-cho, Sakyo-ku, Kyoto 606-8502, Japan}
\affiliation[b]{Institute for Cosmic Ray Research, The University of Tokyo,\\
5-1-5 Kashiwa-no-ha, Kashiwa City, Chiba 277-8582, Japan} 
\affiliation[c]{Kavli Institute for the Physics and Mathematics of the Universe (WPI), Todai Institutes for Advanced Study, The University of Tokyo,\\
5-1-5 Kashiwa-no-ha, Kashiwa City, Chiba 277-8582, Japan}
\affiliation[d]{Graduate School of Science, Nagoya University,\\
Furo-cho, Chikusa-ku, Nagoya City, Aichi 464-8602, Japan}

\emailAdd{hiramatz@yukawa.kyoto-u.ac.jp}
\emailAdd{kawasaki@icrr.u-tokyo.ac.jp}
\emailAdd{saikawa@icrr.u-tokyo.ac.jp}
\emailAdd{sekiguti@a.phys.nagoya-u.ac.jp} 
 
\abstract{We investigate the cosmological constraints on axion models where the domain wall number is greater than one.
In these models, multiple domain walls attached to strings are formed, and they survive for a long time.
Their annihilation occurs due to the effects of explicit symmetry breaking term which might be raised by Planck-scale physics.
We perform three-dimensional lattice simulations and compute the spectra of axions and gravitational waves produced by long-lived domain walls.
Using the numerical results, we estimated relic density of axions and gravitational waves.
We find that the existence of long-lived domain walls leads to the overproduction of cold dark matter axions,
while the density of gravitational waves is too small to observe at the present time.
Combining the results with other observational constraints, we find that
the whole parameter region of models are excluded unless an unacceptable fine-tuning exists.}


\begin{document}
\subheader{\\ ICRR-REPORT-620-2012-9\\ IPMU12-0140\\ YITP-12-58}
\maketitle

\section{\label{sec1}Introduction}
Axion~\cite{Weinberg:1977ma} is a hypothetical particle which arises as a consequence of the Peccei-Quinn (PQ) mechanism~\cite{Peccei:1977hh}, the most attractive solution to the 
strong CP problem of quantum chromodynamics (QCD)~\cite{Kim:1986ax,Kim:2008hd}.
It arises as a Goldstone boson when a global $U(1)_{\mathrm{PQ}}$ symmetry (so called PQ symmetry) is spontaneously broken at some high-energy scale denoted as $F_a$.
The phenomenological searches indicate that axions are ``invisible" due to the weak coupling with matter which is suppressed by $1/F_a$.
This invisibleness is realized by two kinds of models. One is the KSVZ model~\cite{Kim:1979if}, which is realized by introducing additional quarks to the standard model.
Another is the DFSZ model~\cite{Dine:1981rt}, which is realized without introducing additional fermions by assuming two Higgs doublets.
These invisible axions lead to rich implications to astrophysics and cosmology, such that they are good candidates of dark matter filled in the universe~\cite{Preskill:1982cy}.

The remarkable feature of the axion models is that they predict the formation of topological defects in the early universe [see~\cite{2000csot.book.....V} for reviews].
When the PQ symmetry is spontaneously broken, vortex-like defects, called strings, are formed.
These strings are attached by surface-like defects, called domain walls, when the temperature of the universe becomes comparable to the QCD scale.
The structure of the domain walls is determined by an integer number $N_{\mathrm{DW}}$ which is referred as the ``domain wall number".
The value of $N_{\mathrm{DW}}$ is related to the $U(1)_{\mathrm{PQ}}$-$SU(3)_c$-$SU(3)_c$ anomaly which is given by the formula~\cite{Sikivie:1982qv,Georgi:1982ph,Kim:1986ax}
\begin{equation}
N_{\mathrm{DW}} = \left|2\sum_{i=L}\mathrm{Tr}Q_{\mathrm{PQ}}(q_i)T_a^2(q_i)-2\sum_{i=R}\mathrm{Tr}Q_{\mathrm{PQ}}(q_i)T_a^2(q_i)\right|, \label{eq1-1}
\end{equation}
where $Q_{\mathrm{PQ}}(q_i)$ is the $U(1)_{\mathrm{PQ}}$ charge for quark species $q_i$, $\sum_{i=L(R)}$ represents the sum over the left(right)-handed fermions,
and $T_a$ are the generators of $SU(3)_c$ normalized such that $\mathrm{Tr}T_aT_b=I\delta_{ab}$ where $I=1/2$ for fundamental representation of $SU(3)_c$.
Here we choose the unit $Q_{\mathrm{PQ}}=1$ for minimum non-vanishing magnitude of PQ quantum numbers.
In the KSVZ models we obtain $N_{\mathrm{DW}}=1$, while in the DFSZ models it is predicted that $N_{\mathrm{DW}}=2N_g$ where $N_g$ is the number of generations.
As we discuss in Sec.~\ref{sec2}, the cosmological scenario is different between models with $N_{\mathrm{DW}}=1$ and $N_{\mathrm{DW}}>1$.
Recently, we studied the cosmological implication of the models with $N_{\mathrm{DW}}=1$ through the careful numerical simulations~\cite{Hiramatsu:2012gg}.
In this paper, we turn our research into the models with $N_{\mathrm{DW}}>1$.

For the case $N_{\mathrm{DW}}>1$, it is well known that domain walls are stable and they eventually overclose the universe, which conflicts with the standard cosmology~\cite{Zeldovich:1974uw,Sikivie:1982qv}.
So far, several solutions to this problem are proposed. One is to introduce a small symmetry breaking parameter, called the bias~\cite{Gelmini:1988sf},
in the Lagrangian, which makes the walls unstable and leads to their annihilation at late times.
Another solution is to embed the discrete subgroup of $U(1)_{\mathrm{PQ}}$ into another continuous group~\cite{Lazarides:1982tw}.
In such models, the degenerate vacua are connected by symmetry transformations, leading to the same cosmological property with the case for $N_{\mathrm{DW}}=1$.
It is also possible to avoid the domain wall problem by assuming that inflation has occurred after the PQ symmetry breaking. In this case, the inflationary scale is constrained by
the observation of isocurvature fluctuations in the cosmic microwave background (CMB)~\cite{Lyth:1989pb}.
Among these possibilities, here we focus on the first solution, the biased domain walls.

It was pointed out that the bias term, which causes the annihilation of domain walls, can be found in effects of gravity~\cite{Rai:1992xw}.
If we regard the low energy field theory as an effective theory induced by a Planck-scale physics, we expect that the global symmetry
is violated by higher dimensional operators suppressed by the Planck mass $M_P$. This small symmetry violating operators cause the decay of domain walls.
However, in the case of PQ symmetry, the situation is more complicated.
It was argued that these Planck-scale induced terms easily violate the PQ symmetry and cause the CP violation~\cite{Kamionkowski:1992mf,Barr:1992qq,Dine:1992vx}.
In particular, the operators with dimension smaller than $d=10$ are forbidden~\cite{Barr:1992qq} by a requirement of CP conservation.
If it is true, the biased domain walls should be long-lived, and disappear at late time due to the effect of highly suppressed operators.

In our previous numerical study~\cite{Hiramatsu:2010yn}, we found that the late-time annihilation of domain walls actually occurs.
We also pointed out that the gravitational waves produced by domain walls may be detectable in the future experimental searches.
However, the previous results strongly depended on the production rate of axions and gravitational waves, which was treated as free parameters.
In this work, extending the previous study~\cite{Hiramatsu:2010yn}, we aim to determine the amount of radiated axions and gravitational waves precisely.
This requires the estimation of the spectra of radiated particles from domain walls, which can be investigated by performing field-theoretic numerical simulation of scalar fields.

In this paper, we study the formation and evolution of string-wall networks in the models with $N_{\mathrm{DW}}>1$
and calculate the spectra of axions and gravitational waves radiated from them.
This work might be regarded as the first study in which the particle radiations from long-lived domain wall networks are estimated  quantitatively.
Although we focus on the axionic models, part of our findings might be applicable to other particle physics models
which predict the existence of long-lived domain walls, such as domain walls formed by discrete R-symmetry breaking~\cite{Dine:2010eb},
thermal inflationary models~\cite{Moroi:2011be}, and $Z_3$ symmetry breaking in NMSSM~\cite{Hamaguchi:2011nm}.

The organization of this paper is as follows.
We introduce the model and review the property of topological defects in Sec.~\ref{sec2}.
In Sec.~\ref{sec3}, we show the results of numerical simulations.
Using the numerical results, we calculate the relic density of axions and gravitational waves in Sec.~\ref{sec4}.
We show that the overabundance of axions produced by domain walls gives severe constraints on the model parameters.
We also review other observational constraints. Finally, we make a summary in Sec.~\ref{sec5}.
The analysis method that we used in the numerical studies are described in Appendices~\ref{secA},~\ref{secB},~\ref{secC}, and~\ref{secD}.
We make a brief comment on the determination of the surface mass density of domain walls in Appendix~\ref{secE}.
 
In this paper, we work in spatially flat Friedmann-Robertson-Walker (FRW) universe with a metric
\begin{equation}
ds^2 = dt^2 - R^2(t)[dx^2+dy^2+dz^2], \nonumber
\end{equation}
where $R(t)$ is the scale factor of the universe.
We denote the cosmic time as $t$ and the conformal time as $\tau$, where $d\tau=dt/R(t)$.
A dot represents a derivative with respect to the cosmic time, while a prime represents a derivative with respect to the conformal time,
i.e. $\dot{ }=\partial/\partial t$, and $'=\partial/\partial\tau$.
\section{\label{sec2} Axion cosmology and topological defects}
\subsection{\label{sec2-1}Model of domain walls bounded by strings}
We consider the model of a complex scalar field $\Phi$ with the Lagrangian density given by
\begin{equation}
{\cal L} = \frac{1}{2}\partial_{\mu}\Phi^*\partial^{\mu}\Phi - V(\Phi), \label{eq2-1-1}
\end{equation}
where the scalar potential is given by
\begin{equation}
V(\Phi) = \frac{\lambda}{4}(|\Phi|^2-\eta^2)^2 + \frac{m^2\eta^2}{N_{\mathrm{DW}}^2}(1-\cos N_{\mathrm{DW}}\theta). \label{eq2-1-2}
\end{equation}
Here, $\theta$ is the angular part of the scalar field (i.e. $\Phi=|\Phi|e^{i\theta}$).
The first term in Eq.~(\ref{eq2-1-2}) is the usual mexican-hat potential, which causes spontaneous breaking of $U(1)_{\mathrm{PQ}}$ symmetry.
In the early universe, the $U(1)_{\mathrm{PQ}}$ symmetry is broken when the temperature of the universe falls below the energy scale $\sim\eta$.
At this time, the scalar field gets a vacuum expectation value $|\langle\Phi\rangle|^2 = \eta^2$, and cosmic strings are formed.
Then, the angular part of the scalar field can be identified as axion field $a$, $\Phi=\eta e^{i a/\eta}$.
The second term in Eq.~(\ref{eq2-1-2}) may arise due to the non-perturbative effect of QCD.
Because of the presence of this term, the $U(1)_{\mathrm{PQ}}$ symmetry is explicitly broken into its subgroup $Z_{N_{\mathrm{DW}}}$.
This term becomes non-negligible when the mass of the axion $m$ becomes greater than the friction term which comes from the cosmic expansion,
$m\gtrsim H$, where $H$ is the Hubble parameter. Then, the axion field begins to roll down to one of $N_{\mathrm{DW}}$ degenerate minima represented by
\begin{equation}
\theta = \frac{2\pi k}{N_{\mathrm{DW}}},\qquad k = 0, 1, \dots, N_{\mathrm{DW}}-1. \label{eq2-1-3}
\end{equation}
At this stage, domain walls are formed, separating the $N_{\mathrm{DW}}$ vacua~\cite{Sikivie:1982qv}.
These domain walls are attached to cosmic strings around their boundaries~\cite{2000csot.book.....V}.

We note that there are two different energy scales in this model. One is the scale of the PQ symmetry breaking, given by the axion decay constant
\begin{equation}
F_a \equiv \frac{\eta}{N_{\mathrm{DW}}}. \label{eq2-1-4}
\end{equation}
Phenomenologically, it takes a value $F_a\simeq 10^9$-$10^{12}\mathrm{GeV}$ (see e.g.~\cite{Kim:2008hd}).
Another is the QCD scale, $\Lambda_{\mathrm{QCD}}\simeq {\cal O}(100)\mathrm{MeV}$, which induces the second term of
the effective potential~(\ref{eq2-1-2}). The mass of the axion is given by $m\sim{\cal O}(0.1)\Lambda_{\mathrm{QCD}}^2/F_a$.
Since there is a large hierarchy between these two scales $F_a\gg \Lambda_{\mathrm{QCD}}$, the height of the potential barrier of the first term
$\sim\eta^4$ in Eq.~(\ref{eq2-1-2}) is much larger than that of the second term $\sim\Lambda_{\mathrm{QCD}}^4$.

Because of the existence of this large hierarchy, the formation of strings occurs much earlier than the formation of domain walls.
Once strings are formed, they evolve into the ``scaling" regime, in which the population of strings in the Hubble volume tends to remain in ${\cal O}(1)$.
In order to keep this scaling property, the strings lose their energy by emitting closed loops, and these loops decay by radiating
axions~\cite{Davis:1986xc,Davis:1989nj,Hagmann:1990mj,Hagmann:2000ja,Yamaguchi:1998gx,Hiramatsu:2010yu}.
Subsequently, domain walls are formed when the Hubble parameter becomes comparable to the mass of the axion $H\sim m$. Shortly after that, the condition
\begin{equation}
\sigma_{\mathrm{wall}}=\frac{\mu_{\mathrm{string}}(t_w)}{t_w} \label{eq2-1-5}
\end{equation}
is satisfied at the time $t_w$, where $\sigma_{\mathrm{wall}}=9.23mF_a^2$ is the surface mass density of domain walls [see Eq.~(\ref{eqE-7})],
$\mu_{\mathrm{string}}(t)\simeq \pi \eta^2\ln(t/\delta_s)$ is the mass energy of the strings per unit length, and $\delta_s\simeq 1/\sqrt{\lambda}\eta$
is the width of strings. After the time $t_w$, the dynamics is dominated by the tension of domain walls.

The fate of the string-wall networks is different between the case with $N_{\mathrm{DW}}=1$ and the case with $N_{\mathrm{DW}}>1$.
If $N_{\mathrm{DW}}=1$, domain walls bounded by strings are unstable, and they disappear immediately after the formation~\cite{Barr:1986hs}.
On the other hand, if $N_{\mathrm{DW}}>1$, the string-wall networks are stable, and they eventually dominate the energy density of the universe.
If it occurs at sufficiently late time in the universe, it may conflict with standard cosmology~\cite{Zeldovich:1974uw}.
\subsection{\label{sec2-2}Domain wall problem and its solution}
Now, let us look closer at the model with $N_{\mathrm{DW}}>1$.
After the time $t_w$ given by Eq.~(\ref{eq2-1-5}), domain walls are straightened by their tension force up to the horizon scale.
In the past numerical studies~\cite{Ryden:1989vj}, it was found that these networks of domain walls bounded by strings evolve into the scaling regime,
which is similar to the property of the networks of strings.
In this regime the typical length scales of defects, such as the wall curvature radius and the distance of two neighboring walls,
are given by the Hubble radius $H^{-1} = 2t$. Then, the energy density of domain walls behaves as
\begin{equation}
\rho_{\mathrm{wall}} \simeq \frac{\sigma_{\mathrm{wall}}}{H^{-1}} \propto t^{-1}. \label{eq2-2-1}
\end{equation}

Since the dilution of the energy density $\rho_{\mathrm{wall}}$ is slower than that of matters $\sim 1/R^3$ and radiations $\sim1/R^4$,
domain walls eventually dominate the energy density of the universe. This occurs at the time
\begin{equation}
t_{\mathrm{WD}} = \frac{3}{16\pi G\sigma_{\mathrm{wall}}}, \label{eq2-2-2}
\end{equation}
where $G$ is the Newton's gravitational constant.
In other words, the wall domination occurs when the temperature of the universe becomes less than $T\sim{\cal O}(10)\mathrm{keV}$~\cite{Hiramatsu:2010yn}.

In order to avoid this overclosure problem, Sikivie~\cite{Sikivie:1982qv} phenomenologically introduced a term
\begin{equation}
\delta V = -\Xi \eta^3(\Phi e^{-i\delta}+{\mathrm{h.c.}}) \label{eq2-2-3}
\end{equation}
in the potential~(\ref{eq2-1-2}), where $\Xi$ is a dimensionless parameter which is assumed to be much less than unity\footnote{The parameter
$\Xi$ corresponds to what we called as $\xi$ in our previous study~\cite{Hiramatsu:2010yn}. Here we do not use this old notation
in order to avoid the confusion with the scaling parameter defined by Eq.~(\ref{eq3-1-1}).}.
This term, called a bias~\cite{Gelmini:1988sf}, explicitly breaks the discrete symmetry and lifts degenerate vacua represented by Eq.~(\ref{eq2-1-3}).
If this kind of term exists in the potential, it affects as a pressure on walls, which eventually annihilate them.
We can naively estimate the lifetime of the networks $t_{\mathrm{dec}}$ by the ratio between the surface mass density of domain
walls $\sigma_{\mathrm{wall}}$ and the difference of the potential energy between two neighboring vacua~\cite{Hiramatsu:2010yn},
\begin{equation}
t_{\mathrm{dec}} \sim \frac{\sigma_{\mathrm{wall}}}{\Xi\eta^4/N_{\mathrm{DW}}}. \label{eq2-2-4}
\end{equation}
Requiring that the decay of walls occurs before the wall domination (i.e. $t_{\mathrm{dec}}<t_{\mathrm{WD}}$),
we obtain a lower bound on $\Xi$
\begin{equation}
\Xi > 7.2\times10^{-60}\times N_{\mathrm{DW}}^{-3}\left(\frac{m}{6\times 10^{-4}\mathrm{eV}}\right)^2. \label{eq2-2-5}
\end{equation}

In our previous study~\cite{Hiramatsu:2010yn}, we found that string-wall networks actually disappears in the time scale given by Eq.~(\ref{eq2-2-4})
and pointed out that significant amounts of gravitational waves might be produced by these defect networks.
We also found that the cosmological abundance of cold axions produced by string-wall networks gives another constraint on the model parameters.
In this work, we are going to give a concrete analysis of gravitational waves and axion radiations from these string-wall networks by numerically computing the spectrum of these radiations.
\section{\label{sec3} Numerical simulations}
We performed field-theoretic lattice simulations with the model of PQ scalar field $\Phi$ whose potential given by Eq.~(\ref{eqA-1-1}).
We use the three dimensional homogeneous grid with the size $N^3=512^3$.
In the numerical studies, we normalize the dimensionful quantities in the unit of $\eta$. For example, $x\to x\eta$, $\Phi\to\Phi/\eta$, etc.
We choose the comoving size of the simulation box as $L=80$ in this unit. Then, the spatial resolution is estimated as $\Delta x = L/N\simeq0.16$.
We give the time evolution in terms of the conformal time $\tau$ rather than the cosmic time $t$.
We set the initial time as $\tau_i=2.0$ and the final time as $\tau_f=40.0$. The dynamical range of the simulation is therefore $\tau_f/\tau_i=20$.
We normalize the scale factor at the initial time $R(\tau_i)=1$.
The other parameters are summarized in table~\ref{tab1}.
The technical details on the analysis method are described in Appendices~\ref{secA},~\ref{secB},~\ref{secC}, and~\ref{secD}.

\begin{table}[h]
\begin{center} 
\caption{Parameters used in numerical simulations.}
\vspace{3mm}
\begin{tabular}{c c}
\hline\hline
Grid size ($N$) & 512 \\
Box size ($L$) & 80 \\
Total number of time steps & 1900 \\
Time interval ($\Delta\tau$) & 0.02 \\
$\lambda$ & 0.1 \\
$m$ & 0.1 \\
$N_{\mathrm{DW}}$ & varying \\
Initial time ($\tau_i$) & 2.0 \\
Final time ($\tau_f$) & 40.0 \\
\hline\hline
\label{tab1}
\end{tabular}
\end{center}
\end{table}

With the parameters shown in table~\ref{tab1}, the physical scale of the Hubble radius $H^{-1}$, the width of the wall $\delta_w=m^{-1}$, and the width of the string $\delta_s=\lambda^{-1/2}$
divided by the lattice spacing $\Delta x_{\mathrm{phys}}=R(t)\Delta x=R(t)L/N$ are given by
\begin{equation}
\frac{H^{-1}}{\Delta x_{\mathrm{phys}}} = \frac{N\tau}{L}, \qquad \frac{\delta_w}{\Delta x_{\mathrm{phys}}} = \frac{N}{Lm}\left(\frac{\tau_i}{\tau}\right), \qquad \mathrm{and}
\qquad \frac{\delta_s}{\Delta x_{\mathrm{phys}}} = \frac{N}{L\lambda^{1/2}}\left(\frac{\tau_i}{\tau}\right). \label{eq3-1}
\end{equation}
At the end of the simulation $\tau_f = 40$, these ratios become $H^{-1}/\Delta x_{\mathrm{phys}} \simeq256<N$, $\delta_w/\Delta x_{\mathrm{phys}} \simeq3.2$, and $\delta_s/\Delta x_{\mathrm{phys}}\simeq1.01$.
Therefore, these length scales are marginally resolvable at the final time of the simulation.

We naively guess that domain walls formed in the model with $N_{\mathrm{DW}}=2$ have a common property with that formed in
the model of real scalar field with spontaneous breaking of a discrete $Z_2$ symmetry. 
Based on this observation, we also consider the real scalar field theory with the Lagrangian density given by
\begin{eqnarray}
{\cal L} &=& \frac{1}{2}\partial_{\mu}\phi\partial^{\mu}\phi - V(\phi), \label{eq3-2}\\
V(\phi) &=& \frac{m^2}{3\eta^2}\left(\phi^2-\frac{3}{2}\eta^2\right)^2, \label{eq3-3}
\end{eqnarray}
where $\phi$ is the real scalar field. In Appendix~\ref{secE}, we show that this model mimics the axionic model with $N_{\mathrm{DW}}=2$,
in the sense that the surface mass density and the width of domain walls are same.

It should be emphasized that we do not perform the simulation including the bias term~(\ref{eq2-2-3}).
This means that we do not simulate the annihilation process of topological defects.
In order to simulate the annihilation process, much larger dynamical range is required, since each of $N_{\mathrm{DW}}$ vacua is not homogeneously percolated
in the simulation with the small dynamical range~\cite{Hiramatsu:2010yn}.
In our previous study~\cite{Hiramatsu:2010yn}, we could improve dynamical range by performing two-dimensional lattice simulations where the size of data is given by $N^2$.
In the present work, our aim is to investigate the spectrum of axions and gravitational waves produced by defect networks, while
it is impossible to calculate the gravitational waves properly in the two-dimensional field theoretic simulations.
Therefore, we just calculate the spectra in the onset of the scaling regime, and extrapolate the results for the case in which string-wall networks
survive for a long time. Then we use the results of two dimensional simulations of the annihilation process of the networks~\cite{Hiramatsu:2010yn}
when we estimate the decay time of them [see Eq.~(\ref{eq4-1-16})].
\subsection{\label{sec3-1} Evolution of string-wall networks}
Now let us show the results of numerical simulations.
We show the visualization of one realization of the simulation in Fig.~\ref{fig1}.
Using the identification scheme described in Appendix~\ref{secB}, we confirmed that $N_{\mathrm{DW}}$ domain walls attached to strings
are formed at late time of the simulations. The structure of defect networks becomes more complicated when we increase the number $N_{\mathrm{DW}}$.
Figure~\ref{fig2} shows the spatial distribution of the phase of the scalar field $\theta$ in the model with $N_{\mathrm{DW}}=3$ at selected time slices.
Since we give the initial configuration of the scalar field as Gaussian random amplitude (see Appendix~\ref{secA-3}),
the scalar field randomly oscillates around the minima $|\Phi|^2=\eta^2$ of the potential, given by the first term of Eq.~(\ref{eq2-1-2}).
Strings are formed when this initial random configuration becomes relaxed. Subsequently, the value of the phase of the scalar field $\theta$
is separated into $N_{\mathrm{DW}}=3$ domains. Domain walls are located around the boundaries of these domains.

\begin{figure}[htbp]
\centering
$\begin{array}{cc}
\subfigure[$N_{\mathrm{DW}}=2$]{
\includegraphics[width=0.45\textwidth]{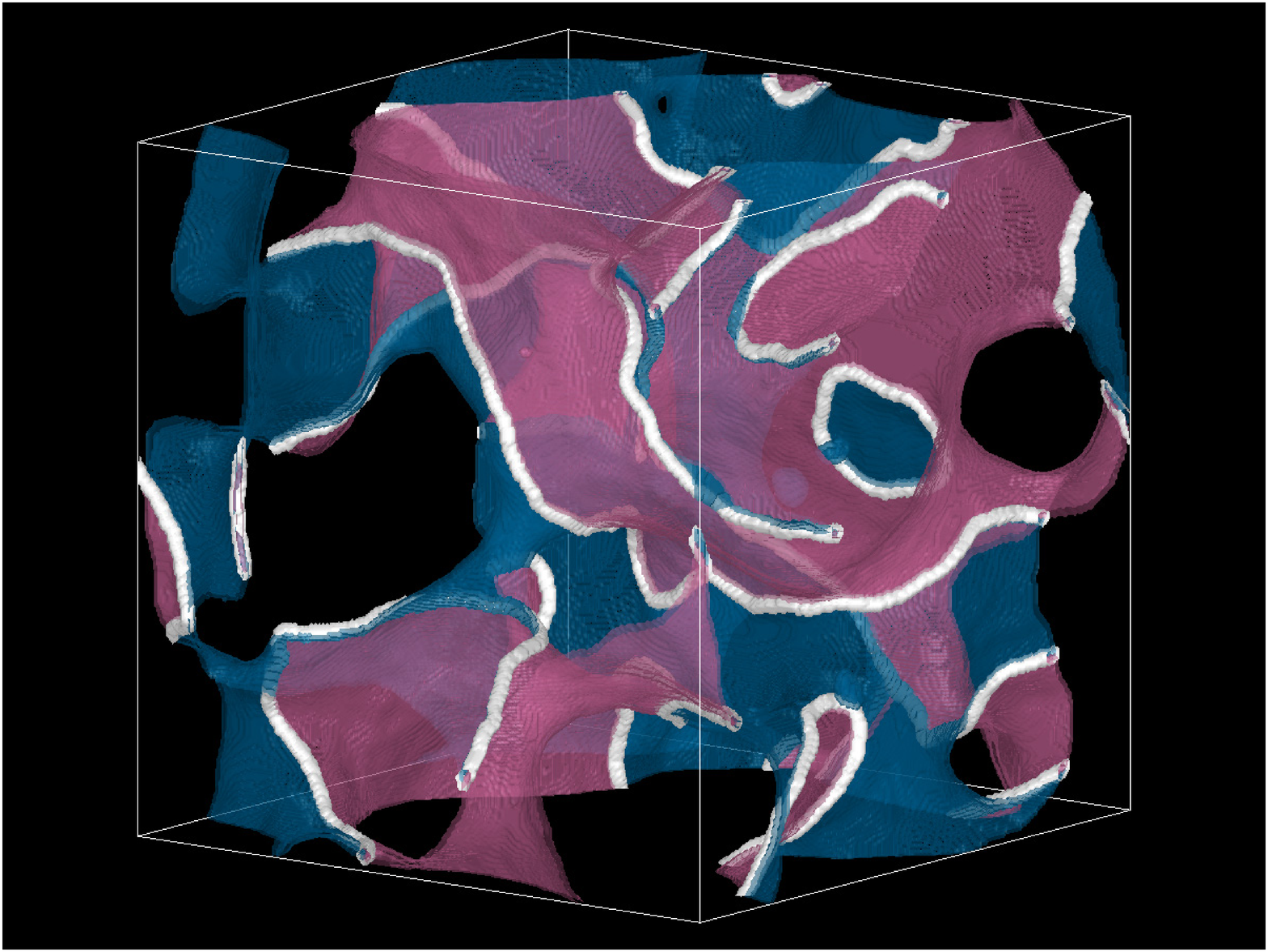}}
\hspace{20pt}
\subfigure[$N_{\mathrm{DW}}=3$]{
\includegraphics[width=0.45\textwidth]{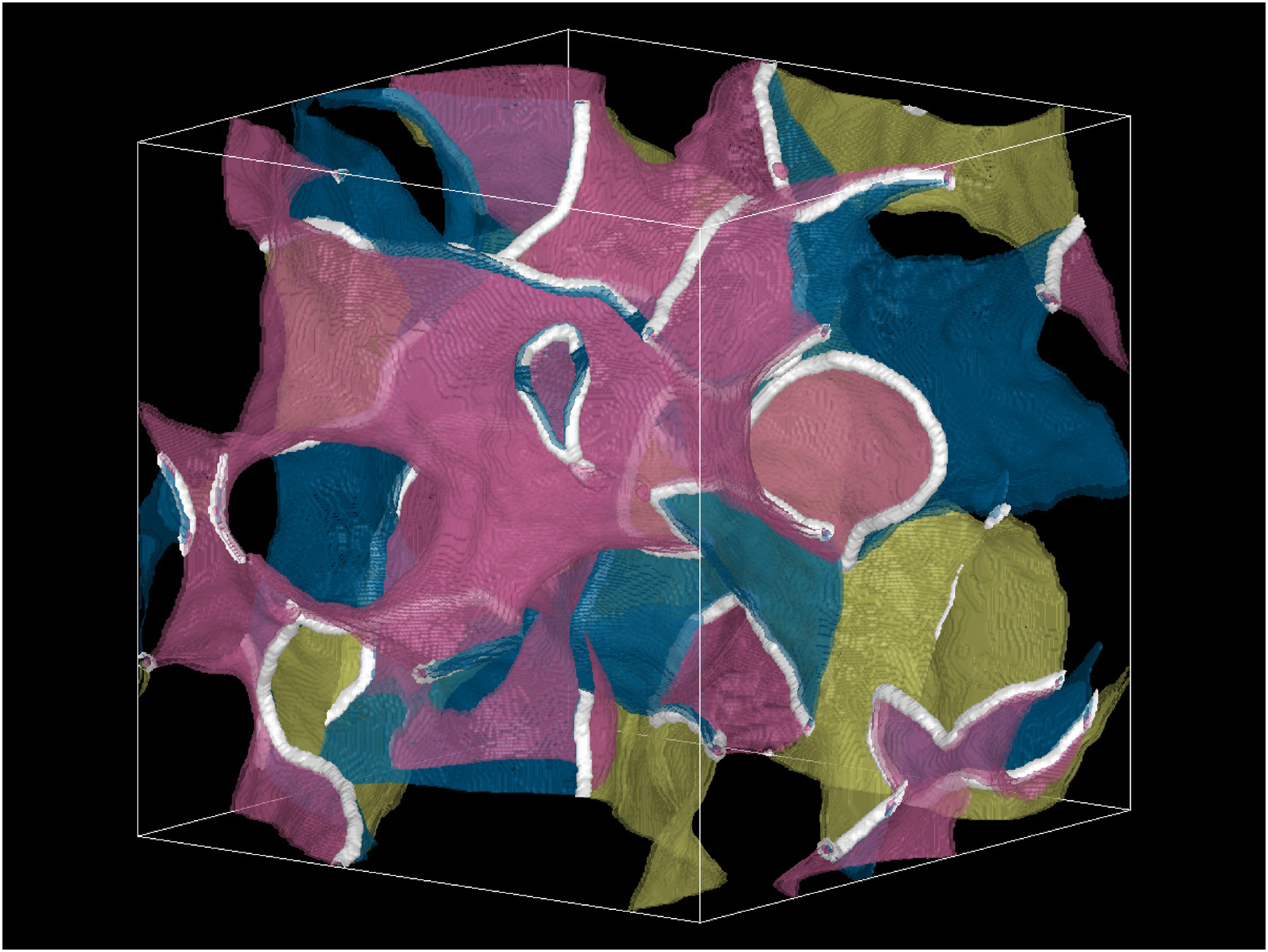}}
\\
\subfigure[$N_{\mathrm{DW}}=4$]{
\includegraphics[width=0.45\textwidth]{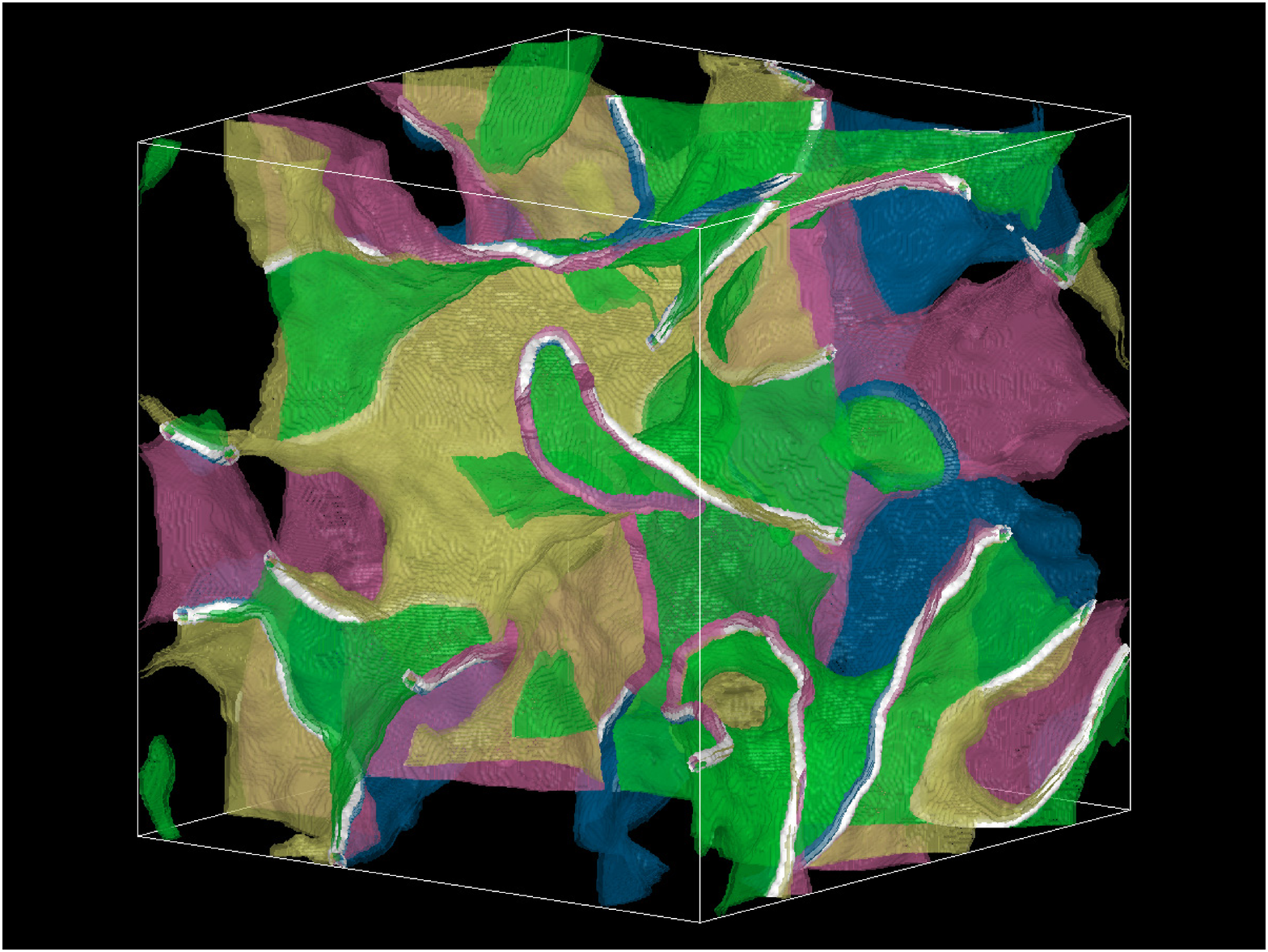}}
\hspace{20pt}
\subfigure[$N_{\mathrm{DW}}=5$]{
\includegraphics[width=0.45\textwidth]{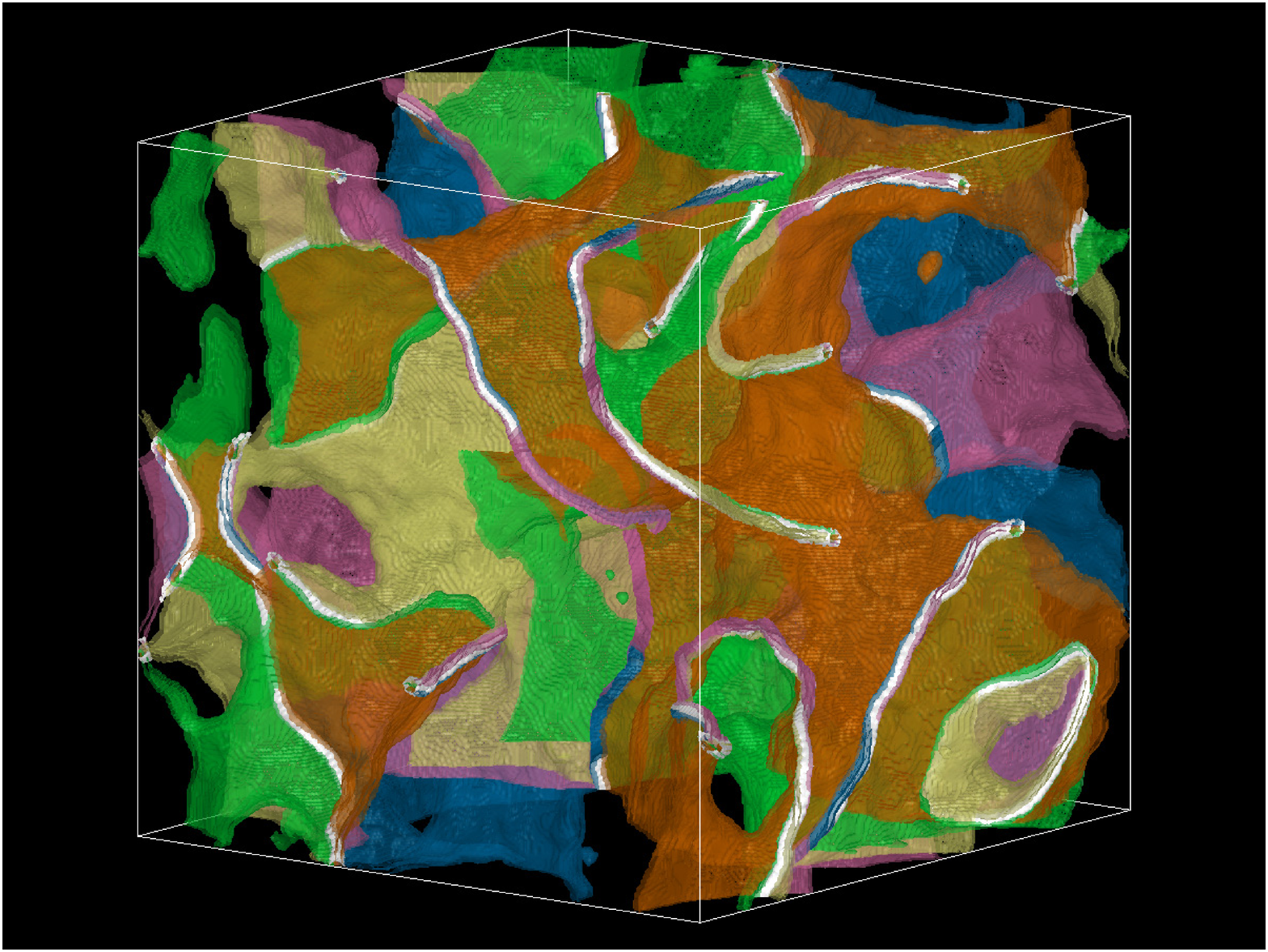}}
\end{array}$
\subfigure[$N_{\mathrm{DW}}=6$]{
\includegraphics[width=0.45\textwidth]{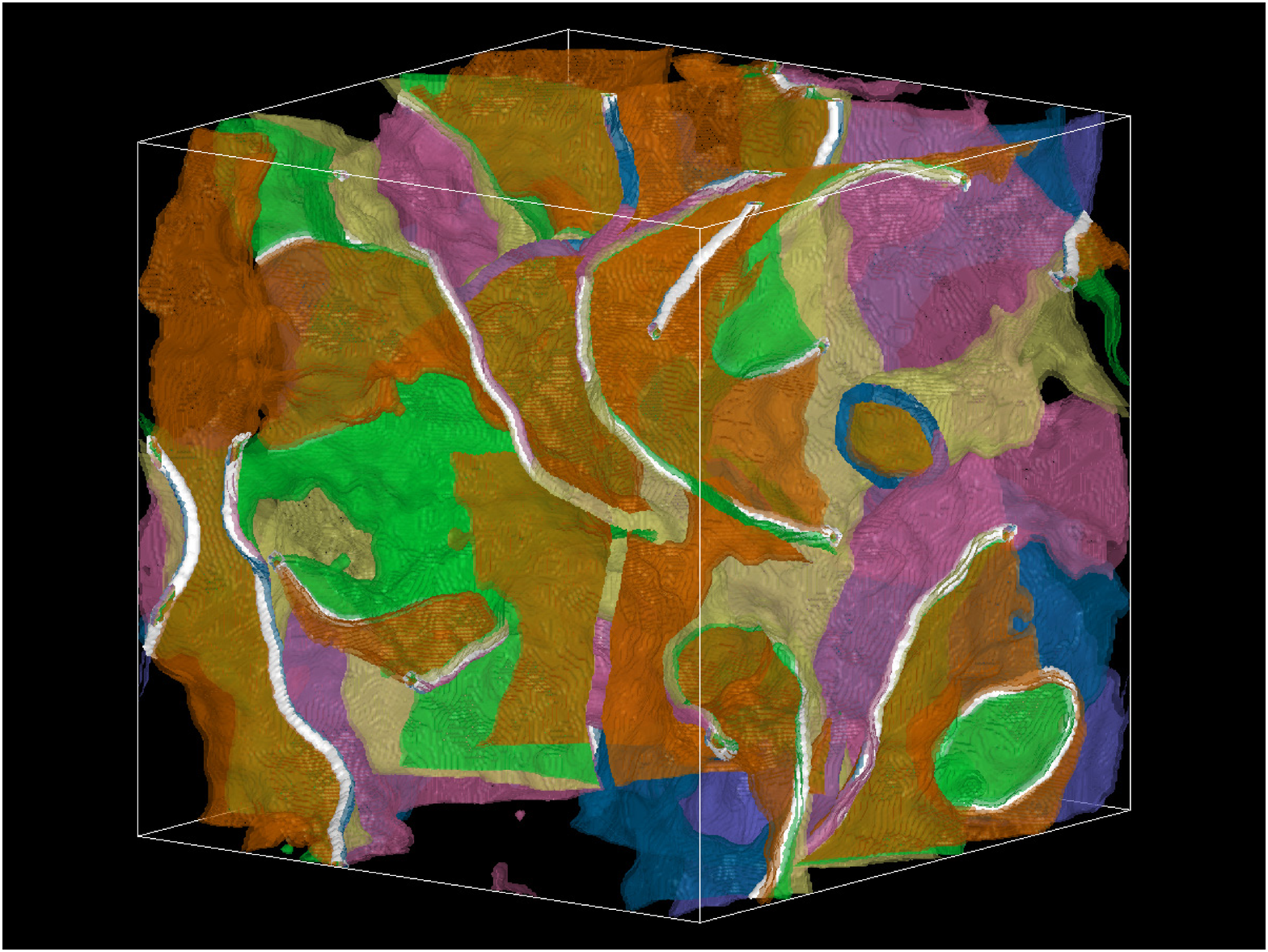}}
\caption{Visualization of the simulations with (a) $N_{\mathrm{DW}}=2$, (b) $N_{\mathrm{DW}}=3$, (c) $N_{\mathrm{DW}}=4$,
(d) $N_{\mathrm{DW}}=5$, and (e) $N_{\mathrm{DW}}=6$. In this figure, we take the box size as $L=60$ and $N=256$, which is smaller than
that shown in table~\ref{tab1}. Each figure shows the spatial configurations of topological defects at the time $\tau=25$.
The white lines correspond to the position of the core of strings, which is identified by using the method described in Appendix~\ref{secB-1}.
$N_{\mathrm{DW}}$ domain walls are represented by surfaces with various colors, which are identified by using the method described in Appendix~\ref{secB-2}.}
\label{fig1}
\end{figure}

\begin{figure}[htbp]
\centering
$\begin{array}{cc}
\subfigure[$\tau=7$]{
\includegraphics[width=0.45\textwidth]{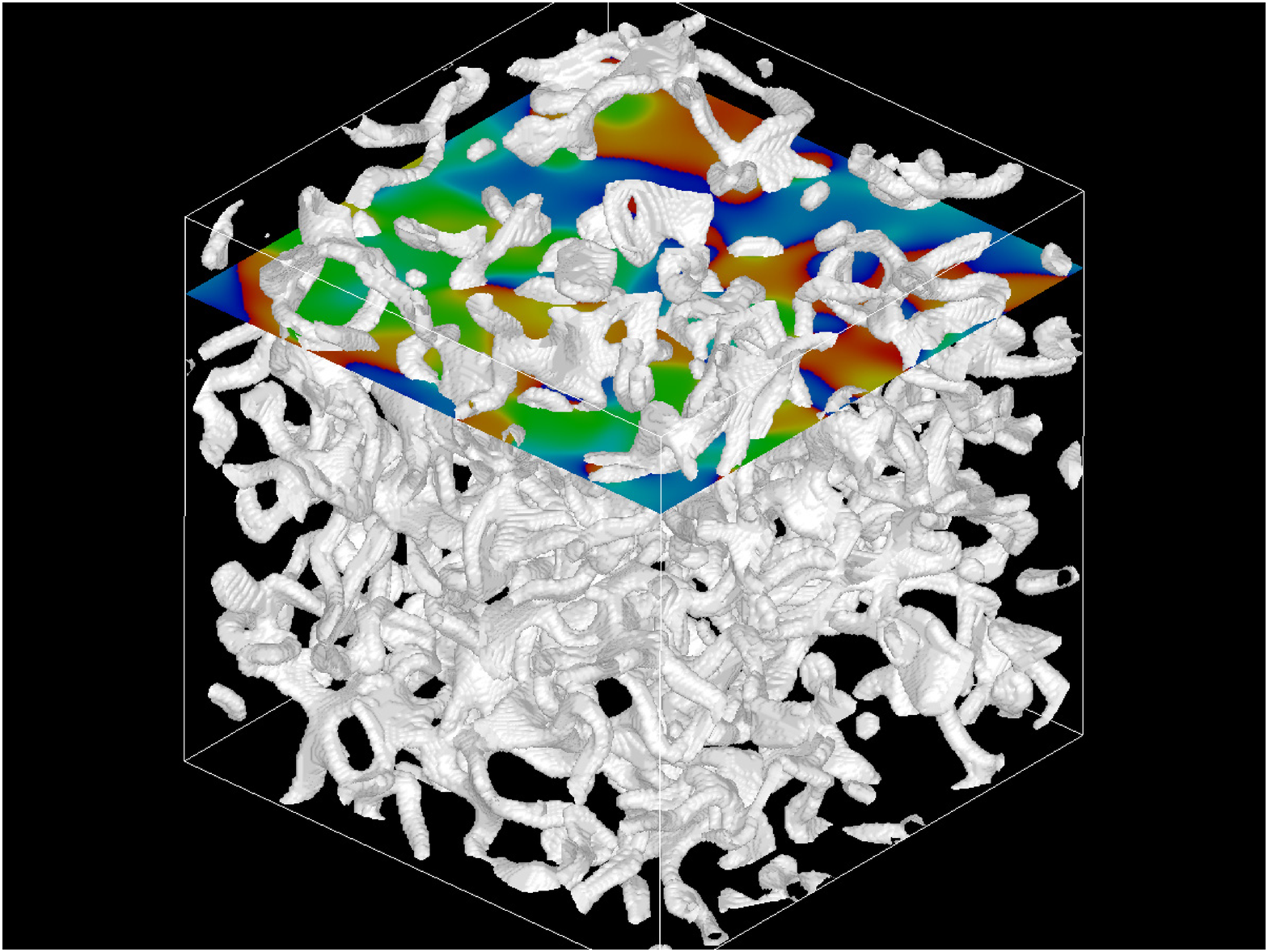}}
\hspace{20pt}
\subfigure[$\tau=13$]{
\includegraphics[width=0.45\textwidth]{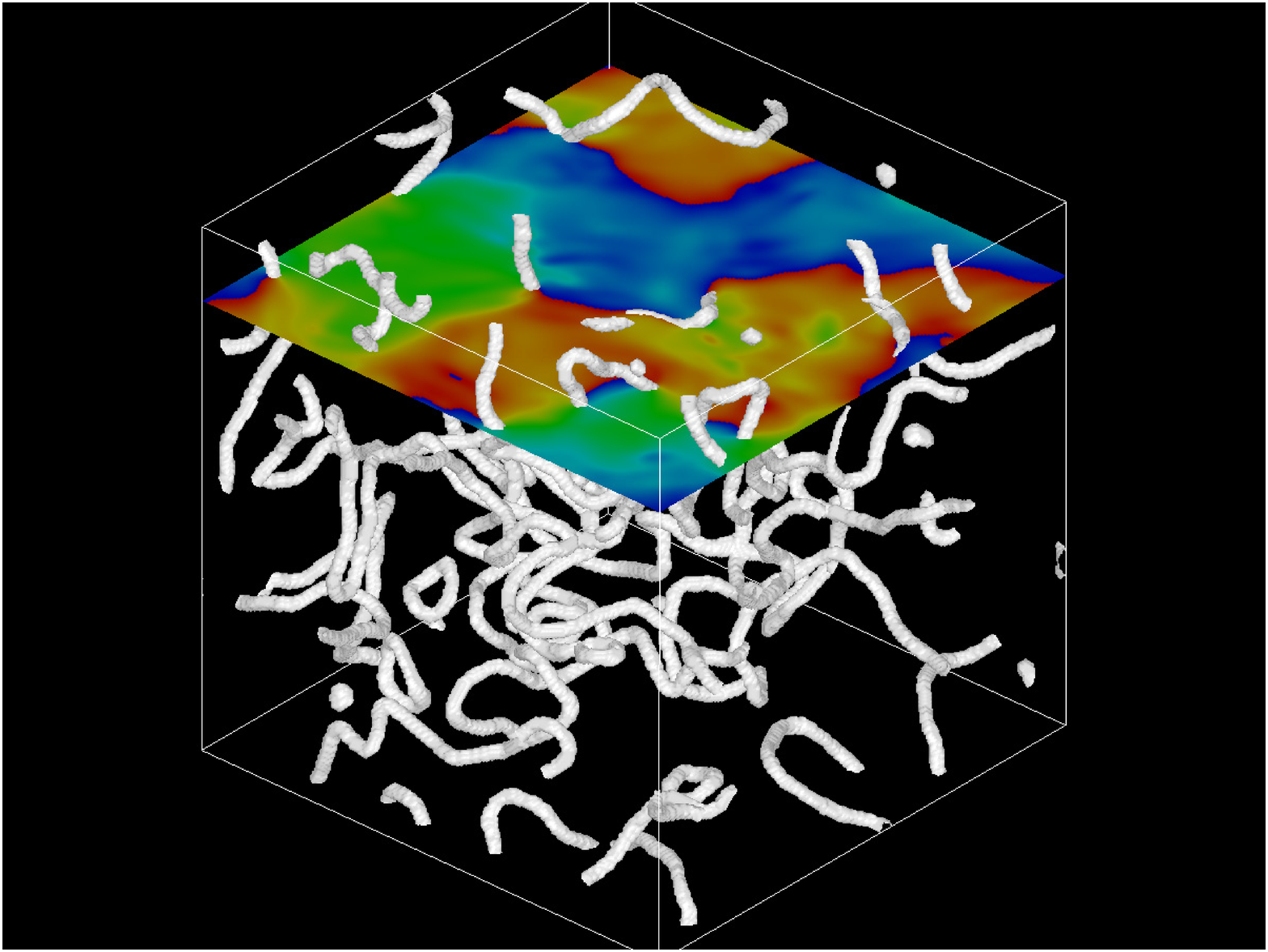}}
\\
\subfigure[$\tau=19$]{
\includegraphics[width=0.45\textwidth]{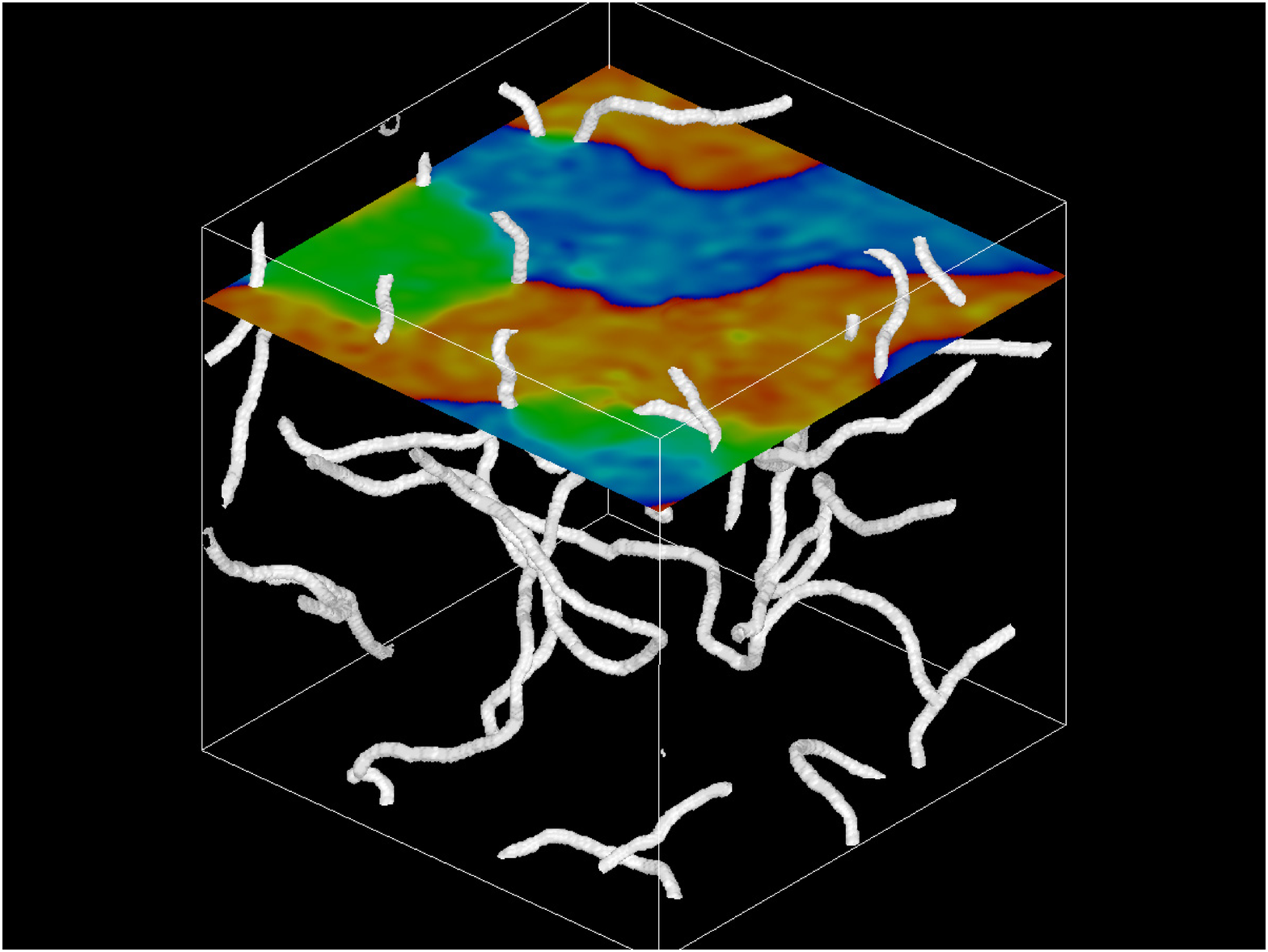}}
\hspace{20pt}
\subfigure[$\tau=25$]{
\includegraphics[width=0.45\textwidth]{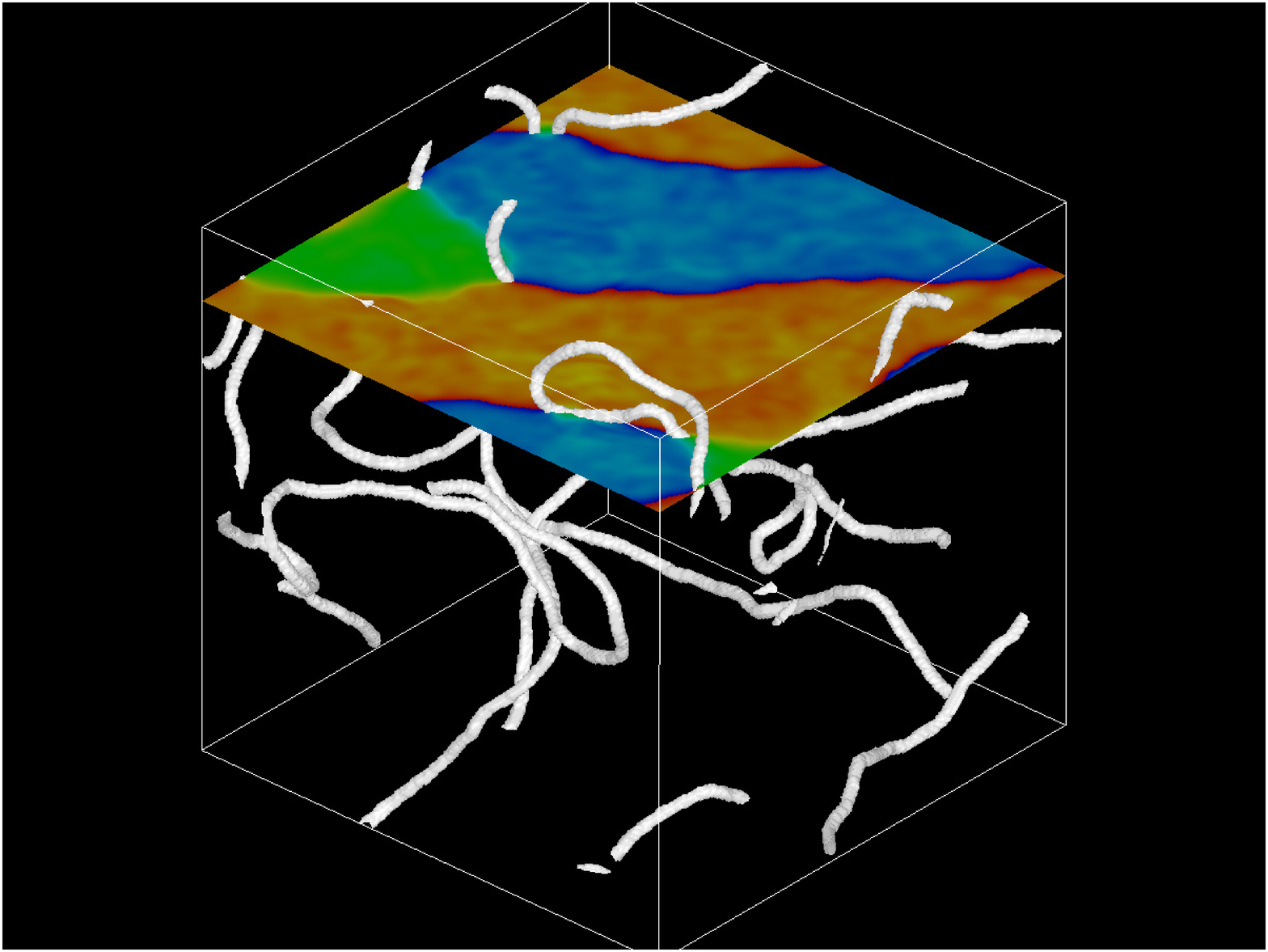}}
\end{array}$
\caption{The distribution of the phase of the scalar field $\theta$ on the two-dimensional slice of the simulation box.
Each figure shows the visualization at different times: (a) $\tau=7$, (b) $\tau=13$, (c) $\tau=19$, and (d) $\tau=25$.
In these figures, we used the same data that are used to visualize the result with $N_{\mathrm{DW}}=3$ in figure~\ref{fig1}.
The value of $\theta$ varies from $-\pi$ (blue) to $\pi$ (red).
At late times, the value of $\theta$ is separated into three domains represented by blue, red, and green region.
Domain walls are located around the boundary of these three regions, $\theta=\pi/3$, $\theta=-\pi/3$, and $\theta=\pm\pi$.
Strings, which are represented by white lines, pass through the point where three regions meet each other.}
\label{fig2}
\end{figure}

We note that the initial stage of simulations does not correctly follow the history of the universe, since domain walls are formed immediately after
the formation of strings. As we noted in Sec.~\ref{sec2-1}, in the realistic case, strings are formed much earlier than the formation of domain walls.
This is due to the hierarchy between two different scales, $\Lambda_{\mathrm{QCD}}/F_a\sim 10^{-11}$.
On the other hand, in our choice of the parameters in numerical simulations, this ratio becomes as small as $\Lambda_{\mathrm{QCD}}/F_a\sim \sqrt{m}\sim{\cal O}(0.1)$
(recall that we used the unit $\eta=1$).
Because of the limitation of the dynamical range of the numerical simulations, we cannot choose this ratio as an arbitrary small value.
Therefore, we concentrate on the evolution at late times where the string-wall networks evolve along to the scaling regime,
and ignore the effect of the initial relaxation regime. We will discuss this point again in Secs.~\ref{sec3-2} and \ref{sec3-3}.

\begin{figure}[htbp]
\begin{center}
\begin{tabular}{c c}
\resizebox{75mm}{!}{\includegraphics[angle=0]{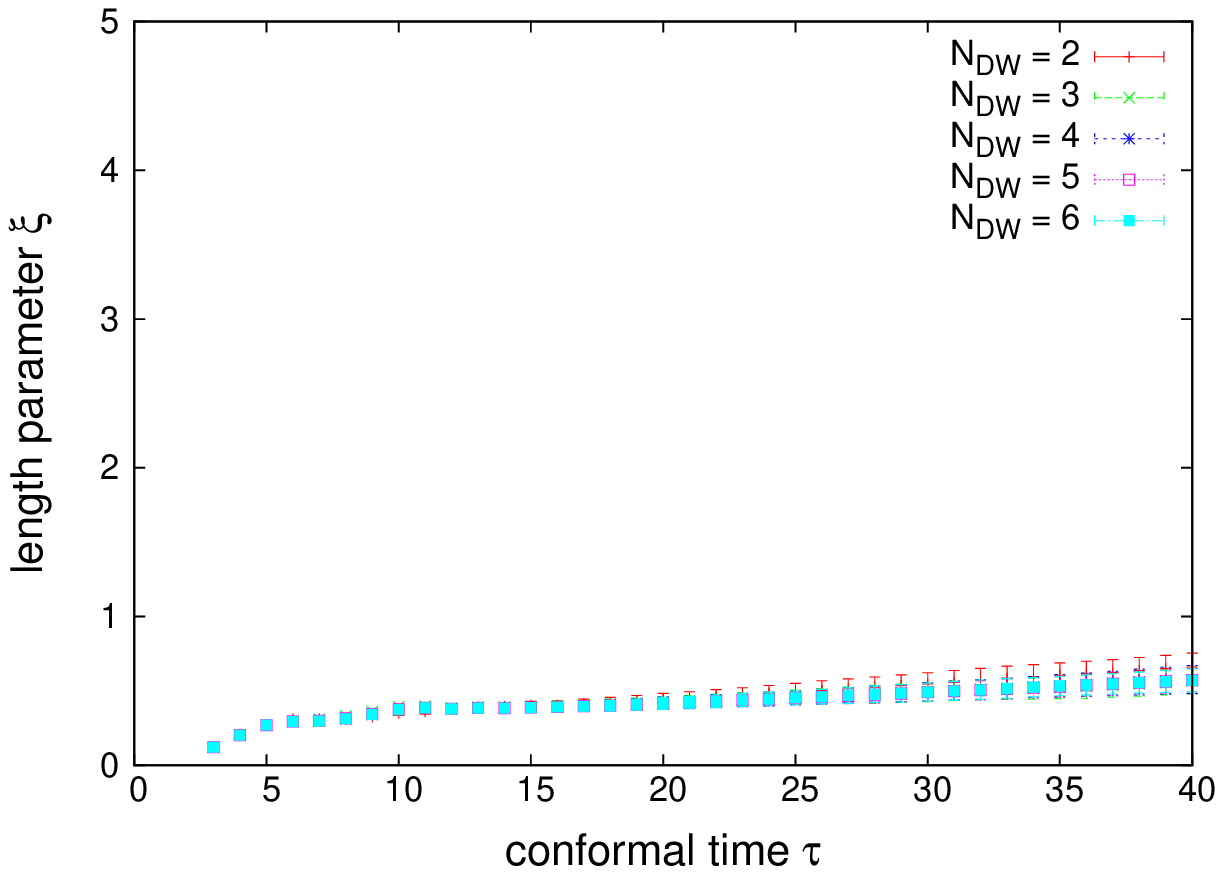}} &
\resizebox{75mm}{!}{\includegraphics[angle=0]{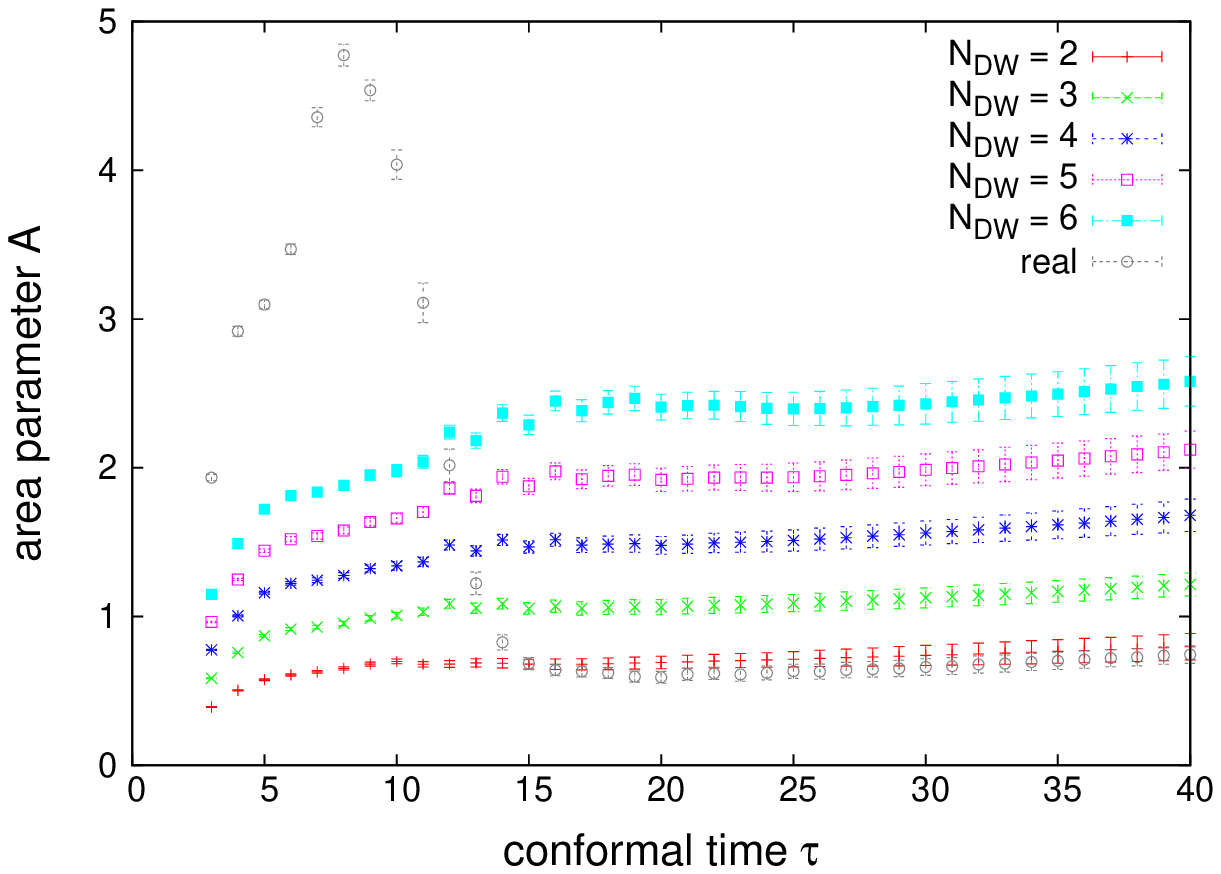}} \\
\end{tabular}
\end{center}
\caption{Time evolution of the length parameter $\xi$ (left panel) and the area parameter ${\cal A}$ (right panel)
for various values of $N_{\mathrm{DW}}$. The points denoted as ``real" in the right panel show the evolution of ${\cal A}$ in the model with
real scalar field defined by Eqs.~(\ref{eq3-2}) and (\ref{eq3-3}).}
\label{fig3}
\end{figure}

We define the length parameter $\xi$ of strings
\begin{equation}
\xi \equiv \frac{\rho_{\mathrm{string}}}{\mu_{\mathrm{string}}}t^2, \label{eq3-1-1}
\end{equation}
and the area parameter ${\cal A}$ of domain walls
\begin{equation}
{\cal A} \equiv \frac{\rho_{\mathrm{wall}}}{\sigma_{\mathrm{wall}}}t. \label{eq3-1-2}
\end{equation}
These parameters take a constant value when the defect networks enter the scaling regime (i.e. $\rho_{\mathrm{string}}\sim1/t^2$ and $\rho_{\mathrm{wall}}\sim 1/t$).
Henceforth we refer to these parameters as the ``scaling parameters". We plot the time evolution of these parameters in Fig.~\ref{fig3}.
From this figure, we see that the defect networks are in scaling regime in the time scale $\tau\gtrsim 15$.
The length parameter $\xi$ does not strongly depend on $N_{\mathrm{DW}}$, and its value is consistent with $\xi\simeq 0.5$-$1.3$ obtained
in previous studies~\cite{Yamaguchi:1998gx,Yamaguchi:1999yp,Hiramatsu:2010yu,Hiramatsu:2012gg}.
On the other hand, the value of the area parameter ${\cal A}$ increases for large $N_{\mathrm{DW}}$.
This fact agrees with the intuitive argument that the number of domain walls increases proportionally to $N_{\mathrm{DW}}$ if the number of strings per simulation box is unchanged.

The right panel of Fig.~\ref{fig3} also implies that the area of domain walls in the real scalar field theory given by Eqs.~(\ref{eq3-2}) and (\ref{eq3-3})
evolves similarly to the axionic model with $N_{\mathrm{DW}}=2$.
This means that the behavior of the networks for the case with $N_{\mathrm{DW}}=2$ is similar to that produced by the model with $Z_2$ symmetry,
except that they contain strings on the surface of walls.
We note that the bump-like behavior of ${\cal A}$ in the real scalar field model at the initial stage of the simulation should be regarded as a numerical fake.
We identify the position of domain walls as the point on which scalar field changes its sign in the model with the real scalar field,
which is different from the identification scheme used for the model with the complex scalar field (see Appendix~\ref{secB-2}).
These different identification schemes cause the different behavior of ${\cal A}$ at the initial stage, where domain walls do not relax into the scaling regime.

\subsection{\label{sec3-2} Production of axions}
From data of the scalar field obtained by numerical simulations, we compute the power spectrum $P(k)$ of axions radiated by string-wall networks, which is defined by
\begin{equation}
\frac{1}{2}\langle\dot{a}(t,{\bf k})^*\dot{a}(t,{\bf k}')\rangle = \frac{(2\pi)^3}{k^2}\delta^{(3)}({\bf k}-{\bf k}')P(k,t), \label{eq3-2-1}
\end{equation}
where $\dot{a}(t,{\bf k})$ is the Fourier component of the time derivative of the axion field, and $\langle\dots\rangle$ represents an ensemble average.
Here, we mask the position where topological defects exist, and extract the pure radiation component in the axion field produced by the defect networks.
See Appendix~\ref{secC} for details on the analysis method.

\begin{figure}[htbp]
\begin{center}
\includegraphics[scale=1.0]{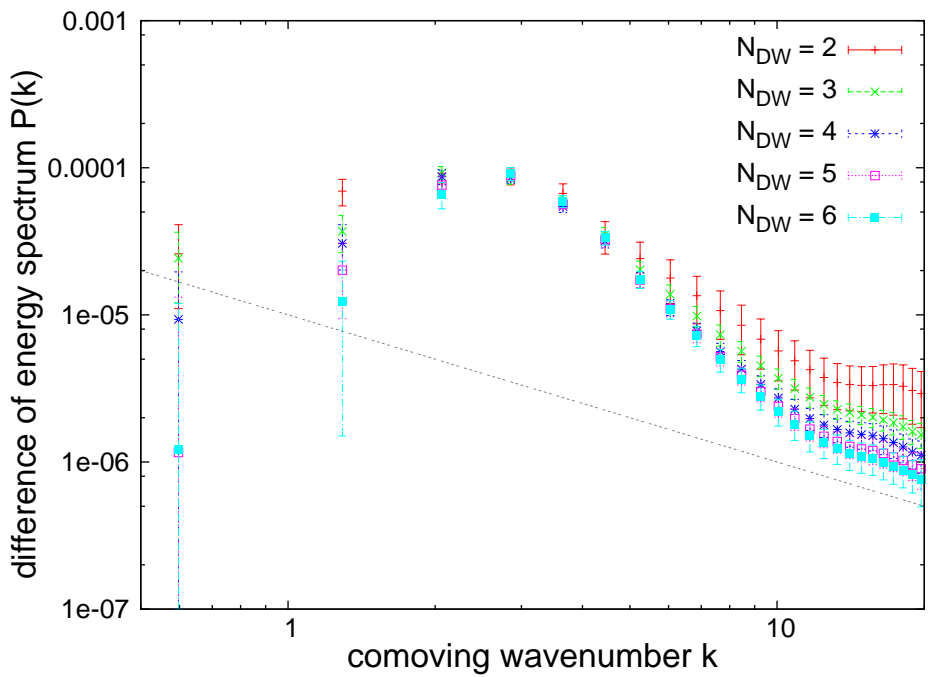}
\end{center}
\caption{The spectra of axions produced by string-wall networks for various values of $N_{\mathrm{DW}}$.
These spectra are obtained by using the masking analysis method described in Appendix~\ref{secC}.
The dotted line represents the power law $P(k) \propto 1/k$.
The spectra shown in this figure are the difference of power spectra evaluated at two different time steps $\tau_1$ and $\tau_2$ [see Eq.~(\ref{eqC-12})].
Here, we choose $\tau_1=14$ and $\tau_2=40$.}
\label{fig4}
\end{figure}

Figure~\ref{fig4} shows the results for $P(k,t_f)$ for various values of $N_{\mathrm{DW}}$ evaluated at the final time of the simulation.
We find that the spectrum has a peak around the momentum scale determined by the mass of the axion, $k/R(t_f)\sim m$, or $k\sim R(t_f)m\sim2$.
This implies that most of the axions are produced by the self-interaction of domain walls whose width is given by the inverse of the mass of the axion $\sim m^{-1}$.
However, this peak does not seem to fall off, but makes a tail at higher momenta.
This tail-like feature was also found in another numerical study~\cite{Hiramatsu:2012gg} in which we investigated the spectrum of axions produced by the annihilation of $N_{\mathrm{DW}}=1$ domain walls
bounded by strings. The appearance of this high-frequency tail can be understood in terms of the conjecture made in Refs.~\cite{Harari:1987ht,Chang:1998tb,Hagmann:1990mj,Hagmann:2000ja}.
The strings whose radius smaller than the horizon size quickly decay due to the tension of domain walls in the time scale comparable to (or less than) the Hubble time.
This fast process does not change much the energy spectrum of the axion field
composing the string configuration, which has a form $dE/dk\sim 1/k$ with a high-momentum cutoff of order $\delta_s^{-1}$
(the width of strings) and a low-momentum cutoff of order $t^{-1}$ (the Hubble radius)~\cite{Hagmann:1990mj}.
This $1/k$ behavior makes a tail in the spectrum of radiated axions, as we see in Fig.~\ref{fig4}.
This argument cannot be applied for the spectrum of axions produced by global strings~\cite{Yamaguchi:1998gx,Hiramatsu:2010yu}, which has a sharp peak at the horizon scale
since in the absence of domain walls the strings lose their energy in the time scale much longer than the Hubble time.
The high-momentum cutoff which is given by the width of the string $\sim \delta_s^{-1}$ is not clearly shown in Fig.~\ref{fig4} because of the poor resolution at the final time of the simulation.

Note that the amplitude of the high-frequency tail becomes small if we increase the domain wall number $N_{\mathrm{DW}}$.
In the case of large $N_{\mathrm{DW}}$, the decay of loops of strings with small curvature radius might be ``frustrated", because of the tension of domain walls
acting on the strings from various directions. Then, the radiation of hard axions with spectrum $\sim 1/k$ is suppressed, compared with the case of small $N_{\mathrm{DW}}$.

We emphasize that our conclusion will be different from that made in~\cite{Chang:1998tb} in support of~\cite{Hagmann:1990mj,Hagmann:2000ja}.
In Ref.~\cite{Chang:1998tb} it was claimed that the spectrum of axions radiated by domain walls behaves as $\sim1/k$ in the whole frequency domain, and hence
the mean energy of radiated axions is proportional to $\ln(\sqrt{\lambda}\eta/m)\sim 60$, which suppresses the relic abundance of cold dark matter axions.
Our result, however, shows that the spectrum does not behave like $\sim 1/k$ except for the high-frequency modes,
and most axions are mildly relativistic, making a peak at the low frequency.
Hence there is no correction like $\ln(\sqrt{\lambda}\eta/m)\sim 60$, and the relic abundance of axions turns out to be large compared with the result of~\cite{Chang:1998tb}.
This fact leads to severe constraints on the axion models, as we will see in Sec.~\ref{sec4}.

The above observation can be confirmed by estimating the energy of radiated axions.
Roughly speaking, the product $P(k)k$ represents the fraction of the energy density of axions with comoving wavenumber $k$ against the total energy of axions.
Hence the ratio of the energy between hard axions which contribute to the tail of the power spectrum and the energy of axions which contribute to the peak of the power spectrum can be estimated as 
${\cal E}_{\mathrm{hard}}/{\cal E}_{\mathrm{peak}}\sim [P(k_{\mathrm{hard}})k_{\mathrm{hard}}]/[P(k_{\mathrm{peak}})k_{\mathrm{peak}}]\sim [{\cal O}(10^{-6})\times{\cal O}(10)]/[{\cal O}(10^{-4})\times{\cal O}(1)]\sim 10^{-1}$.
On the other hand, the ratio between the energy of domain walls and the energy of strings at the final time of the simulations becomes
${\cal E}_{\mathrm{string}}/{\cal E}_{\mathrm{wall}}\sim[t_f\mu_{\mathrm{string}}]/[t_f^2\sigma_{\mathrm{wall}}]\sim 10^{-1}$.
Therefore, strings have $10\%$ of the energy of domain walls at the final time, which contribute to the hard component $\sim 1/k$ of the power spectrum of radiated axions.
Since the fraction ${\cal E}_{\mathrm{string}}/{\cal E}_{\mathrm{wall}}$ decreases with time, we expect that the amplitude of the high-frequency tail would be
negligible compared with the height of the peak of the spectrum if we follow the evolution for much larger dynamical ranges.

Since the population of the axions radiated by string-wall networks is dominated by low-momentum modes which has the frequency comparable to the mass of the axion $m$,
we expect that the mean energy of radiated axions is determined by the parameter $m$.
Here, using the result of $P(k,t_2)$, where $t_2$ is chosen at the late time of the simulation [see Eq.~\eqref{eqC-12}],
we compute the mean comoving momentum of radiated axions
\begin{equation}
\bar{k}(t_2) \equiv \frac{\int\frac{dk}{2\pi^2}P(k,t_2)}{\int\frac{dk}{2\pi^2}\frac{1}{k}P(k,t_2)}, \label{eq3-2-2}
\end{equation}
and its ratio to the axion mass
\begin{equation}
\epsilon_a \equiv \frac{\bar{k}(t_2)/R(t_2)}{m}. \label{eq3-2-3}
\end{equation}
The results are shown in table~\ref{tab2}. We see that the value of the mean momentum is not strongly depend on the number $N_{\mathrm{DW}}$.
However, the uncertainty becomes large for the case with large $N_{\mathrm{DW}}$. This is because a large portion of the simulation box is masked
when we evaluate the power spectrum, which causes a large systematic errors when we evaluate the spectrum in the large
scale\footnote{We find that there is a systematic error in our masking analysis method. See Appendix~\ref{secC} for details.}, as shown in Fig.~\ref{fig4}.

\begin{table}[h]
\begin{center} 
\caption{The values of mean comoving momentum $\bar{k}$ and parameter $\epsilon_a$ defined in Eqs.~(\ref{eq3-2-2}) and (\ref{eq3-2-3}) for various values of $N_{\mathrm{DW}}$.}
\vspace{3mm}
\begin{tabular}{c c c}
\hline\hline
$N_{\mathrm{DW}}$ & $\bar{k}$ & $\epsilon_a$ \\
\hline
2 & 2.38$\pm$0.28 & 1.19$\pm$0.14 \\
3 & 2.38$\pm$0.25 & 1.19$\pm$0.12 \\
4 & 2.63$\pm$0.29 & 1.31$\pm$0.15 \\
5 & 2.97$\pm$0.45 & 1.49$\pm$0.22 \\
6 & 3.07$\pm$0.42 & 1.54$\pm$0.21 \\
\hline\hline
\label{tab2}
\end{tabular}
\end{center}
\end{table}

The claim is that the parameter defined by Eq.~(\ref{eq3-2-3}) does not depend on the time $t_2$.
To confirm it, we calculate the parameters $\bar{k}$ and $\epsilon_a$ at different time steps, while fixing the reference time $\tau_1=14$ in Eq.~(\ref{eqC-12}).
The results are shown in table~\ref{tab3}. We find that the value of $\bar{k}$ changes proportionally to $R(t_2)$.
On the other hand, the ratio $\epsilon_a$ between the \emph{physical} momentum $\bar{k}/R$ and the axion mass $m$ remains constant.
Therefore, we use the parameter $\epsilon_a$ as a constant of ${\cal O}(1)$ when we perform the analytic investigations in Sec.~\ref{sec4}.

\begin{table}[h]
\begin{center} 
\caption{The values of $\bar{k}$ and $\epsilon_a$ evaluated at different times $\tau_2$ for the case with $N_{\mathrm{DW}}=3$.}
\vspace{3mm}
\begin{tabular}{c c c}
\hline\hline
$\tau_2$ & $\bar{k}$ & $\epsilon_a$ \\
\hline
24 & 1.42$\pm$0.22 & 1.19$\pm$0.18 \\
28 & 1.62$\pm$0.13 & 1.15$\pm$0.09 \\
32 & 1.88$\pm$0.19 & 1.18$\pm$0.12 \\
36 & 2.04$\pm$0.22 & 1.13$\pm$0.12 \\
40 & 2.38$\pm$0.25 & 1.19$\pm$0.12 \\
\hline\hline
\label{tab3}
\end{tabular}
\end{center}
\end{table}

\begin{figure}[htbp]
\begin{center}
\begin{tabular}{c c}
\resizebox{75mm}{!}{\includegraphics[angle=0]{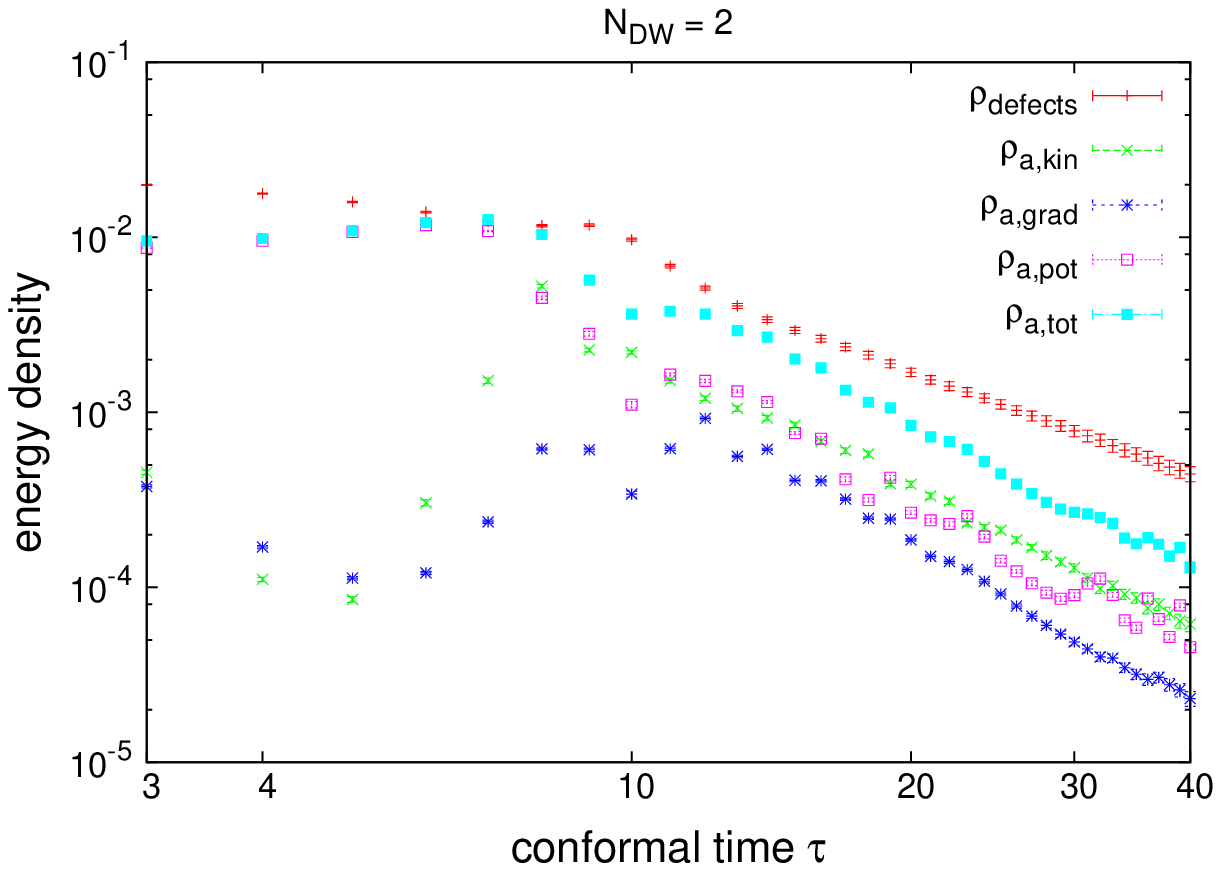}} &
\resizebox{75mm}{!}{\includegraphics[angle=0]{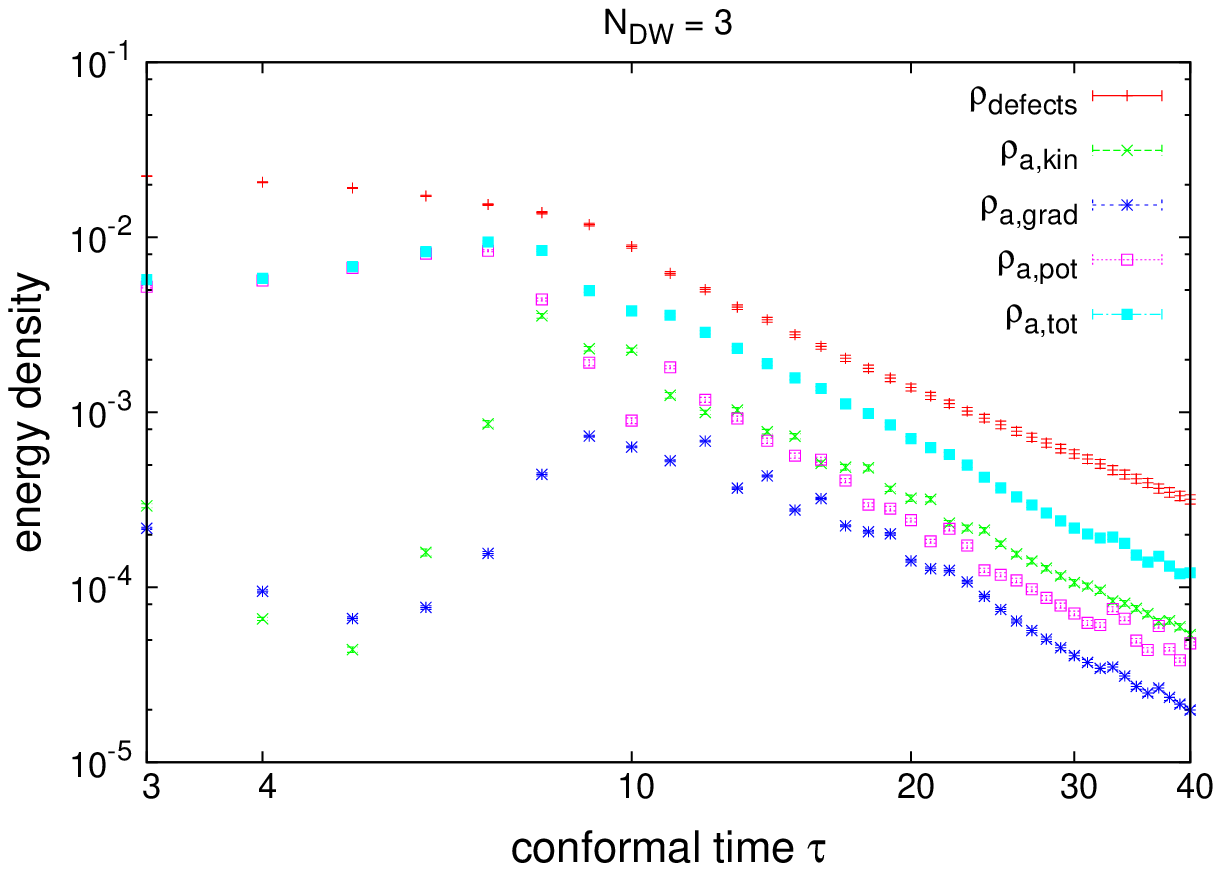}} \\
\resizebox{75mm}{!}{\includegraphics[angle=0]{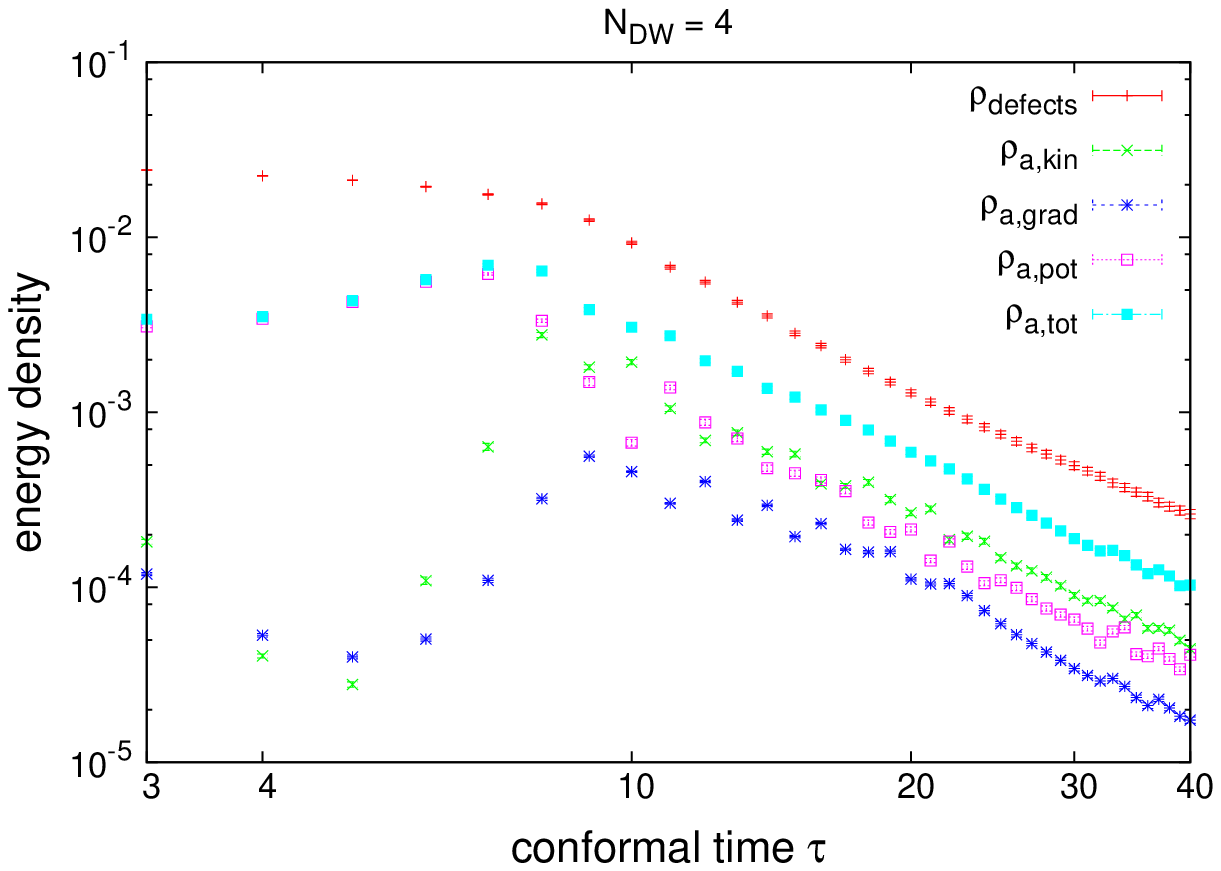}} &
\resizebox{75mm}{!}{\includegraphics[angle=0]{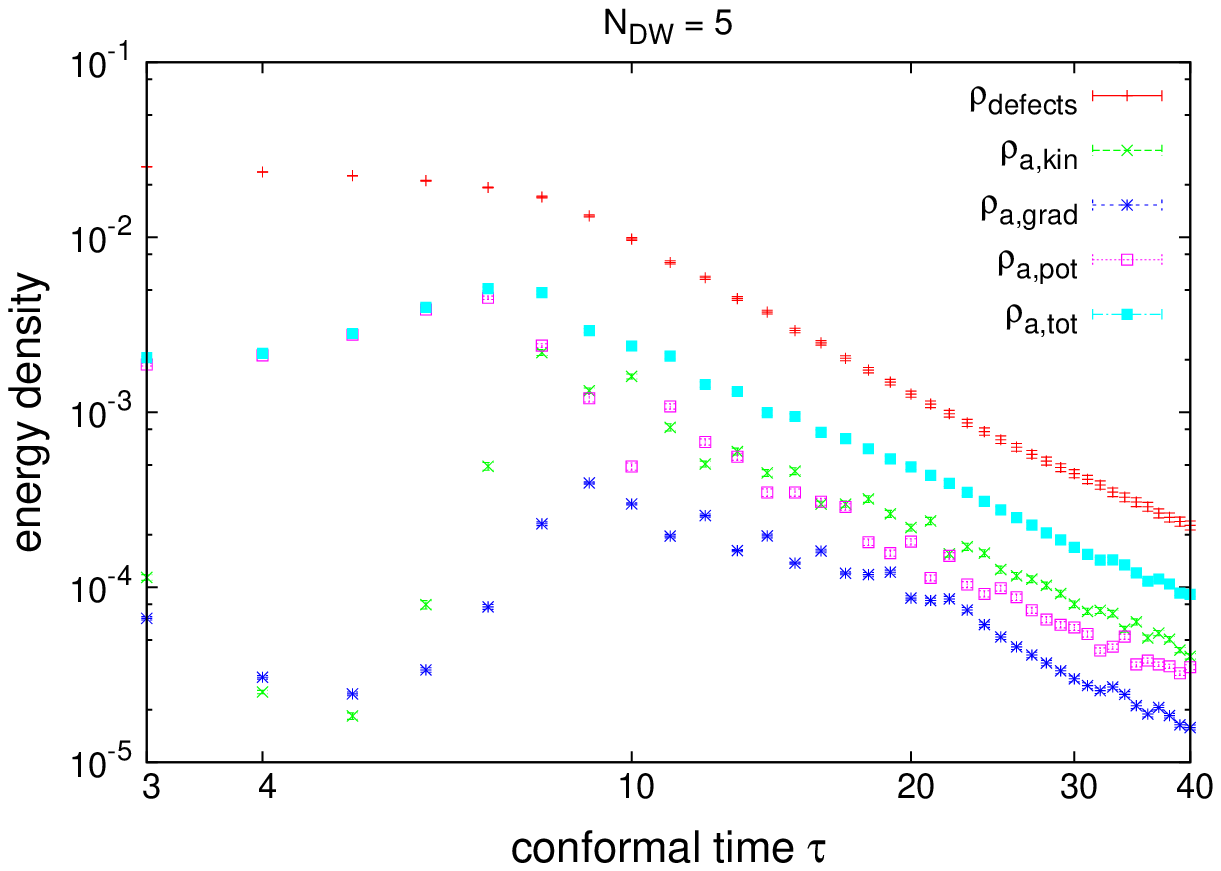}} \\
\end{tabular}
\resizebox{75mm}{!}{\includegraphics[angle=0]{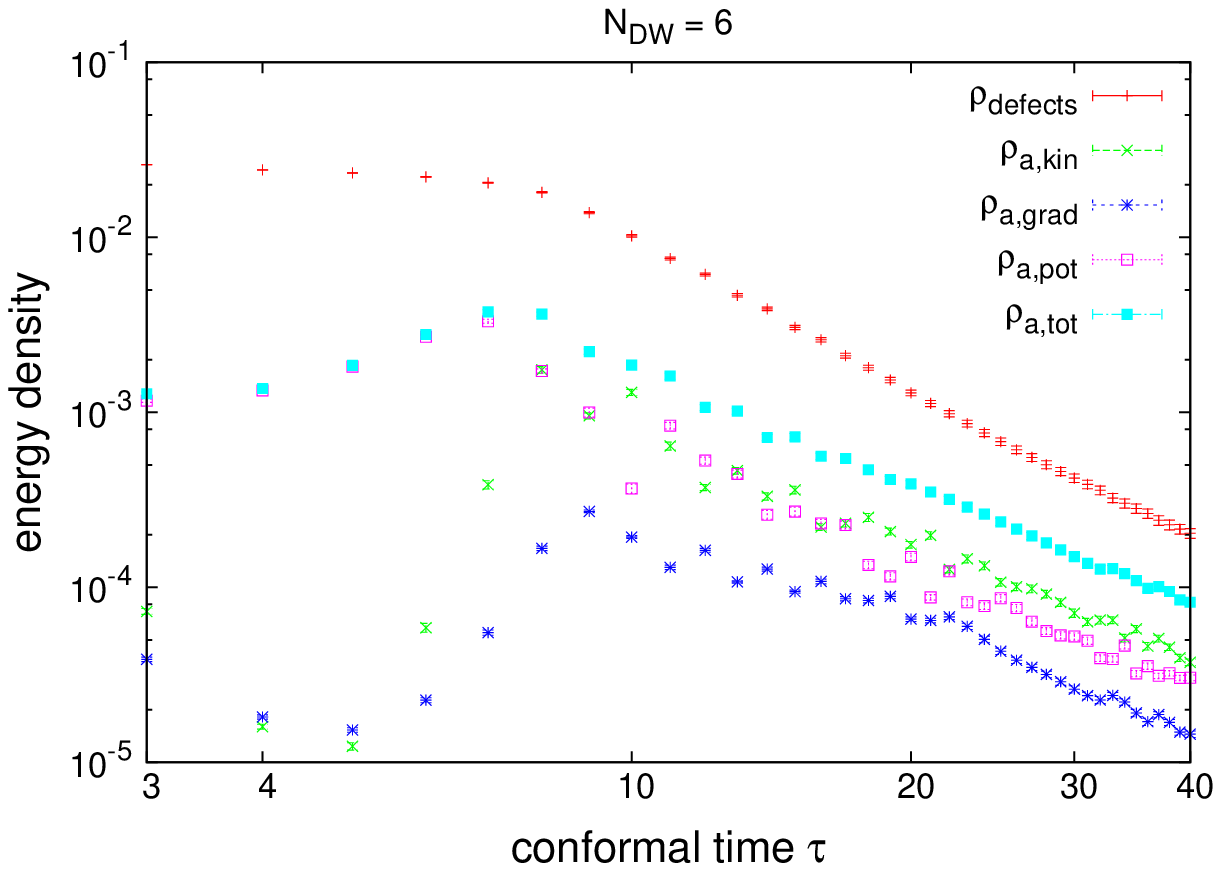}}
\end{center}
\caption{Time evolution of energy densities for various values of $N_{\mathrm{DW}}$.
$\rho_{\mathrm{defects}}$ represents the energy density of the string-wall networks which is evaluated by summing up the data of scalar field
placed near the defects whose position is identified by using the algorithm described in Appendix~\ref{secB}.
$\rho_{a,\mathrm{kin}}$, $\rho_{a,\mathrm{grad}}$, and $\rho_{a,\mathrm{pot}}$ are the kinetic, gradient, and potential energy density of radiated axions, respectively.
$\rho_{a,\mathrm{tot}}$ is the sum of $\rho_{a,\mathrm{kin}}$, $\rho_{a,\mathrm{grad}}$, and $\rho_{a,\mathrm{pot}}$.
These energy densities of axions are evaluated by subtracting the data of scalar field which contribute to $\rho_{\mathrm{defects}}$.}
\label{fig5}
\end{figure}

We close this subsection by showing Fig.~\ref{fig5}, the time evolution of the total energy density of axions and topological defects in the simulation box.
From this figure, we see that the domain wall networks evolve as $\rho_{\mathrm{wall}}\sim1/t$, or $\rho_{\mathrm{wall}}\sim1/\tau^2$ in conformal time.
This behavior is expected from the property of the scaling solution, given by Eq.~(\ref{eq2-2-1}).
We also find that the energy density of axions radiated by string-wall networks behaves similarly to that of topological defects at the late time.
We will confirm this behavior in Sec.~\ref{sec4-1}.

\subsection{\label{sec3-3} Production of gravitational waves}
Next, we investigate the production of gravitational waves from string-wall networks.
The spectrum of gravitational waves is represented by the quantity
\begin{equation}
\Omega_{\mathrm{gw}}(k,t) = \frac{1}{\rho_c(t)}\frac{d\rho_{\mathrm{gw}}(k,t)}{d\ln k}, \label{eq3-3-1}
\end{equation}
where $\rho_c(t)$ is the critical density of the universe at the time $t$, and $\rho_{\mathrm{gw}}$ is the energy density of gravitational waves.
We compute $\Omega_{\mathrm{gw}}$ using the data of the scalar field $\Phi$ obtained from the numerical simulation.
The method of the calculation is summarized in Appendix~\ref{secD}.

Figure~\ref{fig6} shows the results for $\Omega_{\mathrm{gw}}(k,t)$ for various values of $N_{\mathrm{DW}}$.
The slope of the spectrum changes at two characteristic scales. One is the Hubble radius ($k\sim2\pi R(\tau_f)H(\tau_f)\sim 0.1$)
and another is the width of the domain wall ($k\sim R(\tau_f)m\sim2$).
This result agrees with the general argument made by two of the present authors (M. K. and K. S.) in~\cite{Kawasaki:2011vv}.
However, the amplitude of $\Omega_{\mathrm{gw}}$ at high momenta is enhanced for the case with large $N_{\mathrm{DW}}$.
This can be interpreted as follows. Since the number of domain walls is large in the model with large $N_{\mathrm{DW}}$,
domain walls are ``frustrated", and there are many small scale configurations whose correlation length is shorter than
the Hubble radius. Therefore, the small scale anisotropy becomes large in the model with large domain wall number $N_{\mathrm{DW}}$.
We also plot the spectrum of gravitational waves produced by domain walls in the real scalar field theory given by Eqs.~(\ref{eq3-2}) and (\ref{eq3-3}).
Although the evolution of domain wall networks shown in Fig.~\ref{fig3} is similar between the model with $N_{\mathrm{DW}}=2$
and the model with real scalar field, the form of the spectrum of gravitational waves is slightly different between these two models
at high momenta. This might be caused by the different form of the effective potential which determines the small scale structure of the defects.

\begin{figure}[htbp]
\begin{center}
\includegraphics[scale=1.0]{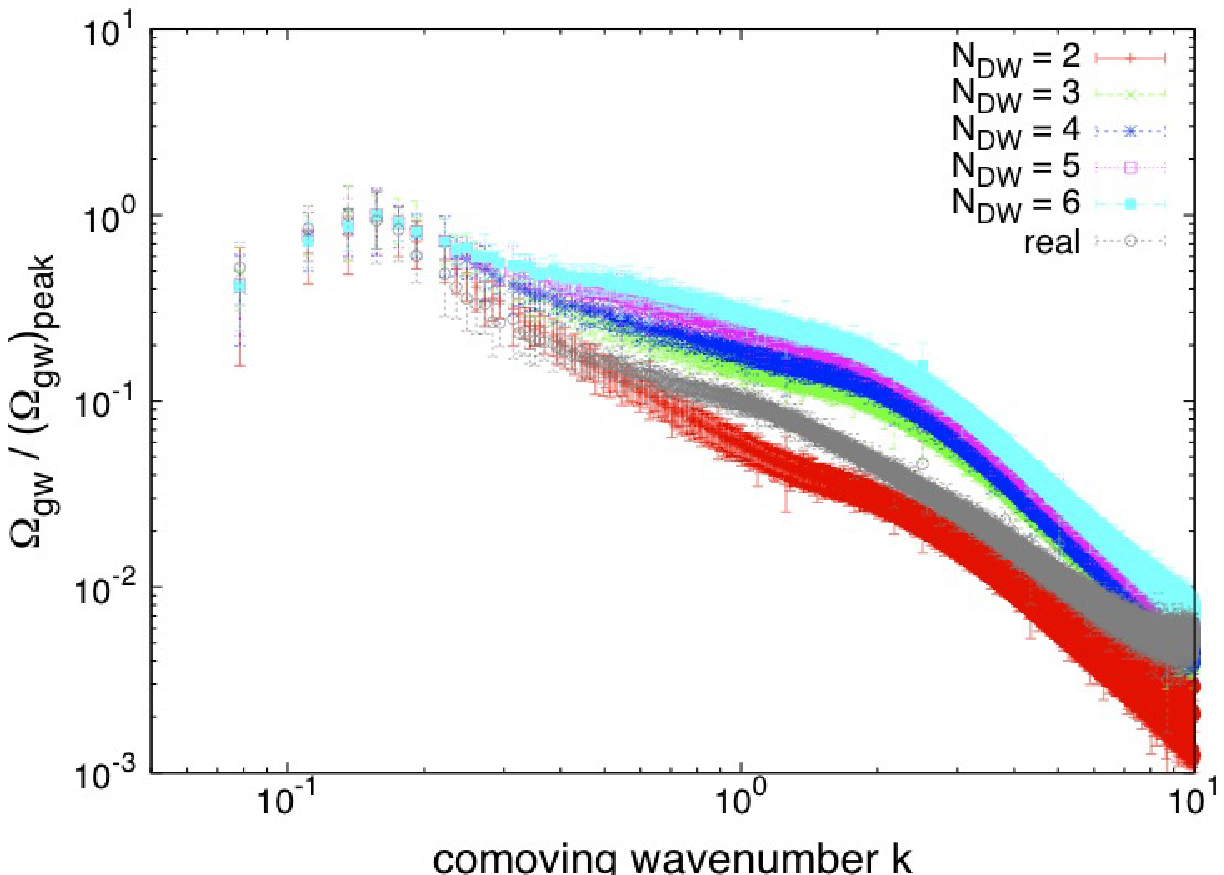}
\end{center}
\caption{The spectra of gravitational waves for various values of $N_{\mathrm{DW}}$.
The plot denoted as ``real" shows the spectrum of gravitational waves produced in the model with real scalar field defined by Eqs.~(\ref{eq3-2}) and (\ref{eq3-3}).
In this figure, we normalized the vertical axis by the amplitude at the peak momentum.}
\label{fig6}
\end{figure}

\begin{figure}[htbp]
\begin{center}
\begin{tabular}{c c}
\resizebox{75mm}{!}{\includegraphics[angle=0]{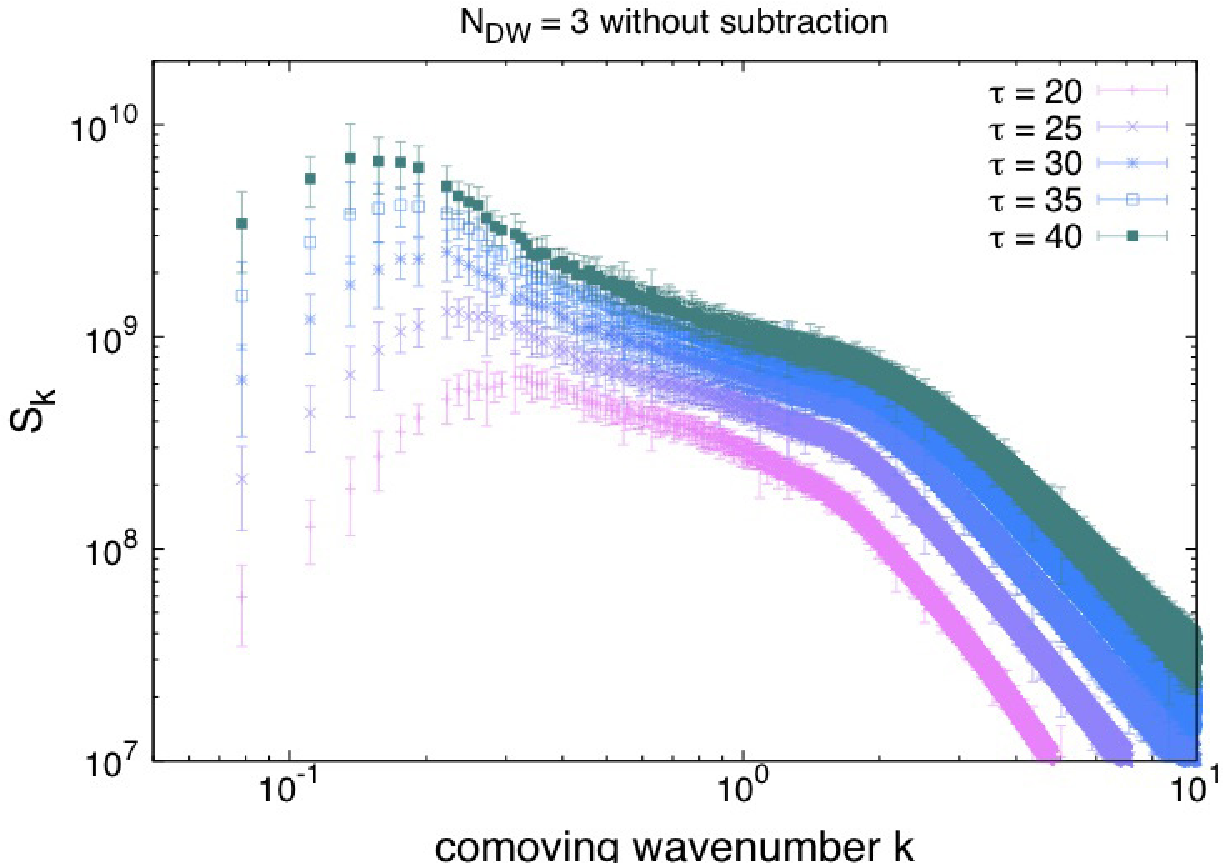}} &
\resizebox{75mm}{!}{\includegraphics[angle=0]{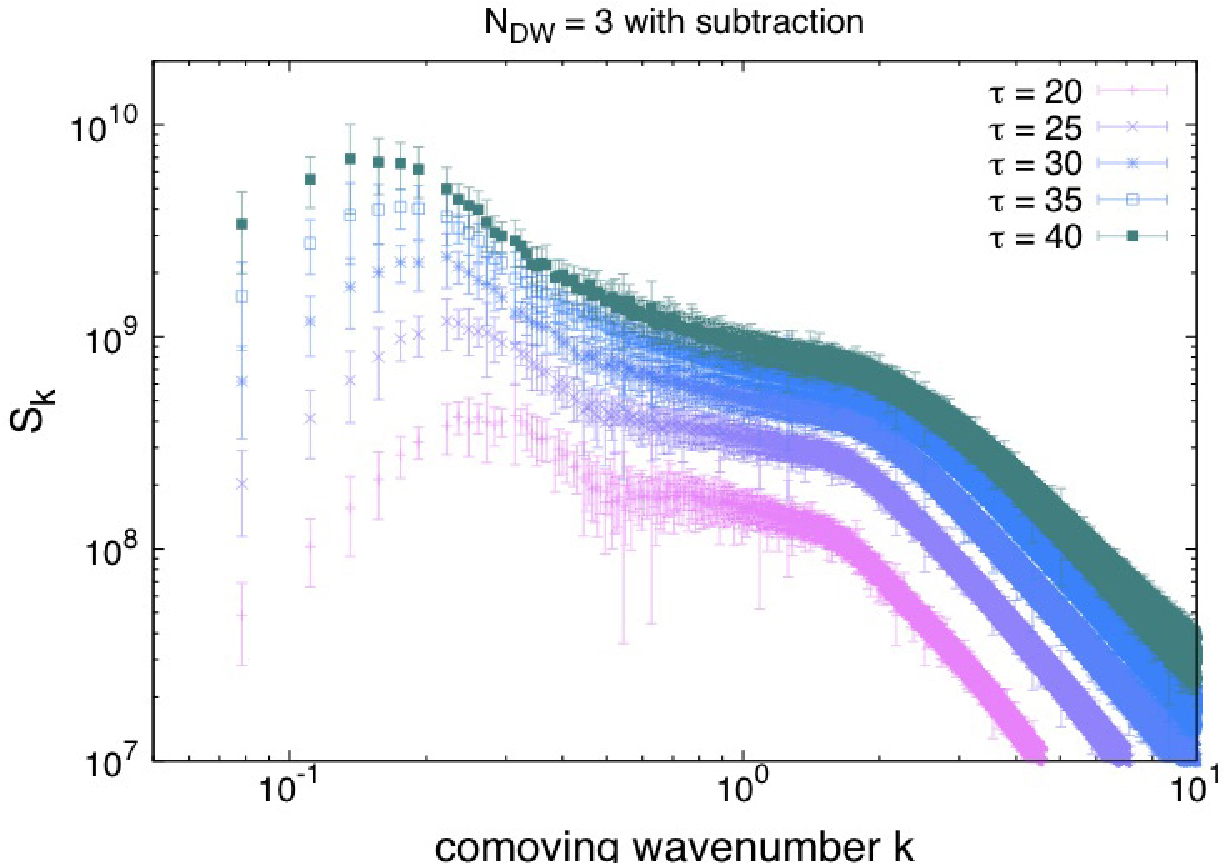}} \\
\end{tabular}
\end{center}
\caption{The time evolution of the spectrum of gravitational waves for the case with $N_{\mathrm{DW}}=3$.
Here, we plot the function $S_k(\tau)$ defined by Eq.~(\ref{eqD-17}) (left panel) and $\Delta S_k(\tau)$ defined by Eq.~(\ref{eqD-19}) (right panel), which is proportional to $\Omega_{\mathrm{gw}}$
in the radiation-dominated background. In the right panel, we subtract the contribution of gravitational waves produced before the time $\tau_g=15$.
The spectra are shown from the conformal time $\tau=20$ (pink) to $\tau=40$ (green) with the interval $\Delta\tau=5$.}
\label{fig7}
\end{figure}

As we discussed in previous subsections, at the early times ($\tau\lesssim 15$)
topological defects do not relax into the scaling regime. However, the nonlinear dynamics of scalar fields during this initial stage
produces a ``burst" of gravitational waves, which may contaminate the spectrum of gravitational waves produced at late times
where the defects networks evolve along to the scaling solution. We subtract this contribution with the same manner that we used for the subtraction
of axions produced at the initial stage [see Eq.~(\ref{eqD-19})].
Figure~\ref{fig7} shows the time evolution of the spectrum of gravitational waves, where we compare the effect of this subtraction procedure. 
We see that the form of the spectrum is different at early times if we make a subtraction.
In particular, there is a bump at the scale given by the Hubble radius $k/R\sim H$, and a plateau which falls off at the scale given by $k/R\sim m$.
This form is different from what we found in our previous studies~\cite{Hiramatsu:2010yz,Kawasaki:2011vv}, where a nearly ``flat" spectrum extends between two characteristic scales.
It seems that this difference in the form of spectra is caused by the contamination of radiations produced at the initial relaxation regime.
We also note that the dynamical range of the past numerical studies ($\tau_f/\tau_i\simeq 12$)~\cite{Hiramatsu:2010yz,Kawasaki:2011vv} is shorter than that of the present studies ($\tau_f/\tau_i=20$).
Therefore, the bump-like feature is less apparent in the past numerical studies with shorter dynamical ranges.
The form of the spectrum at the final time of the simulation is similar between the result with subtraction and that without subtraction, which
indicates that contaminations of the initial stage are diluted due to the cosmic expansion.

\begin{figure}[htbp]
\begin{center}
\includegraphics[scale=1.0]{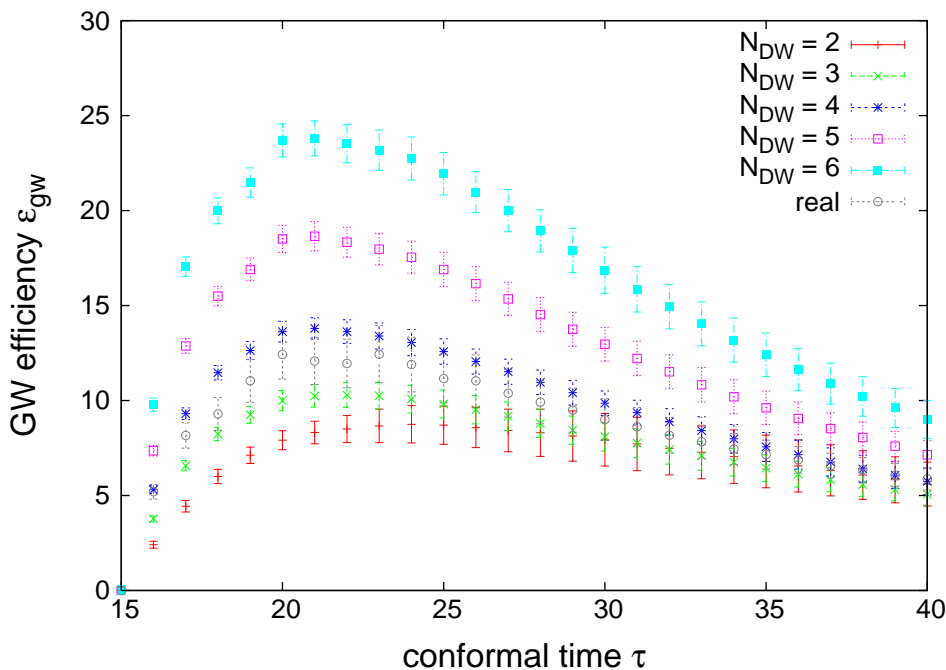}
\end{center}
\caption{The time evolution of the efficiency parameter $\epsilon_{\mathrm{gw}}$ defined by Eq.~(\ref{eq3-3-5})
for various values of $N_{\mathrm{DW}}$. The points denoted as ``real" in the right panel show the evolution of $\epsilon_{\mathrm{gw}}$ in the model with
real scalar field defined by Eqs.~(\ref{eq3-2}) and (\ref{eq3-3}).
In this figure, we subtract the contribution of gravitational waves produced before the time $\tau_g=15$.}
\label{fig8}
\end{figure}

The spectrum shown in Fig.~\ref{fig6} is normalized in terms of the peak amplitude in order to give the relative form of the spectrum between different theoretical parameter $N_{\mathrm{DW}}$.
On the other hand, the total amplitude of gravitational waves is determined by the following arguments.
Since the spectrum of gravitational waves has a peak at the momentum which is given by the Hubble scale,
we regard that the most of the gravitational waves are generated at the length scale $\sim t$.
Suppose that the radiation of gravitational waves occurs with the time scale comparable to the Hubble time.
The power of gravitational waves radiated from domain walls is estimated by using the quadrupole formula~\cite{Maggiore:1900zz}
\begin{equation}
P \sim G\dddot{Q}_{ij}\dddot{Q}_{ij}, \label{eq3-3-2}
\end{equation}
where $Q_{ij}\sim M_{\mathrm{DW}}t^2$ is the quadrupole moment,
and $M_{\mathrm{DW}}\sim \rho_{\mathrm{wall}}t^3\sim\sigma_{\mathrm{wall}}{\cal A}t^2$ is the mass energy of domain walls, where we used Eq.~(\ref{eq3-1-2}).
The energy of gravitational waves is estimated as
\begin{equation}
E_{\mathrm{gw}} \sim P t \sim G{\cal A}^2\sigma_{\mathrm{wall}}^2t^3. \label{eq3-3-3}
\end{equation}
We see that the energy density of gravitational waves does not depend on the time $t$
\begin{equation}
\rho_{\mathrm{gw}} \sim E_{\mathrm{gw}}/t^3 \sim G{\cal A}^2\sigma_{\mathrm{wall}}^2. \label{eq3-3-4}
\end{equation}
Now, we define the efficiency $\epsilon_{\mathrm{gw}}$ of gravitational waves
\begin{equation}
\rho_{\mathrm{gw}}^{(\mathrm{sim})} \equiv \epsilon_{\mathrm{gw}}G{\cal A}^2\sigma_{\mathrm{wall}}^2, \label{eq3-3-5}
\end{equation}
where $\rho_{\mathrm{gw}}^{(\mathrm{sim})}$ is the total energy density of gravitational waves computed from numerical simulations.
We plot the time evolution of the parameter $\epsilon_{\mathrm{gw}}$ in Fig.~\ref{fig8}.
Although the value of $\epsilon_{\mathrm{gw}}$ does not remain in constant exactly, it seems to converge into a universal value of ${\cal O}(5$-$10)$.
 
\begin{figure}[htbp]
\begin{center}
\includegraphics[scale=1.0]{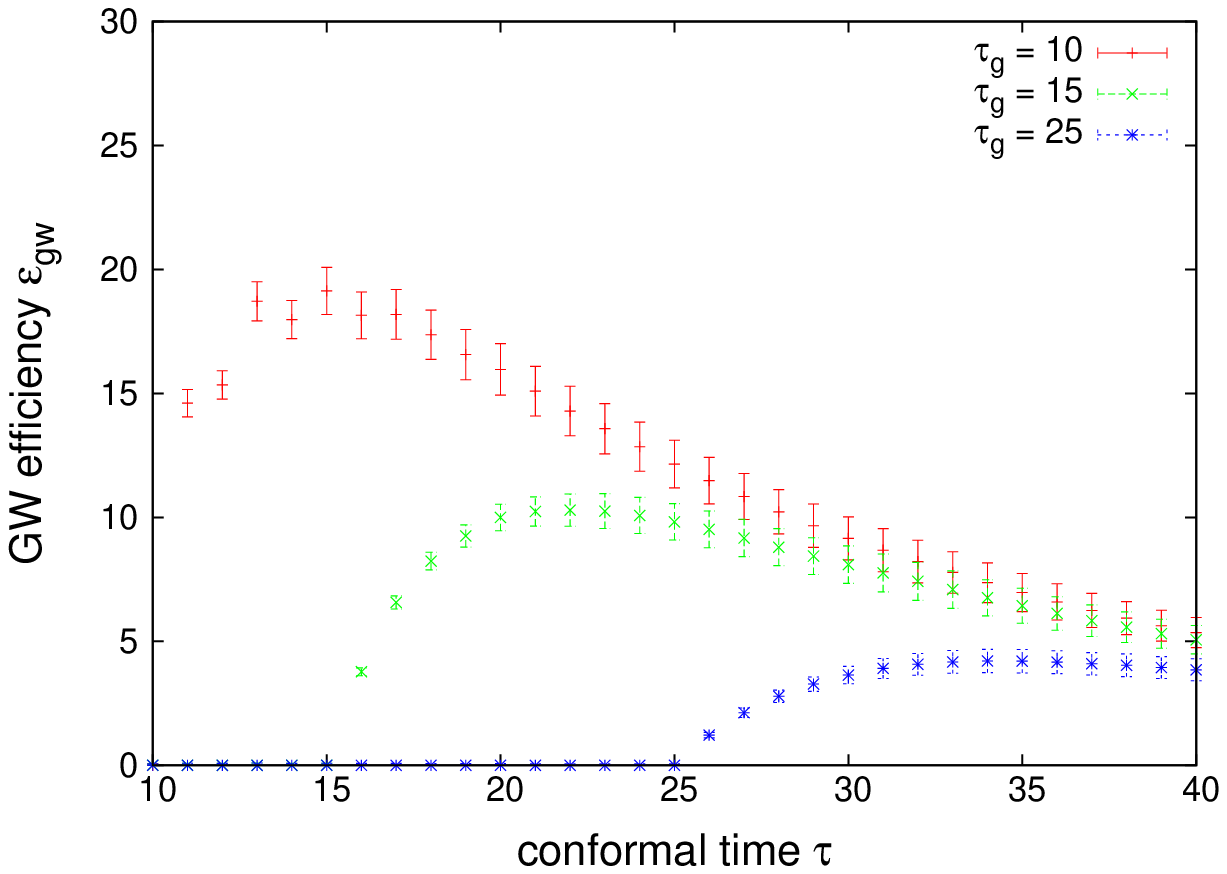}
\end{center}
\caption{The time evolution of $\epsilon_{\mathrm{gw}}$ and its dependence on the choice of $\tau_g$.
In this figure, we plot the result of the model with $N_{\mathrm{DW}}=3$.}
\label{fig9}
\end{figure}

We note that the time dependence of $\epsilon_{\mathrm{gw}}$ is strongly affected by the choice of the time $\tau_g$
at which we subtract the contribution of gravitational waves produced in the initial relaxation regime [see Eq.~(\ref{eqD-19})].
This is shown in Fig.~\ref{fig9}, where we plot the time evolution of $\epsilon_{\mathrm{gw}}$ for various values of $\tau_g$.
If we choose $\tau_g$ at an early time of the simulation, $\rho_{\mathrm{gw}}^{(\mathrm{sim})}$ contains the contribution
of gravitational waves produced during the initial relaxation regime. Hence the behavior of $\rho_{\mathrm{gw}}^{(\mathrm{sim})}$
might be different from what we expect from Eq.~(\ref{eq3-3-5}), where we assumed that the energy density of domain walls is given by 
the scaling formula~(\ref{eq3-1-2}). This additional contribution is not negligible even at the late time of the simulations
because the dynamical range of the numerical simulation is short. This constancy of $\epsilon_{\mathrm{gw}}$ should be tested in future numerical studies with improved dynamical ranges.
Since the time variation of $\epsilon_{\mathrm{gw}}$ is small for large value of $\tau_g$, as shown in the plot of $\tau_g=25$ in Fig.~\ref{fig9},
we assume that $\epsilon_{\mathrm{gw}}$ takes a constant value when string-wall networks enter into the scaling regime,
and use the value $\epsilon_{\mathrm{gw}}\simeq 5$ when we estimate the abundance of gravitational waves in the next section.

\section{\label{sec4} Observational constraints}
In this section, we give constraints on the parameters of axion models with long-lived domain walls, using the results of numerical studies
described in the previous section. First, we calculate the relic abundance of cold axions and gravitational waves produced by
string-wall networks in Sec.~\ref{sec4-1}. Then we briefly review some experimental constraints in Secs.~\ref{sec4-2} and~\ref{sec4-3}.
We combine these constraints, and discuss their implications on the models with $N_{\mathrm{DW}}>1$ in Sec.~\ref{sec4-4}.
We find that the overabundance of the cold axions produced by domain walls gives a stringent bound on the model parameters, which
excludes the most part of models, except for tiny loopholes. Finally we comment on some exceptions in Sec.~\ref{sec4-5}.
\subsection{\label{sec4-1} Axion cold dark matter abundance}
Here we assume that the energy of strings becomes negligible compared with that of domain walls, and axions and gravitational waves
are only produced by domain walls. In this approximation, the evolution of energy densities for domain walls, axions, and gravitational waves
is described by the following coupled equations
\begin{eqnarray}
\frac{d\rho_{\mathrm{wall}}}{dt} &=& -H\rho_{\mathrm{wall}} - \left.\frac{d\rho_{\mathrm{wall}}}{dt}\right|_{\mathrm{emission}}, \label{eq4-1-1}\\
\frac{d\rho_a}{dt} &=& -3H\rho_a + \frac{d\rho_{w\to a}}{dt}, \label{eq4-1-2}\\
\frac{d\rho_{\mathrm{gw}}}{dt} &=& -4H\rho_{\mathrm{gw}} + \frac{d\rho_{w\to g}}{dt}, \label{eq4-1-3}
\end{eqnarray}
where
\begin{equation}
\left.\frac{d\rho_{\mathrm{wall}}}{dt}\right|_{\mathrm{emission}} = \frac{d\rho_{w\to a}}{dt} + \frac{d\rho_{w\to g}}{dt} \label{eq4-1-4}
\end{equation}
is the rate of radiation from domain walls. Following from Eqs.~(\ref{eq3-1-2}) and (\ref{eq3-3-5}),
we put an ansatz on the behavior of the energy densities of domain walls and gravitational waves
\begin{equation}
\rho_{\mathrm{wall}} = {\cal A}\frac{\sigma_{\mathrm{wall}}}{t},\qquad\mathrm{and}\qquad \rho_{\mathrm{gw}} = \epsilon_{\mathrm{gw}}G{\cal A}^2\sigma^2_{\mathrm{wall}}, \label{eq4-1-5}
\end{equation}
where ${\cal A}$ and $\epsilon_{\mathrm{gw}}$ are constants.
Substituting these expressions to the above equations, we obtain
\begin{eqnarray}
\left.\frac{d\rho_{\mathrm{wall}}}{dt}\right|_{\mathrm{emission}} &=& {\cal A}\frac{\sigma_{\mathrm{wall}}}{2t^2}, \label{eq4-1-6}\\
\frac{d\rho_{w\to g}}{dt} &=& 2\epsilon_{\mathrm{gw}}\frac{G{\cal A}^2\sigma_{\mathrm{wall}}^2}{t}. \label{eq4-1-7}
\end{eqnarray} 
From Eqs. (\ref{eq4-1-4}), (\ref{eq4-1-6}), and (\ref{eq4-1-7}), we find
\begin{equation}
\frac{d\rho_{w\to a}}{dt} = {\cal A}\frac{\sigma_{\mathrm{wall}}}{2t^2} - 2\epsilon_{\mathrm{gw}}\frac{G{\cal A}^2\sigma_{\mathrm{wall}}^2}{t}. \label{eq4-1-8}
\end{equation}
Note that the sign of the right hand side of this equation becomes negative for $t\gtrsim (G\sigma_{\mathrm{wall}})^{-1}$.
This is not problematic, since domain walls have to collapse before the time $t\sim (G\sigma_{\mathrm{wall}})^{-1}$ otherwise they overclose the universe [see Eq.~(\ref{eq2-2-2})].

Let us define the energy and number of axions per comoving box
\begin{eqnarray}
E_a(t) \equiv R^3(t)\rho_a(t), \label{eq4-1-9}\\
N_a(t) \equiv R^3(t)n_a(t), \label{eq4-1-10}
\end{eqnarray}
where $n_a(t)$ is the number density of axions at the time $t$.
Integrating Eqs.~(\ref{eq4-1-2}) and (\ref{eq4-1-8}), we find
\begin{equation}
E_a(t) = \int^t_{t_r}dt'R^3(t')\left[\frac{{\cal A}\sigma_{\mathrm{wall}}}{2}t^{-2}-2\epsilon_{\mathrm{gw}}G\sigma_{\mathrm{wall}}^2{\cal A}^2t^{-1}\right], \label{eq4-1-11}
\end{equation}
where $t_r$ is the time when domain walls begin to radiate axions, which might be chosen as the time of the QCD phase transition.
In the regime $t\gg t_r$, it reduces to
\begin{equation}
E_a(t) \simeq R^3(t)\left({\cal A}\frac{\sigma_{\mathrm{wall}}}{t} - \frac{4}{3}\epsilon_{\mathrm{gw}}G\sigma_{\mathrm{wall}}^2{\cal A}^2\right). \label{eq4-1-12}
\end{equation}
For sufficiently early times $t\ll(G\sigma_{\mathrm{wall}})^{-1}$, the first term of Eq.~(\ref{eq4-1-12}) dominates over the second term,
which leads to the behavior $\rho_a\propto1/t$. This agrees with what we observed in numerical simulations, shown in Fig.~\ref{fig5}.
The ratio between $E_a(t)$ and $N_a(t)$ can be determined by the result of numerical simulations
\begin{equation}
\frac{E_a(t)}{N_a(t)} = \sqrt{1+\epsilon_a^2}m, \label{eq4-1-13}
\end{equation}
where $\epsilon_a$ is defined by Eq.~(\ref{eq3-2-3}). Suppose that domain walls disappear at the time $t_{\mathrm{dec}}$ and the radiation of axions and gravitational waves is
terminated at that time. In this case, the energy densities of axions and gravitational waves at the present time $t_0$ are given by
\begin{eqnarray}
\rho_{a,\mathrm{dec}}(t_0) &=& \frac{mN_a(t_{\mathrm{dec}})}{R^3(t_0)} \nonumber \\
&=& \frac{1}{\sqrt{1+\epsilon_a^2}}\left(\frac{R(t_{\mathrm{dec}})}{R(t_0)}\right)^3\left({\cal A}\frac{\sigma_{
\mathrm{wall}}}{t_{\mathrm{dec}}} - \frac{4}{3}\epsilon_{\mathrm{gw}}G\sigma_{\mathrm{wall}}^2{\cal A}^2\right), \label{eq4-1-14}\\
\rho_{\mathrm{gw}}(t_0) &=& \left(\frac{R(t_{\mathrm{dec}})}{R(t_0)}\right)^4\epsilon_{\mathrm{gw}}G{\cal A}^2\sigma_{\mathrm{wall}}^2, \label{eq4-1-15}
\end{eqnarray}
where the subscript ``dec" in the left hand side of Eq.~(\ref{eq4-1-14}) represents that the axions produced by the decay of domain walls.

The annihilation of domain walls occurs in the time scale estimated by Eq.~(\ref{eq2-2-4})
\begin{equation}
t_{\mathrm{dec}} = \alpha \frac{m}{N_{\mathrm{DW}}\Xi\eta^2}. \label{eq4-1-16}
\end{equation}
The numerical coefficient can be fixed to $\alpha\simeq 18$, according to our previous study~\cite{Hiramatsu:2010yn}.
Note that the relation
\begin{equation}
\frac{R(t_{\mathrm{dec}})}{R(t_0)} = 5.45\times 10^{-9}\times N_{\mathrm{DW}}^{-3/2}\left(\frac{10^{10}\mathrm{GeV}}{F_a}\right)^{3/2}\left(\frac{10^{-58}}{\Xi}\right)^{1/2}, \label{eq4-1-17}
\end{equation}
where we used the expression for the mass of the axion
\begin{equation}
m = 6\times 10^{-4}\mathrm{eV}\left(\frac{10^{10}\mathrm{GeV}}{F_a}\right). \label{eq4-1-18}
\end{equation}
Substituting Eqs.~(\ref{eq4-1-16}) and (\ref{eq4-1-17}) into Eqs.~(\ref{eq4-1-14}) and (\ref{eq4-1-15}),
we obtain the density parameter of axions and gravitational waves
\begin{eqnarray}
\Omega_{a,\mathrm{dec}}(t_0)h^2 &=& \frac{\rho_{a,\mathrm{dec}}(t_0)}{\rho_c(t_0)/h^2} \nonumber \\
&=& 1.01\times 10^3 \times \frac{\cal A}{\sqrt{1+\epsilon_a^2}}N_{\mathrm{DW}}^{-3/2}\left(\frac{10^{-58}}{\Xi}\right)^{1/2}\left(\frac{10^{10}\mathrm{GeV}}{F_a}\right)^{1/2} \nonumber \\
&&\times\left[1-5.35\times 10^{-3}\times\epsilon_{\mathrm{gw}}{\cal A}N_{\mathrm{DW}}^{-3}\left(\frac{10^{-58}}{\Xi}\right)\left(\frac{10^{10}\mathrm{GeV}}{F_a}\right)^2\right], \label{eq4-1-19}\\
\Omega_{\mathrm{gw}}(t_0)h^2 &=& \frac{\rho_{\mathrm{gw}}(t_0)}{\rho_c(t_0)/h^2} \nonumber \\
&=& 2.24\times 10^{-8}\times\epsilon_{\mathrm{gw}}{\cal A}^2N_{\mathrm{DW}}^{-6}\left(\frac{10^{-58}}{\Xi}\right)^2\left(\frac{10^{10}\mathrm{GeV}}{F_a}\right)^4, \label{eq4-1-20}
\end{eqnarray}
where $h$ is the reduced Hubble parameter today: $H_0=100h$km$\cdot$sec$^{-1}$Mpc$^{-1}$.
We can fix the numerical values of ${\cal A}$, $\epsilon_a$ and $\epsilon_{\mathrm{gw}}$ from the result of numerical simulations.
Then, the relic abundance of axions and gravitational waves is determined by three theoretical parameters: $N_{\mathrm{DW}}$, $\Xi$, and $F_a$.
The second term in the square bracket in Eq.~(\ref{eq4-1-19}) represents the effect of gravitational radiation, which is negligible at early times but becomes relevant when
domain walls survive for a long time. In particular, when domain wall survives enough time such that they overclose the universe, $t_{\mathrm{dec}}\sim (G\sigma_{\mathrm{wall}})^{-1}$,
the contribution of the second term in the square bracket in Eq.~(\ref{eq4-1-19}) becomes ${\cal O}(1)$.
In such a case, the gravitational field around walls cannot be regarded as a small perturbation~\cite{Vilenkin:1981zs}, and we must use the full general relativistic treatment, which is out of the scope of this paper.
Hence, the expression~(\ref{eq4-1-19}) is valid only for the time $t\ll t_{\mathrm{WD}}\sim (G\sigma_{\mathrm{wall}})^{-1}$.
This condition is automatically satisfied as long as we consider the parameter region where domain walls do not overclose the universe.

We note that there are other contributions to the axion cold dark matter abundance, in addition to $\Omega_{a,\mathrm{dec}}$.
One is the coherent oscillation of the homogeneous axion field (zero mode)~\cite{Preskill:1982cy}
and another is the production of axions from global strings~\cite{Davis:1986xc}. Their contributions are estimated as~\cite{Hiramatsu:2012gg}
\begin{eqnarray}
\Omega_{a,\mathrm{0}}(t_0)h^2 &\simeq& 1.2\times \left(\frac{F_a}{10^{12}\mathrm{GeV}}\right)^{1.19}, \label{eq4-1-21}\\
\Omega_{a,\mathrm{string}}(t_0)h^2 &\simeq& 2.0\times \left(\frac{F_a}{10^{12}\mathrm{GeV}}\right)^{1.19}, \label{eq4-1-22} 
\end{eqnarray}
where the subscripts ``0" and ``string" represent that they are
the contributions of zero modes and global strings, respectively\footnote{The result for $\Omega_{a,\mathrm{string}}(t_0)h^2$
in~\cite{Hiramatsu:2012gg} should be multiplied by a factor 1/2
since we used the wrong expression of the mass energy of the strings per unit length $\mu_{\mathrm{string}}$.
The correct expression is $\mu_{\mathrm{string}}\simeq \pi\eta^2\ln(t/\delta_s)$, which is smaller than that we used in~\cite{Hiramatsu:2012gg} by a factor of 1/2.
Also, the value of $\Omega_{a,0}(t_0)h^2$ in~\cite{Hiramatsu:2012gg} should be corrected by a factor of 2, if we include the anharmonic correction at the initial time of the coherent oscillation.}.
In the above expressions, we fixed the QCD scale such that $\Lambda_{\mathrm{QCD}}=400$MeV.
The total abundance of cold dark matter axions is given by the sum of Eqs.~(\ref{eq4-1-19}), (\ref{eq4-1-21}) and (\ref{eq4-1-22}),
\begin{equation}
\Omega_ah^2 =  \Omega_{a,0}h^2 + \Omega_{a,\mathrm{string}}h^2 + \Omega_{a,\mathrm{dec}}h^2. \label{eq4-1-23}
\end{equation}
We require that $\Omega_ah^2$ should not exceed the observed value of the abundance of cold dark matter $\Omega_{\mathrm{CDM}}h^2=0.11$~\cite{Komatsu:2010fb}.
This gives an upper bound on $F_a$ and a lower bound on $\Xi$.
\subsection{\label{sec4-2} Neutron electric dipole moment}
Next we consider the effect of the $\Xi$ term on the degree of strong CP-violation, following Refs.~\cite{Kamionkowski:1992mf,Barr:1992qq}.
The QCD Lagrangian contains a term
\begin{equation}
{\cal L} = \frac{\bar{\theta}}{32\pi^2}G^{a\mu\nu}\tilde{G}^a_{\mu\nu}, \label{eq4-2-1}
\end{equation}
where $G^{a\mu\nu}$ is the gluon field strength, $\tilde{G}^a_{\mu\nu}$ is the dual of it, and $\bar{\theta}$ is a dimensionless parameter.
This term contributes to the neutron electric dipole moment (NEDM) with the amplitude $d_n = 4.5\times 10^{-15}\bar{\theta}e$cm~\cite{Kim:2008hd}.
The recent experimental bound on the NEDM gives $|d_n|<2.9\times 10^{-26}e$cm~\cite{Baker:2006ts}, which requires
\begin{equation}
\bar{\theta} < 0.7\times 10^{-11}. \label{eq4-2-2}
\end{equation}
The original idea of Peccei and Quinn is to set $\bar\theta=0$ by introducing a symmetry~\cite{Peccei:1977hh}.
However, if the additional term (\ref{eq2-2-3}) exists in the effective potential, it shifts the CP-conserving minimum with a magnitude controlled by the parameters $\Xi$ and $\delta$.

Substituting the parametrization $\Phi=\eta e^{ia/\eta}$ into the effective potential given by Eqs.~(\ref{eq2-1-2}) and (\ref{eq2-2-3}),
and expanding for small $a/\eta$, we obtain the effective potential for $a$
\begin{equation}
V(a)\simeq \frac{1}{2}m_{\mathrm{phys}}^2 a^2 - m_{\mathrm{phys}}^2F_a\bar{\theta}a, \label{eq4-2-3}
\end{equation}
where $m_{\mathrm{phys}}$ is the effective mass of the axion
\begin{equation}
m_{\mathrm{phys}}^2 = m^2+m_*^2 \equiv m^2 + 2\Xi N_{\mathrm{DW}}^2 F_a^2\cos\delta, \label{eq4-2-4}
\end{equation}
and $\bar{\theta}$ is the shifted minimum
\begin{equation}
\bar{\theta} = \frac{\langle a\rangle}{F_a} = \frac{2\Xi N_{\mathrm{DW}}^3 F_a^2\sin\delta}{m^2 + 2\Xi N_{\mathrm{DW}}^2 F_a^2\cos\delta}. \label{eq4-2-5}
\end{equation}
Requiring that this should not exceed the experimental bound~(\ref{eq4-2-2}), we obtain an upper bound for $\Xi$.
\subsection{\label{sec4-3} Astrophysical bounds}
It is known that various astrophysical observations give a lower bound on $F_a$.
Here we just summarize the constraints obtained so far. See~\cite{Raffelt:1990yz,Raffelt:2006cw} for more comprehensive reviews.

First of all, the emission of axions in the sun gives a constraint on the axion-photon coupling, $g_{a\gamma\gamma}\equiv(\alpha_{\mathrm{em}}/2\pi F_a)c_{a\gamma\gamma}$,
where $\alpha_{\mathrm{em}}$ is the fine-structure constant, and $c_{a\gamma\gamma}$ is a numerical factor of ${\cal O}(1)$.
The helioseismological consideration on the anomalous solar energy loss due to the axion emission gives a bound $g_{a\gamma\gamma}<10^{-9}\mathrm{GeV}^{-1}$~\cite{Schlattl:1998fz}.
There are also direct experimental searches. The Tokyo Axion Helioscope~\cite{Inoue:2008zp} gives a bound $g_{a\gamma\gamma}<(5.6$-$13.4)\times10^{-10}\mathrm{GeV}^{-1}$ for $0.84\mathrm{eV}<m<1.00\mathrm{eV}$.
Phase I of the CERN Axion Solar Telescope (CAST)~\cite{Andriamonje:2007ew} gives $g_{a\gamma\gamma}<8.8\times10^{-11}\mathrm{GeV}^{-1}$ for $m\lesssim0.02\mathrm{eV}$.
It is improved in Phase II~\cite{Arik:2008mq} as $g_{a\gamma\gamma}<2.2\times10^{-10}\mathrm{GeV}^{-1}$ for $m\lesssim0.4\mathrm{eV}$.
The sensitivity is expected to be improved up to $g_{a\gamma\gamma}<\mathrm{few}\times10^{-12}\mathrm{GeV}^{-1}$ in the next generation helioscopes~\cite{Irastorza:2011gs}.

The observation of globular clusters gives another bound.
The helium-burning lifetimes of horizontal branch (HB) stars give a bound for axion-photon coupling $g_{a\gamma\gamma}\lesssim 0.6\times 10^{-10}\mathrm{GeV^{-1}}$~\cite{Raffelt:1999tx}.
Furthermore, the delay of helium ignition in red-giant branch (RGB) stars due to the axion cooling gives $g_{aee}<2.5\times 10^{-13}$~\cite{Raffelt:1994ry},
where $g_{aee}\equiv C_em_e/F_a$ is the axion-electron coupling, $m_e$ is the mass of the electron, $C_e=\cos^2\beta/3$, and $\tan\beta$ is the ratio of two Higgs vacuum
expectation values.

The axion-electron coupling is also constrained by the observation of white-dwarfs.
The cooling time of white-dwarfs due to the axion emission gives a bound $g_{aee}<4\times 10^{-13}$~\cite{Raffelt:1985nj}.
Recently, it is reported that the fitting of the luminosity function of white-dwarfs
and the pulsation period of a ZZ Ceti star is improved due to the axion cooling, which implies the axion-electron coupling
$g_{aee}\simeq$(0.6-1.7)$\times 10^{-13}$~\cite{Isern:2008nt} or $g_{aee}\simeq$(0.8-2.8)$\times 10^{-13}$~\cite{1992ApJ...392L..23I}.

Finally, the axion-nucleon coupling is constrained from the energy loss rate of the supernova 1987A~\cite{Raffelt:2006cw}.
This gives the strongest astrophysical bound on the axion decay constant compared with other bounds described above, 
\begin{equation}
F_a > 4\times 10^8\mathrm{GeV}. \label{eq4-3-1}
\end{equation}
Hence in the following we take it as a lower bound on $F_a$ when we combine it with other observational constraints.
Note that the estimation of the axion emission rate suffers from various numerical uncertainties which may modify the bound by a factor of ${\cal O}(1)$~\cite{Raffelt:2006cw}.
\subsection{\label{sec4-4} Implication for models with $N_{\mathrm{DW}}>1$}
\begin{figure}[htbp]
\begin{center}
\includegraphics[scale=1.0]{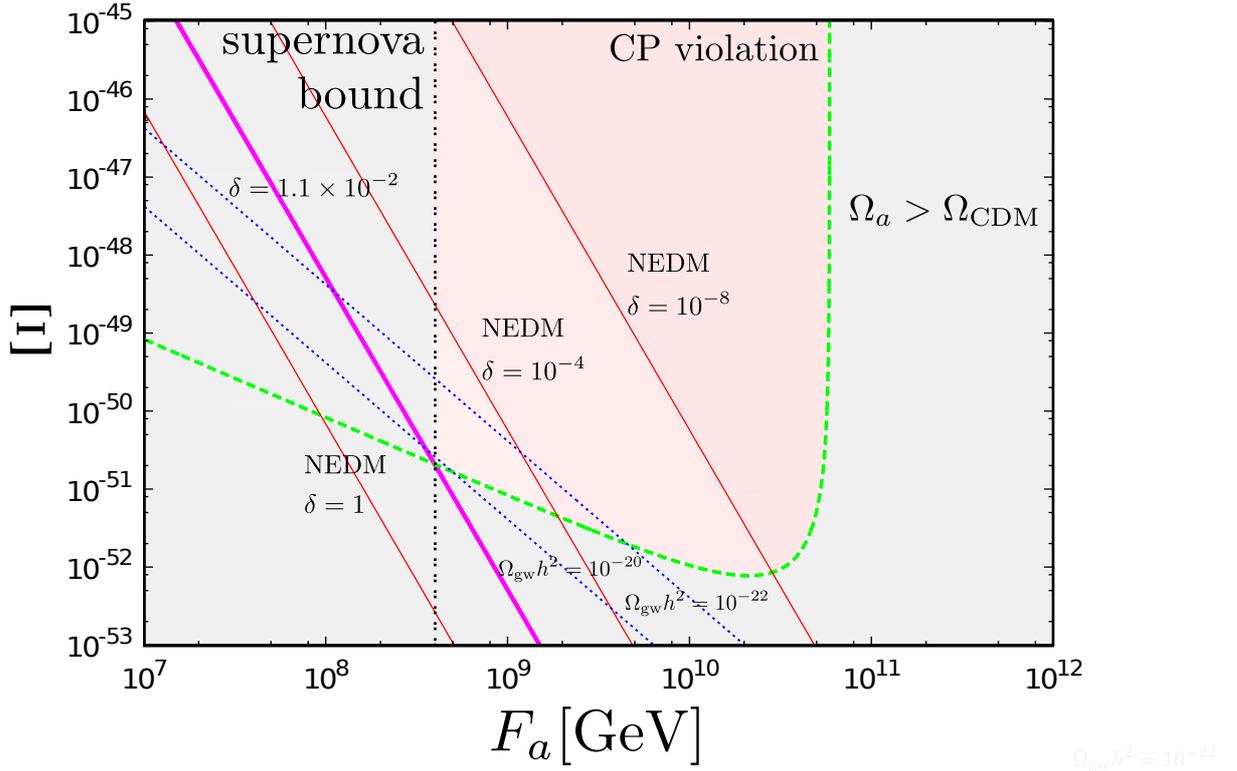}
\end{center}
\caption{The various observational constraints in the parameter space of $F_a$ and $\Xi$.
The green dashed-line represents the parameter region estimated by Eq.~(\ref{eq4-1-23}) where $\Omega_a=\Omega_{\mathrm{CDM}}$ is satisfied,
and the region below this line is excluded since the relic abundance of axions exceeds the cold dark matter abundance observed today.
The vertical dotted-line represents the bound given by Eq.~(\ref{eq4-3-1}) which comes from the observation of supernova 1987A.
The red solid-lines represent the NEDM bound given by Eqs.~(\ref{eq4-2-2}) and (\ref{eq4-2-5}) for $\delta=1$, $10^{-4}$, and $10^{-8}$.
The region above these lines is excluded since it leads to an experimentally unacceptable amount of CP-violation.
The pink line represents the NEDM bound for the critical value $\delta=1.1\times 10^{-2}$.
There are still allowed regions if the value of $\delta$ is smaller than this critical value.
The blue dotted-lines represent the parameter regions on which the amplitude of gravitational waves, given by Eq.~(\ref{eq4-1-20}),
becomes $\Omega_{\mathrm{gw}}h^2=10^{-20}$ and $10^{-22}$.
In this figure, we fixed other parameters as $N_{\mathrm{DW}}=6$, $\epsilon_a=1.5$, ${\cal A}=2.6$, and $\epsilon_{\mathrm{gw}}=5$.}
\label{fig10}
\end{figure}

In Fig.~\ref{fig10}, we plot the observational constraints described in the previous subsections.
Based on the estimation given by Eq.~(\ref{eq4-1-23}), we find that the overabundance of cold axions gives a stringent lower bound
on the bias parameter, $\Xi\gtrsim 10^{-50}$-$10^{-52}$.
This lower bound is much stronger than that obtained from the overclosure of domain walls (\ref{eq2-2-5}).
Combined with the NEDM constraint, it completely excludes the parameter region if we assume that the phase $\delta$ of the bias term~(\ref{eq2-2-3}) is ${\cal O}(1)$.
This bound might be weakened if we assume the extremely small value of $\delta$, but this assumption spoils the genius of the original PQ solution to the strong CP problem.
We must fine-tune $\delta$ in order to solve the fine-tuning problem of $\bar{\theta}$.

We find that the amplitude of gravitational waves produced by domain walls is weak $\Omega_{\mathrm{gw}}h^2\lesssim 10^{-20}$
in the parameter space considered above. This is irrelevant to any observations, since even the ultimate phase of DECIGO~\cite{Seto:2001qf},
which is a space-borne interferometer planned to launch in the future, will have a sensitivity $\Omega_{\mathrm{gw}}h^2\sim 10^{-18}$~\cite{Kudoh:2005as}.
The weakness of the signal of gravitational waves is due to the fact that domain walls annihilate in early epoch in order to avoid the overclosure of axions produced from them,
and the dilution factor of cosmic expansion becomes large.

The difference of this result from our previous study~\cite{Hiramatsu:2010yn} arises due to the different estimation on the mean momentum of radiated axions.
In the present analysis described in Sec.~\ref{sec3-2}, it turns out that the radiated axions are mildly relativistic with the momentum comparable to their mass,
$k/R\simeq m$, or $\epsilon_a\simeq 1$. However, in the previous study~\cite{Hiramatsu:2010yn}, we assumed that the radiated axions have a mean energy
$\omega\simeq \gamma m$ where $\gamma\simeq 60$, according to the result of the past numerical study\footnote{We note that there is
another numerical study which indicates $\gamma\simeq 3$~\cite{Nagasawa:1994qu}.
Our present results support this estimation rather than $\gamma\simeq 60$ obtained in~\cite{Chang:1998tb}.}~\cite{Chang:1998tb}.
This leads to an underestimation of the number of axions produced by domain walls by a factor $\sqrt{\epsilon_a^2+1}/\gamma\sim 10^{-2}$.
In this case, the bound which comes from the overclosure of cold axions is weakened.
The large factor of $\gamma$ originated in the assumption made in~\cite{Chang:1998tb}, that the radiated axions have a hard spectrum $dE/dk\sim1/ k$
which leads to a logarithmic correction $\gamma\sim\ln(\sqrt{\lambda}\eta/m)\sim 60$.
In Sec.~\ref{sec3-2} we have shown that this contribution is negligible compared with the contribution of the peak located at the axion mass.

We note that there is a parameter region around $F_a\sim 10^6\mathrm{GeV}$, called the ``hadronic axion window"~\cite{Chang:1993gm,Moroi:1998qs,Buckley:2007tm}, 
that is not excluded by observations. This occurs due to the ambiguity in light quark masses which leads to the cancellation in the axion-photon coupling $g_{a\gamma\gamma}$ for KSVZ models.
In such a case, the astrophysical bounds on $g_{a\gamma\gamma}$ do not significantly constrain the value of $F_a$.
Here we do not consider this possibility, as it strongly depends on the quantum number assignment of the PQ-charged fermions~\cite{Buckley:2007tm}.
Also, we note that in this case the axion becomes a candidate of hot dark matter~\cite{Moroi:1998qs} which can be constrained by observation of the large-scale structure~\cite{Hannestad:2005df}.

From the results shown in Fig.~\ref{fig10}, we conclude that the axion models with $N_{\mathrm{DW}}>1$ are excluded
if the PQ symmetry is broken after inflation. The exception is the case in which the value of $\delta$ is suppressed.
As we see in Fig.~\ref{fig10}, the degree of tuning in $\delta$ is $<1.1\times 10^{-2}$ in order to avoid all observational constraints.
Furthermore, by using Eqs.~\eqref{eq4-1-19} and \eqref{eq4-2-5}, one can show that the degree of tuning does not strongly depend on the value of $N_{\mathrm{DW}}$.
Hence, about 1\% fine tuning is required regardless of the value of $N_{\mathrm{DW}}$.
\subsection{\label{sec4-5} Scenario with extremely small $\delta$}
Although there is no theoretical reason to consider the extremely small value of $\delta$, it is possible,
by accident or design of some fundamental physics, to yield $\delta$ whose value is small enough to avoid the NEDM bound.
Let us comment on such a situation. In this case, the value of $\Xi$ can take larger value than that we considered so far.
In particular, if $\Xi$ becomes as large as
\begin{equation}
\Xi > 2\times 10^{-45}N_{\mathrm{DW}}^{-2}\left(\frac{10^{10}\mathrm{GeV}}{F_a}\right)^4, \label{eq4-4-1}
\end{equation}
the axion mass is not determined by QCD instanton effect, but by the correction term $m_*$ [see Eq.~(\ref{eq4-2-4}) and also Ref.~\cite{Barr:1992qq}].
Then, the cosmological history is significantly modified.

When the condition (\ref{eq4-4-1}) is satisfied, the magnitude of the bias term (\ref{eq2-2-3}) exceeds that of the QCD potential [the second term of Eq.~(\ref{eq2-1-2})].
Hence, ``domain walls" with the tension $\sigma_{\mathrm{wall}}\simeq8m_*F_a^2$ are formed when the Hubble parameter becomes $H\simeq m_*$.
The subsequent history is categorized into two possibilities, analogously to the scenarios described in Sec.~\ref{sec2}.
One possibility is that strings are get attached by domain walls at the time $t\sim m_*$, and they quickly disappear due to the tension of walls.
Let us call this scenario case I. On the other hand, if the structure of the term (\ref{eq2-2-3}) is the form $\propto\Phi^n$, where $n$ is an integer,
$n$ domain walls are attached to strings, and they survive for a long time. Let us call this scenario case II.
These string-wall networks annihilate due to the effect of a ``bias" term which arises from the QCD instanton effect, $\delta V\sim \Lambda_{\mathrm{QCD}}^4\sim m^2F_a^2$.
This occurs at the time $t_{\mathrm{dec},*}\sim \sigma_{\mathrm{wall}}/\Lambda_{\mathrm{QCD}}^4\sim 8m_*/m^2$.
Requiring that domain walls should disappear before the epoch of big bang nucleosynthesis (BBN), $t_{\mathrm{dec},*}<t_{\mathrm{BBN}}\sim 1\mathrm{sec}$, we obtain
\begin{equation}
\Xi <2\times 10^{-23}\times N_{\mathrm{DW}}^{-2}\left(\frac{F_a}{10^{10}\mathrm{GeV}}\right)^{-6}. \label{eq4-4-2}
\end{equation}

In case I, the abundance of axions is estimated in the usual way, as a sum of the coherent oscillation, strings, and decay of domain walls bounded by strings.
The number of axions is fixed at $t\sim t_*\equiv m_*^{-1}$. At that time, the energy density of axions is given by $\rho_a(t_*)\sim 10\times m_*^2F_a^2$,
where the factor $10$ arises from the fact that the abundance of axions becomes larger by an order of magnitude if we include the contribution of strings and domain walls~\cite{Hiramatsu:2010yu,Hiramatsu:2012gg}.
The present energy density becomes $\rho_a(t_0)\sim 10(R(t_*)/R(t_0))^3m_*^2F_a^2$, which leads
\begin{equation}
\Omega_{a\mathrm{(case\ I)}}^{\mathrm{nt}}h^2 \simeq 1.3\times 10^{-5}\times N_{\mathrm{DW}}^{1/2}\left(\frac{\Xi}{10^{-45}}\right)^{1/4}\left(\frac{F_a}{10^{10}\mathrm{GeV}}\right)^{5/2}, \label{eq4-4-3}
\end{equation}
where the subscript ``nt" indicates that they are produced non-thermally.

Axions are also produced from thermal bath of the primordial plasma.
The difference from the usual scenario is that the axion mass is given by $m_*$ so that their relic energy density can be large.
The number of axions is fixed when their interactions with the thermal bath with the process mediated by gluons freeze out.
Their number at the present time is given by $n_a^{\mathrm{th}}(t_0)\simeq 7.5\mathrm{cm}^{-3}$~\cite{Masso:2002np}, which leads
\begin{equation}
\Omega_a^{\mathrm{th}}h^2 \simeq 3.2\times 10^{-7}\times N_{\mathrm{DW}}\left(\frac{F_a}{10^{10}\mathrm{GeV}}\right)\left(\frac{\Xi}{10^{-45}}\right)^{1/2}, \label{eq4-4-4}
\end{equation}
where the subscript ``th" indicates that they are produced thermally.
For sufficiently large value of $\Xi$, the contribution of thermal component~(\ref{eq4-4-4}) exceeds the non-thermal conponent~(\ref{eq4-4-3}).
Requiring that it should not exceed the present cold dark matter abundance $\Omega_a^{\mathrm{th}}h^2<0.11$, we obtain
\begin{equation}
\Xi < 1.2 \times 10^{-34}\times N_{\mathrm{DW}}^{-2}\left(\frac{F_a}{10^{10}\mathrm{GeV}}\right)^{-2}\quad \mathrm{case\ I}. \label{eq4-4-5}
\end{equation}

On the other hand, in case II, axions are copiously produced from long-lived domain walls.
The relic abundance of axions produced by domain walls can be estimated by using the argument analogous to that leads Eq.~(\ref{eq4-1-19}).
Then we obtain
\begin{equation}
\Omega_{a\mathrm{(case\ II)}}^{\mathrm{nt}}h^2 \simeq 9.1\times 10^{-5}\times N_{\mathrm{DW}}^{3/2}\left(\frac{\Xi}{10^{-45}}\right)^{3/4}\left(\frac{F_a}{10^{10}\mathrm{GeV}}\right)^{9/2}. \label{eq4-4-6}
\end{equation}
In this case, the thermal component does not exceeds the non-thermal component. Requiring that $\Omega_{a\mathrm{(case\ II)}}^{\mathrm{nt}}h^2<0.11$, we obtain
\begin{equation}
\Xi < 1.3 \times 10^{-41}\times N_{\mathrm{DW}}^{-2}\left(\frac{F_a}{10^{10}\mathrm{GeV}}\right)^{-6}\quad \mathrm{case\ II}. \label{eq4-4-7}
\end{equation}
This bound is more severe than that obtained from the lifetime of domain walls, given by Eq.~(\ref{eq4-4-2}).

However, if $\Xi$ is much larger than the values considered above, axions become heavy and no longer stable.
Their dominant  decay channel is a decay into two photons. The lifetime is given by
\begin{eqnarray}
t_{\gamma} &=& \Gamma_{\gamma}^{-1} = \frac{64\pi}{m_*^3g_{a\gamma\gamma}^2} \nonumber\\
&=& 1.1\times 10^{41} \mathrm{sec}\times N_{\mathrm{DW}}^{-3}\left(\frac{\Xi}{10^{-45}}\right)^{-3/2}\left(\frac{10^{10}\mathrm{GeV}}{F_a}\right), \label{eq4-4-8}
\end{eqnarray}
where $\Gamma_{\gamma}$ is the decay late of the process $a\to\gamma\gamma$, and we fixed the numerical coefficient $c_{a\gamma\gamma}=1$ in the axion-photon coupling for simplicity.
Since the axion mass is not given by $m$, we must treat $m_*$ and $g_{a\gamma\gamma}$ as different parameters.
Cosmological bounds on such models are investigated in~\cite{Cadamuro:2011fd}. The heavy axions which decay at early times lead to various effects on the cosmological observations,
such as the distortion in the spectrum of CMB, and the disagreement of the theoretical prediction with the observed light element abundance.
Each of the observations gives a severe constraints on the coupling $g_{a\gamma\gamma}$. In order to avoid these constraints,
the lifetime of axions must be short enough to decay earlier than the onset of the BBN, $t_{\gamma}\lesssim10^{-2}\mathrm{sec}$.
This requirement leads a bound
\begin{equation}
\Xi > 4.6\times 10^{-17}\times N_{\mathrm{DW}}^{-2}\left(\frac{F_a}{10^{10}\mathrm{GeV}}\right)^{-2/3}. \label{eq4-4-9}
\end{equation}

In summary, if the value of $\Xi$ is sufficiently large, the mass of axions is determined by the term proportional to $\Xi$, and this parameter is again constrained by various observational considerations.
In order to avoid the observational bounds, axions should be light enough to avoid the overclosure of the universe, whose condition is given by Eq.~(\ref{eq4-4-5}) or (\ref{eq4-4-7}),
or heavy enough to decay before the BBN epoch, whose condition is given by Eq.~(\ref{eq4-4-9}).
\section{\label{sec5} Summary and discussions}
In this paper, we have investigated the axion cosmology with $N_{\mathrm{DW}}>1$, where domain walls survive for a long time in the early universe.
We performed three-dimensional lattice simulations to follow the evolution of string-wall networks in the scaling regime, and
computed the spectra of axions and gravitational waves produced from them.
The radiated axion is mildly relativistic with the momentum comparable to the axion mass, which is similar to the result of the model with $N_{\mathrm{DW}}=1$~\cite{Hiramatsu:2012gg}.
The ratio of the mean momentum of axions to their mass, given by Eq.~(\ref{eq3-2-3}), has a value $\epsilon_a\simeq 1$.
The spectrum of gravitational waves has a peak at the momentum comparable to the Hubble scale $k/R\sim H$,
with a plateau extending up to the momentum comparable to the inverse of the width of domain walls $k/R\sim m$.
The magnitude of gravitational waves is given by Eq.~(\ref{eq3-3-5}) with the numerical coefficient $\epsilon_{\mathrm{gw}}\simeq 5$.

Using the numerical results, we calculated relic densities of axions and gravitational waves produced by long-lived string-wall networks.
We found that the significant amounts of axions are produced from topological defects, and the requirement that they should not overclose the universe
gives a stringent constraint on the model parameters. Combined with bounds given by astrophysical observations and the experimental result on the NEDM,
we showed that the most part of parameter regions are excluded unless the value of the phase of the bias term $\delta$ is extremely fine-tuned.
In such parameter regions, the amplitude of gravitational waves is too weak to observe even in future gravitational wave detectors with high sensitivities.

Cosmological domain wall problem leads to serious constraints on the axion models.
The options that we can choose to avoid this problem are listed as follows:
\begin{enumerate}
\item The axion is realized such that $N_{\mathrm{DW}}=1$. This can be obtained by introducing a heavy quark\footnote{This can also be realized
by embedding the discrete subgroup $Z_{N_{\mathrm{DW}}}$ into a continuous group~\cite{Lazarides:1982tw}.
However, this requires another continuous (gauge or global) group in addition to the standard model symmetry groups.}~\cite{Kim:1979if}.
Domain walls disappear shortly after their formation, and there is no cosmological problem. In this case, axion mass is constrained into $m\simeq 10^{-3}$-$10^{-2}\mathrm{eV}$~\cite{Hiramatsu:2012gg}.
\item The PQ symmetry is broken during inflation. In this case, the population of topological defects is wiped away beyond the horizon scale, and domain walls are not problematic.
However, the observation of the isocurvature fluctuations in anisotropies of CMB gives constraints
on the model parameters~\cite{Lyth:1989pb}.
\item The existence of $\Xi$ term annihilates domain walls. This possibility is allowed only if the phase $\delta$ of the bias term is extremely suppressed. However, the value of $\Xi$ is still constrained by cosmology,
such as Eqs.~(\ref{eq4-4-5}), (\ref{eq4-4-7}), and (\ref{eq4-4-9}).
\end{enumerate}

If the PQ symmetry is a global symmetry arising accidentally in the low energy effective theory, it is easily violated by higher dimensional operators
which is suppressed by powers of the Planck mass $M_P$~\cite{Dine:1992vx}. These operators might be identified as the $\Xi$ term, and their magnitude must be suppressed
in order not to shift the value of $\bar{\theta}$. The existence of long-lived domain walls makes a problem much worse, since the value of $\Xi$ cannot be arbitrary small.
Hence we conclude that the axion models with $N_{\mathrm{DW}}>1$ seems to be excluded from cosmological considerations,
except for the cases 2 and 3 enumerated above.
However, in the cases 2 and 3, some tunings on other theoretical parameters such as the inflationary scale (for case 2)
or the value of $\delta$ (for case 3) are required. Furthermore, even in the case 1,
there still remains a question of how $\Xi$ is suppressed, which should be addressed from some fundamental theories.
\acknowledgments{
Numerical computation in this work was carried out at the 
Yukawa Institute Computer Facility. 
This work is supported by Grant-in-Aid for
Scientific research from the Ministry of Education, Science, Sports, and
Culture (MEXT), Japan, No.14102004 and No.21111006 (M.~K.~)  and also by
World Premier International Research Center Initiative (WPI Initiative), MEXT, Japan.
K.~S.~and T.~S.~are supported by the Japan Society for the Promotion of Science (JSPS) through research fellowships.
T.~H.~was supported by JSPS Grant-in-Aid for Young Scientists (B)
No.23740186 and also by MEXT HPCI STRATEGIC PROGRAM.}
\appendix
\section{\label{secA} Numerical schemes}
In this appendix, we describe the analysis method which we used in the numerical studies.
Our numerical code is the combined version of that we used in past numerical studies~\cite{Hiramatsu:2010dx,Hiramatsu:2010yn,Hiramatsu:2010yu,Hiramatsu:2012gg}.
\subsection{\label{secA-1} Field equations}
The equation of motion for $\Phi$ is obtained by varying the Lagrangian density given by Eq.~(\ref{eq2-1-1}) with the potential
\begin{equation}
V(\Phi) = \frac{\lambda}{4}(|\Phi|^2-\eta^2)^2 + \frac{m^2\eta^2}{N_{\mathrm{DW}}^2}\left(1-\frac{|\Phi|}{\eta}\cos N_{\mathrm{DW}}\theta\right). \label{eqA-1-1}
\end{equation}
The second term of Eq.~(\ref{eqA-1-1}) is different from that of Eq.~(\ref{eq2-1-2}).
The original potential (\ref{eq2-1-2}) is not well defined at $\Phi=0$, which causes instabilities in the numerical studies.
We avoid this singularity by multiplying a factor $|\Phi|$ in front of the cosine term.
The difference between Eq.~(\ref{eq2-1-2}) and Eq.~(\ref{eqA-1-1}) is not important in the bulk region on which $|\Phi|=\eta$,
and we observe that the quantitative behavior such as time evolution of the scaling parameter of topological defects is not so much affected by this modification.

In the numerical studies, we follow the evolution of two real scalar fields $\phi_1$ and $\phi_2$ which are defined by $\Phi=\phi_1+i\phi_2$.
We normalize all dimensionful quantities in the unit of $\eta$, i.e. $\eta=1$. With this normalization, the equations of motion for two scalar fields in the FRW background become
\begin{eqnarray}
&&\ddot{\phi}_1 + 3H\dot{\phi}_1 - \frac{\nabla^2}{R^2(t)}\phi_1\nonumber\\
 &&\ = -\lambda\phi_1(\phi_1^2+\phi_2^2-1) + \frac{m^2}{N_{\mathrm{DW}}^2}(\cos\theta\cos N_{\mathrm{DW}}\theta+N_{\mathrm{DW}}\sin\theta\sin N_{\mathrm{DW}}\theta), \label{eqA-1-2} \\
&&\ddot{\phi}_2 + 3H\dot{\phi}_2 - \frac{\nabla^2}{R^2(t)}\phi_2 \nonumber\\
&&\ = -\lambda\phi_2(\phi_1^2+\phi_2^2-1) + \frac{m^2}{N_{\mathrm{DW}}^2}(\sin\theta\cos N_{\mathrm{DW}}\theta-N_{\mathrm{DW}}\cos\theta\sin N_{\mathrm{DW}}\theta). \label{eqA-1-3}
\end{eqnarray}
If we use the conformal time $\tau$ ($d\tau=dt/R(t)$), the above equations become
\begin{eqnarray}
&&f_1''-\nabla^2f_1\nonumber\\
&&\ = -\lambda f_1(|f_1|^2+|f_2|^2-R^2) + \frac{R^3m^2}{N_{\mathrm{DW}}^2}(\cos\theta\cos N_{\mathrm{DW}}\theta+N_{\mathrm{DW}}\sin\theta\sin N_{\mathrm{DW}}\theta), \label{eqA-1-4}\\
&&f_2''-\nabla^2f_2 \nonumber\\
&&\ = -\lambda f_2(|f_1|^2+|f_2|^2-R^2) + \frac{R^3m^2}{N_{\mathrm{DW}}^2}(\sin\cos N_{\mathrm{DW}}\theta-N_{\mathrm{DW}}\cos\theta\sin N_{\mathrm{DW}}\theta), \label{eqA-1-5}
\end{eqnarray}
where $f\equiv R\Phi=f_1+if_2$, and a prime denotes a derivative with respect to the conformal time $\tau$.
Here, we assume the radiation dominated background. This model has three parameters: $\lambda$, $N_{\mathrm{DW}}$ and $m$.
In the numerical studies, we fix the values of $\lambda$ and $m$, and vary the value of $N_{\mathrm{DW}}$.
\subsection{\label{secA-2} Discretization of differential equations}
To execute the time evolution which follows from Eqs.~(\ref{eqA-1-4}) and (\ref{eqA-1-5}), we used the symplectic integration scheme developed in~\cite{Yoshida:1990zz}.
In this numerical scheme the (implicit) Hamiltonian of the system is exactly conserved and there is no secular growth of the conserved quantity.
In practice, we compute the following sequence of mappings,
\begin{eqnarray}
f(\tau,{\bf x}) &\to& f(\tau+c_r\Delta\tau,{\bf x}) = f(\tau,{\bf x}) + c_r\Delta\tau f'(\tau,{\bf x}), \label{eqA-2-1}\\
f'(\tau,{\bf x}) &\to& f'(\tau+d_r\Delta\tau,{\bf x}) = f'(\tau,{\bf x}) + d_r\Delta\tau f''(\tau,{\bf x}), \label{eqA-2-2}
\end{eqnarray}
from $r=1$ to $r=4$, in each time step. The values of coefficients $c_r$ and $d_r$ are given by~\cite{Yoshida:1990zz}
\begin{eqnarray}
c_1=c_4=\frac{1}{2(2-2^{1/3})},\quad c_2=c_3=\frac{1-2^{1/3}}{2(2-2^{1/3})}, \nonumber\\
d_1 = d_3 = \frac{1}{2-2^{1/3}}, \quad d_2 = -\frac{2^{1/3}}{2-2^{1/3}},\quad d_4=0. \label{eqA-2-3}
\end{eqnarray}
It turns out that this integration scheme has fourth-order accuracy in time.

The spatial derivative and the Laplacian of the scalar fields are computed by using the finite difference method.
For example, the Laplacian is given by 
\begin{eqnarray}
(\nabla^2f)_{i,j,k} &=& \frac{1}{12(\Delta x)^2}\left[16(f_{i+1,j,k}+f_{i-1,j,k}+f_{i,j+1,k}+f_{i,j-1,k}+f_{i,j,k+1}+f_{i,j,k-1})\right.\nonumber\\
&& -(f_{i+2,j,k}+f_{i-2,j,k}+f_{i,j+2,k}+f_{i,j-2,k}+f_{i,j,k+2}+f_{i,j,k-2}) \nonumber\\
&&\left. - 90f_{i,j,k}\right], \label{eqA-2-4}
\end{eqnarray}
where $f_{i,j,k}=f(\tau,x_i,y_j,z_k)$, and $\Delta x=L/N$ is the grid spacing.
This formula has fourth-order accuracy in $\Delta x$.
We imposed the periodic boundary condition on the boundaries of the simulation box.
\subsection{\label{secA-3} Initial conditions}
We generate initial conditions in the similar way with Refs.~\cite{Hiramatsu:2010yn,Kawasaki:2011vv}.
We treat $\phi_1$ and $\phi_2$ as two independent real scalar fields, and assume that each of them has massless quantum fluctuations at the initial time.
With this assumption, we generate initial conditions as Gaussian random amplitudes in momentum space.
Then we Fourier transform them into the configuration space to give the spatial distribution of the fields.
We emphasize that this choice of initial conditions does not take account of the correct circumstances in the QCD epoch.
However, we expect that the result in the late time of the simulations are qualitatively unchanged if we used the different initial conditions,
since the evolution of defect networks in the scaling regime is not so much affected by the initial field configurations.
See appendix of Ref.~\cite{Hiramatsu:2010yn} for more discussions.
\section{\label{secB} Identification of topological defects}
Using the lattice algorithm described in Appendix~\ref{secA}, we can solve the time evolution of two real scalar fields
$\phi_1(t,{\bf x})$ and $\phi_2(t,{\bf x})$ in the comoving simulation box.
From these data of scalar fields, we estimate various physical quantities, such as the length of strings, the area of domain walls,
and the spectra of axions and gravitational waves radiated by defect networks.
In order to calculate such physical quantities, we have to identify the position of topological defects in the simulation box.
In this appendix, we summarize our identification method of topological defects, and describe how to calculate scaling
parameters of strings $\xi$ and domain walls ${\cal A}$ defined in Sec.~\ref{sec3-1}.
\subsection{\label{secB-1} Identification of strings}
We used the method developed in~\cite{Hiramatsu:2010yu} to identify the position of strings.
Consider a quadrate which consists of four neighboring grids in the simulation box (see points A, B, C, and D in Fig.~\ref{fig11}).
At each vertex point, we estimate the phase of the complex scalar field $\theta$ from the data of two real scalar fields $\phi_1$ and $\phi_2$.
Then, we obtain a mapping from the quadrate in the real space to the phase distribution in the field space, as shown in Fig.~\ref{fig11}.
Let us denote the minimum range of the phase which contains  all the images of four vertices as $\Delta\theta$.
The string exists within the quadrate if $\Delta\theta>\pi$ is satisfied, as long as the value of the phase changes continuously around the core of strings.

We can also determine the position at which a sting penetrates the quadrate.
At the boundary of the quadrate, we determine the positions of points on which $\phi_1=0$ or $\phi_2=0$ is satisfied, by using
linear interpolation of the values of $\phi_1$ and $\phi_2$.
If the string correctly penetrate the quadrate, there are two points with $\phi_1=0$ and two points with $\phi_2=0$ on the boundary of the quadrate.
Then we determine the position of the string as the point at which the line connecting two points with $\phi_1=0$
intersects with the line connecting two points with $\phi_2=0$ (see left-hand panel of Fig.~\ref{fig11}).

This simple criterion breaks down when the quadrate is penetrated by more than two strings.
However, we observed that such region is at most 1\% of the whole simulation box when the system relaxes into the scaling regime.
Therefore, this scheme enables us to determine the position of strings with at least 99\% accuracy.
\subsection{\label{secB-2} Identification of domain walls}
\begin{figure}[htbp]
\includegraphics[scale=0.46]{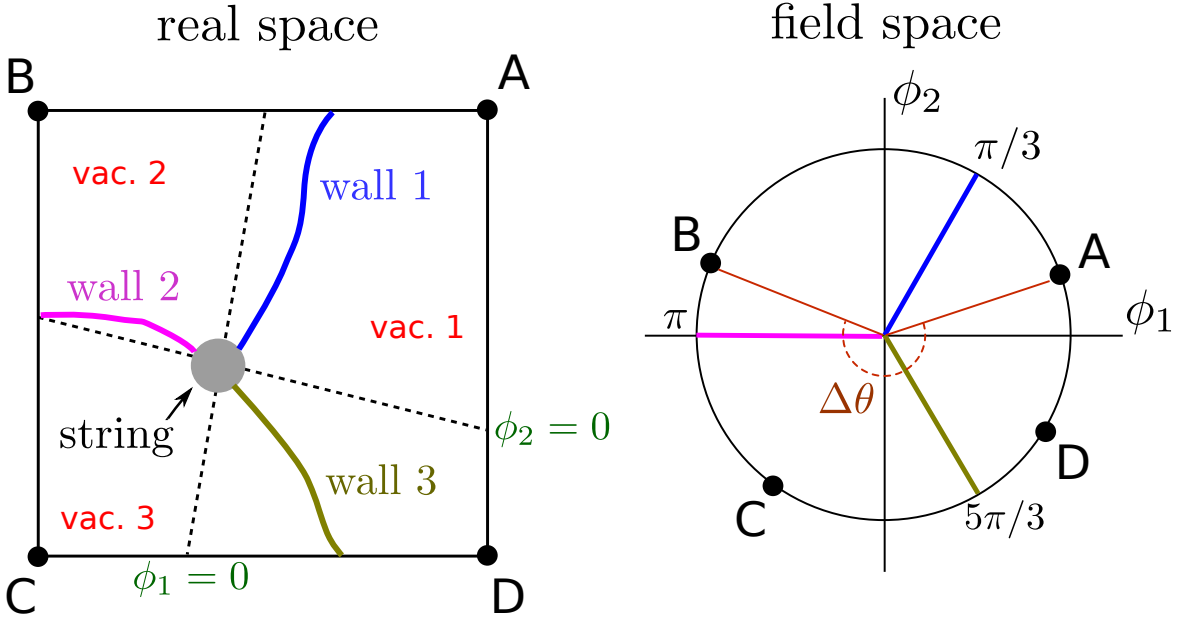}
\caption{Schematics of the identification method of topological defects.
Left panel shows a quadrate with four vertex points (A, B, C, and D) penetrated by a string and domain walls in the real space.
Right panel shows a mapping of them in the field space.
In this figure, we assume the model with $N_{\mathrm{DW}}=3$. Hence, there are three domain walls attached to the string.
The blue (wall~1), pink (wall~2), and yellow (wall~3) lines represent the locations of the center of domain walls, which correspond to
$\theta=\pi/3$, $\theta=\pi$, and $\theta=5\pi/3$, respectively, in the field space.
The region is separated into three domains, vac.~1, vac.~2, and vac.~3, which are surrounded by domain walls on the boundaries.
The dashed lines correspond to the loci of $\phi_1=0$ and $\phi_2=0$, which intersect on the core of the string.
In the field space, we define $\Delta\theta$ as the minimum range of the phase which contains all the images of four vertices.
A string penetrates the quadrate if the condition $\Delta\theta>\pi$ is satisfied.}
\label{fig11}
\end{figure}

The identification of domain walls is simpler than that of strings.
We separate the region of the phase of $\Phi$, $0\le\theta<2\pi$, into $N_{\mathrm{DW}}$ domains.
For example, if $N_{\mathrm{DW}}=3$, we obtain three domains with $0\le\theta<\pi/3$ and $5\pi/3\le\theta<2\pi$ (vac.~1),
$\pi/3\le\theta<\pi$ (vac.~2), and $\pi\le\theta<5\pi/3$ (vac.~3), as shown in Fig.~\ref{fig11}.
At each grid point, we compute the phase of the scalar field and assign the number of the vacuum domain (i.e. vac.~1, vac.~2, or vac.~3)
which contains that point. Let us call the neighboring grid points that differ by a unit lattice spacing $\Delta x=L/N$ as the ``link".
We identify that the domain wall intersects the link if the number of vacuum is different between the two ends of the link.

\subsection{\label{secB-3} Calculation of scaling parameters}
The scaling parameters of strings and domain walls are given by
\begin{equation}
\xi = \frac{\rho_{\mathrm{string}}}{\mu_{\mathrm{string}}}t^2\quad\mathrm{and}\quad{\cal A} = \frac{\rho_{\mathrm{wall}}}{\sigma_{\mathrm{wall}}}t, \label{eqB-3-1}
\end{equation}
where $\rho_{\mathrm{string}}$ is the energy density of strings, $\rho_{\mathrm{wall}}$ is the energy density of domain walls,
$\mu_{\mathrm{string}}$ is the mass energy of strings per unit length, and $\sigma_{\mathrm{wall}}$ is the surface mass density of domain walls.
It might be difficult to determine the values of $\mu_{\mathrm{string}}$ and $\sigma_{\mathrm{wall}}$ directly from the data of scalar fields
since they are quantities obtained by integrating $\rho_{\mathrm{string}}$ and $\rho_{\mathrm{wall}}$ over the coordinates which are perpendicular to
the direction along which the defects extend.
Instead of estimating $\mu_{\mathrm{string}}$ and $\sigma_{\mathrm{wall}}$, we calculate the length $l$ of strings and area $A$ of domain walls in the comoving
coordinates, such that
\begin{equation}
\rho_{\mathrm{string}} = \frac{\mu_{\mathrm{string}}l}{R^2(t)V}\quad \mathrm{and}\quad \rho_{\mathrm{wall}} = \frac{\sigma_{\mathrm{wall}}A}{R(t)V}, \label{eqB-3-2}
\end{equation}
where $V=L^3$ is the comoving volume of the simulation box.
Then, the scaling parameters are given by
\begin{equation}
\xi = \frac{lt^2}{R^2(t)V}\quad\mathrm{and}\quad{\cal A} = \frac{At}{R(t)V}. \label{eqB-3-3}
\end{equation}

The length of strings $l$ can be computed by connecting the loci of strings identified by the method described in the previous subsection.
The computation of the area of domain walls $A$ is not so straightforward, since it depends on the way how we define the segment of $A$
in each lattice. Here, we use the simple algorithm introduced by Press, Ryden, and Spergel~\cite{Press:1989yh}.
Define the quantity $\delta_{\pm}$ which takes the value 1 if the number of vacuum is different between the two ends of the link, and
the value 0 otherwise. We sum up $\delta_{\pm}$ for each of the links with the weighting factor
\begin{equation}
A = \Delta A \sum_{\mathrm{links}}\delta_{\pm}\frac{|\nabla\theta|}{|\theta_{,x}|+|\theta_{,y}|+|\theta_{,z}|}, \label{eqB-3-4}
\end{equation}
where $\Delta A=(\Delta x)^2$ is the area of one grid surface, and $\theta_{,x}$, etc. is a derivative of $\theta({\bf x})$ with respect to $x$.
The weighting factor given by the function of $\nabla\theta$ gives the average number of links per area segment.
This method takes account of the orientation of the surface of walls in the lattice.
See~\cite{Press:1989yh} for details.
\section{\label{secC} Calculation of power spectrum of radiated axions}
The calculation of power spectrum of axions radiated from topological defects is executed by using
the masking analysis method developed in Refs.~\cite{Hiramatsu:2010yu,Hiramatsu:2012gg}.
In this appendix, we describe the main line of the calculation scheme. 
Readers may also refer to~\cite{Hiramatsu:2010yu,Hiramatsu:2012gg} for complete description on the analysis method.

The power spectrum $P(k,t)$ of axions is defined by
\begin{equation}
\frac{1}{2}\langle\dot{a}(t,{\bf k})^*\dot{a}(t,{\bf k}')\rangle = \frac{(2\pi)^3}{k^2}\delta^{(3)}({\bf k}-{\bf k}')P(k,t), \label{eqC-1}
\end{equation}
where $\langle\dots\rangle$ represents an ensemble average.
$\dot{a}(t,{\bf k})$ is the Fourier component of the time derivative of the axion field
\begin{equation}
\dot{a}(t,{\bf k}) = \int d^3{\bf x} e^{i{\bf k}\cdot{\bf x}}\dot{a}(t,{\bf x}), \label{eqC-2}
\end{equation}
where the value of $\dot{a}(t,{\bf x})$ is obtained from the simulated data of $\Phi$ and $\dot{\Phi}$
\begin{equation}
\dot{a}(t,{\bf x}) = \mathrm{Im}\left[\frac{\dot{\Phi}}{\Phi}(t,{\bf x})\right]. \label{eqC-3}
\end{equation}
If $a(t,{\bf x})$ is a free field, it can be shown that $P(k,t)$ is proportional to the energy spectrum of axions~\cite{Hiramatsu:2012gg}.

The data of $\dot{a}$ obtained by numerical simulations contain the contaminations from the core of strings or domain walls.
We mask the contribution of axion field near strings or domain walls by introducing a window function
\begin{equation}
W({\bf x}) = \left\{
\begin{array}{l l}
0 & (\mathrm{near\ defects}) \\
1 & (\mathrm{elsewhere})
\end{array}
\right.
.\label{eqC-4}
\end{equation}
Then, we compute the power spectrum $\tilde{P}(k)$ of the masked axion field
\begin{equation}
\tilde{\dot{a}}({\bf x}) = W({\bf x})\dot{a}({\bf x}). \label{eqC-5}
\end{equation}
Unfortunately, the masked spectrum $\tilde{P}(k)$ is not equivalent to the true spectrum of free axions.
This discrepancy can be resolved by using the pseudo-power spectrum estimator (PPSE)
\begin{equation}
P_{\mathrm{PPSE}}(k) \equiv \frac{k^2}{V}\int\frac{dk'}{2\pi^2}M^{-1}(k,k')\tilde{P}(k'), \label{eqC-6}
\end{equation}
where $M^{-1}(k,k')$ is a matrix which satisfies
\begin{equation}
\int\frac{k'^2dk'}{2\pi^2}M^{-1}(k,k')M(k',k'') = \frac{2\pi^2}{k^2}\delta(k-k''). \label{eqC-7}
\end{equation}
Here, $M(k,k')$ is the window weight matrix defined by
\begin{equation}
M(k,k')\equiv \frac{1}{V^2}\int\frac{d\Omega_k}{4\pi}\frac{d\Omega_{k'}}{4\pi}|W({\bf k}-{\bf k}')|^2, \label{eqC-8}
\end{equation}
where $W({\bf k})$ is the Fourier transform of the window function $W({\bf x})$,
$\Omega_k$ is a unit vector representing the direction of ${\bf k}$, and $d\Omega_k=d\cos\theta d\phi$.
It turns out that the ensemble average of $P_{\mathrm{PPSE}}(k)$ is equivalent to the spectrum of radiated axions~\cite{Hiramatsu:2010yu}.

Naively, $M(k,k^\prime)$ in Eq.~(\ref{eqC-8}) can be evaluated by averaging $|W({\bf k}-{\bf k}^\prime)|^2/V^2$ 
over discrete ${\bf k}$ and ${\bf k}^\prime$ in the lattice. This however requires $\mathcal O(N^2)$
arithmetics, which is practically prohibited when the grid number $N$ is large as is in our analysis. 
Instead, we approximate $M(k,k^\prime)$ by partially taking the continuous limit, which we here 
describe briefly.
With a transformation ${\bf P}={\bf k}-{\bf k}^\prime$ and 
${\bf R}=\frac{{\bf k}+{\bf k}^\prime}2$, Eq.~(\ref{eqC-8}) can be rewritten as 
\begin{eqnarray}
M(k,k^\prime)&&=\frac{1}{(4\pi)^2 V^2k^2{k^\prime}^2}\int d^3{\bf P} |W({\bf P})|^2 \notag \\
&&\times \int d^3{\bf R}
~\delta\left(k-\left|{\bf R}+\frac{\bf P}2\right|\right)
\delta\left(k^\prime-\left|{\bf R}-\frac{\bf P}2\right|\right). \label{eqC-9}
\end{eqnarray}
In the continuous limit, the integration over $\bf R$ can be performed analytically.
\begin{eqnarray}
&&\int d^3{\bf R} 
~\delta\left(k-\left|{\bf R}+\frac{\bf P}2\right|\right)
\delta\left(k^\prime-\left|{\bf R}-\frac{\bf P}2\right|\right) \notag \\
&&\quad=2\pi\int^\infty_0 R^2dR\int^1_{-1}d\mu 
~\delta\left(R-\sqrt{\frac{k^2+{k^\prime}^2}2-\frac{P^2}{4}}\right)
\delta\left(\mu-\frac{k^2-{k^\prime}^2}{2RP}\right)\frac{kk^\prime}{R^2P}\notag \\
&&\quad=\frac{2\pi kk^\prime}{P}, \label{eqC-10}
\end{eqnarray}
where $P=|{\bf P}|$, $R=|{\bf R}|$ and $\mu\equiv\frac{{\bf R}\cdot{\bf P}}{RP}$.
We thus obtain 
\begin{equation}
M(k,k^\prime)=\frac1{8\pi V^2kk^\prime}\int d^3{\bf P} \frac{|W({\bf P})|^2}P. \label{eqC-11}
\end{equation}
Evaluation of Eq.~(\ref{eqC-11}) in a lattice has a good scaling with only $O(N)$ arithmetics.
On the other hand, the approximation can cause a systematic error
in estimation of $P_{\rm PPSE}(k)$, which shall be discussed shortly later.

The power spectrum $P_{\mathrm{PPSE}}(k,t)$ contains the contributions of axions produced before the time $t$.
This means that it contains the contributions of axions radiated at initial epoch where the defects do not relax into the scaling regime.
Since we are interested in the spectrum of axions radiated from topological defects which evolve along the scaling solution,
we must subtract the contribution of axions produced at initial relaxation regime.
This can be achieved by evaluating the power spectra for selected time steps $t_1$ and $t_2>t_1$,
and subtracting the spectrum obtained at $t_2$ from that obtained at $t_1$~\cite{Hiramatsu:2012gg}
\begin{equation}
\Delta P(k,t_2) = P_{\mathrm{PPSE}}(k,t_2) - P_{\mathrm{PPSE}}(k,t_1)\frac{\omega(k,t_2)}{\omega(k,t_1)}\left(\frac{R(t_1)}{R(t_2)}\right)^3, \label{eqC-12}
\end{equation}
where $(\omega(k,t_2)/\omega(k,t_1))(R(t_1)/R(t_2))^3$ is a dilution factor which comes from the effect of the cosmic expansion, and
\begin{equation}
\omega(k,t) = \sqrt{m^2+k^2/R(t)^2} \label{eqC-13}
\end{equation}
is the energy of axions with momentum $k/R(t)$.
Here, $t_1$ should be chosen as the time of the onset of the scaling regime.
Throughout this paper, we use $\tau_1=14$ as the subtraction time, at which the system begins to follow the scaling regime.

\begin{figure}[htbp]
\begin{center}
\includegraphics[scale=1.0]{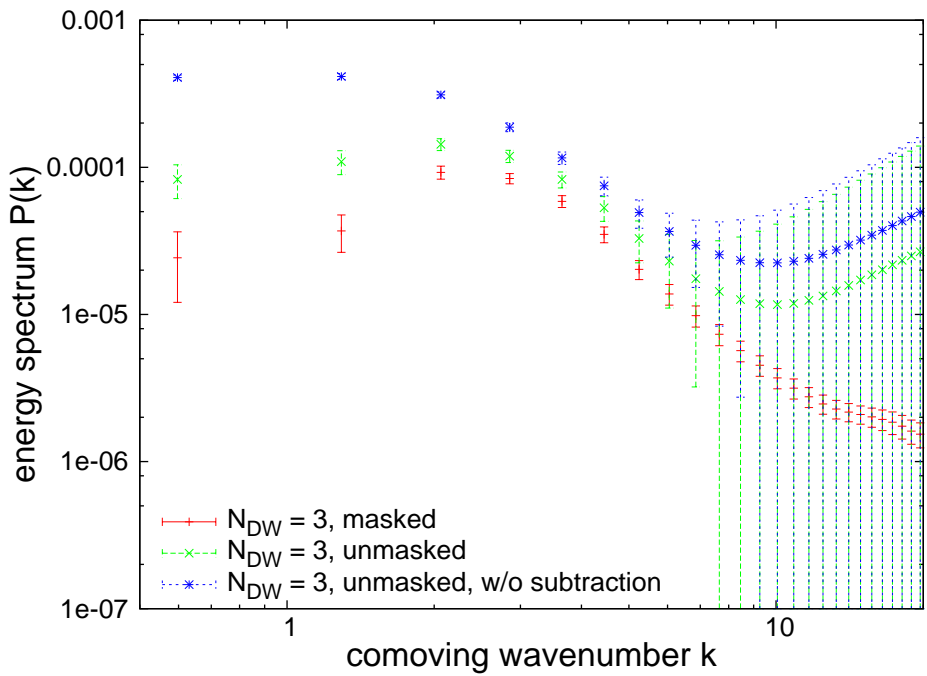}
\end{center}
\caption{The power spectrum of axions evaluated at the time $\tau_2=40$ for the case $N_{\mathrm{DW}}=3$.
The red plot shows the spectrum evaluated with masking the contamination of the core of topological defects and
subtracting the contribution of axions produced at initial stage ($\tau\le 14$).
The green plot shows the spectrum obtained by the same model without masking the contamination of defects.
The blue plot shows the spectrum evaluated without masking and subtraction.}
\label{fig12}
\end{figure}

Figure~\ref{fig12} shows the effect of this subtraction procedure. We see that the position of the peak is falsified, and the amplitude of the spectrum is overestimated
if we do not perform the subtraction. Figure~\ref{fig12} also shows that the spectrum is overestimated at higher momentum scale
when we do not mask the data near the position of topological defects.

\begin{figure}[htbp]
\begin{center}
\includegraphics[scale=1.0]{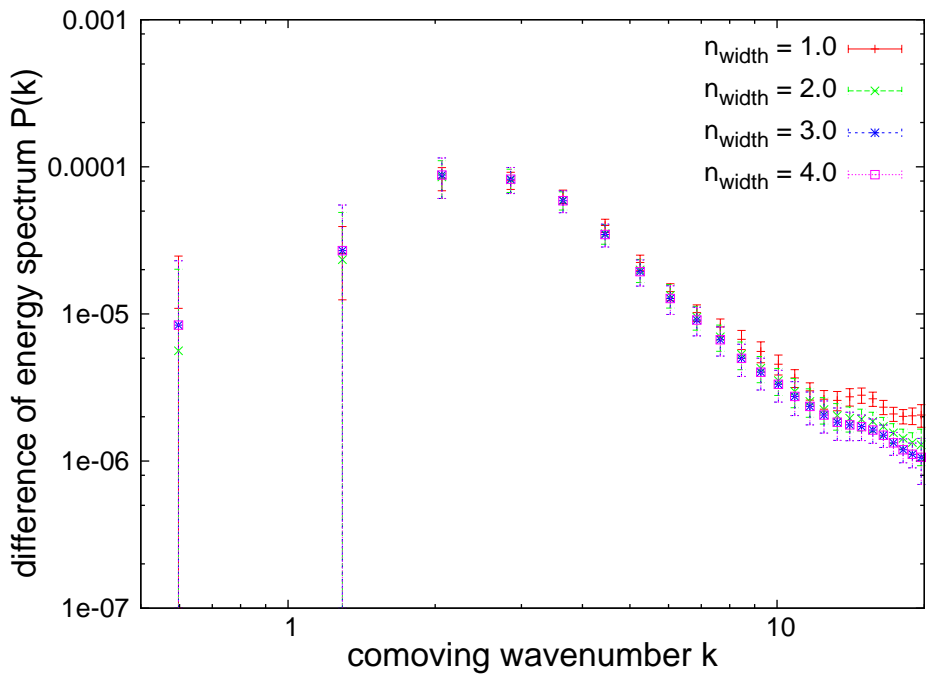}
\end{center}
\caption{The spectra of axions for various values of $n_{\mathrm{width}}$.
In these plots, we used the same parameters as those used in Fig.~\ref{fig4} for $N_{\mathrm{DW}}=3$.
These are the results obtained by a single realization simulation. Hence the error bars are estimated by $\tilde{C}(k,k')$
given by Eq.~(\ref{eqC-16}), but they do not contain the variance between different realizations, which is given by Eq.~(\ref{eqC-18}).}
\label{fig13}
\end{figure}

We note that the form of the power spectrum is affected by the choice of the region where the masking takes place.
In Eq.~(\ref{eqC-4}), we mask the field data over grid points within a distance $d_{\mathrm{width}}$ from the position of the core of defects
which we identified by using the methods described in Appendix~\ref{secB}.
Here we choose the critical distance as $d_{\mathrm{width}} = n_{\mathrm{width}}\delta_s = n_{\mathrm{width}}\lambda^{-1/2}$ for strings and
$d_{\mathrm{width}} = n_{\mathrm{width}}\delta_w = n_{\mathrm{width}}m^{-1}$ for walls, where $\delta_s$ and $\delta_w$ are widths of
strings and domain walls, respectively, and $n_{\mathrm{width}}$ is an arbitrary number.
Figure~\ref{fig13} shows the power spectrum for various values of $n_{\mathrm{width}}$.
We see that the power spectrum in large $k$ is overestimated when we use the smaller value of $n_{\mathrm{width}}$.
Since the spectra agree within the error bars for the results with $n_{\mathrm{width}}\ge 2$, we use $n_{\mathrm{width}}=2$
for numerical analyses presented in the text.

Finally, we comment on the estimation of errors of the power spectrum.
Error is estimated by the following procedure based on the Monte Carlo sampling.
After we compute $P_{\mathrm{PPSE}}(k)$, we generate $\dot{a}({\bf k})$ as Gaussian random field with a variance $P_{\mathrm{PPSE}}(k)/k^2$ in the Fourier space.
Then we execute the inverse Fourier transformation of $\dot{a}({\bf k})$ to obtain the data of $\dot{a}({\bf x})$ in the real space.
Let us denote this spurious field data as $\dot{a}_{\mathrm{MC}}({\bf x})$.
We apply the same masking procedure to the field $\dot{a}_{\mathrm{MC}}({\bf x})$ and obtain the power spectrum $P_{\mathrm{PPSE,MC}}(k)$.
We iterate this procedure by $n_{\mathrm{MC}}$ times, and compute the mean power spectrum
\begin{equation}
\bar{P}_{\mathrm{PPSE,MC}}(k) =\frac{1}{n_{\mathrm{MC}}}\sum^{n_{\mathrm{MC}}}_{i=1} P^i_{\mathrm{PPSE,MC}}(k) \label{eqC-14}
\end{equation}
and the covariance matrix of $P_{\mathrm{PPSE,MC}}(k)$
\begin{equation}
C(k,k') = 
\frac{1}{n_{\mathrm{MC}}}\sum^{n_{\mathrm{MC}}}_{i=1}(P^i_{\mathrm{PPSE,MC}}(k)-\bar{P}_{\mathrm{PPSE,MC}}(k))(P^i_{\mathrm{PPSE,MC}}(k')-\bar{P}_{\mathrm{PPSE,MC}}(k')), \label{eqC-15}
\end{equation}
where $P^i_{\mathrm{PPSE,MC}}(k)$ is the power spectrum obtained at $i$-th sampling step.
Although PPSE should be unbiased as formally shown in Ref.~\cite{Hiramatsu:2010yu}, 
we observed that $\bar{P}_{\mathrm{PPSE,MC}}(k)$ does not converge into $P_{\mathrm{PPSE}}(k)$ obtained by the field-theoretic simulation. This bias $(\bar{P}_{\mathrm{PPSE,MC}}(k)-P_{\mathrm{PPSE}}(k))$ arises mainly from use of the approximation $M(k,k^\prime)$ in Eq.~(\ref{eqC-11}). In our analysis, we regard the bias as 
the systematic error. Then the covariance matrix 
$\tilde C(k,k^\prime)$, which accounts for both the statistical and systematic errors,  can be given by 
\begin{eqnarray}
\tilde C(k,k^\prime)&&=C(k,k^\prime)
+(\bar P_{\rm PPSE,MC}(k)-P_{\rm PPSE}(k))(\bar P_{\rm PPSE,MC}(k^\prime)-P_{\rm PPSE}(k^\prime)) \notag \\
&&=\frac{1}{n_{\mathrm{MC}}}\sum^{n_{\mathrm{MC}}}_{i=1}(P^i_{\mathrm{PPSE,MC}}(k)-{P}_{\mathrm{PPSE}}(k))(P^i_{\mathrm{PPSE,MC}}(k')-{P}_{\mathrm{PPSE}}(k')). \label{eqC-16}
\end{eqnarray}
In addition to these errors in estimation of $P_{\rm PPSE}(k)$ in each realization, $P_{\rm PPSE}(k)$ also varies from one realization to another, variance from which should be also included as a statistical error.
Thus energy spectrum averaged over realizations and its covariance matrix can be finally given by
\begin{eqnarray}
\hat P_{\rm PPSE}(k)&=&\frac{1}{N_r}\sum^{N_r}_{j=1} P^{(j)}_{\rm PPSE}(k), \label{eqC-17}\\
\hat C(k,k^\prime)&=&\frac1{N_r}\sum^{N_r}_{j=1} \tilde C^{(j)}(k,k^\prime) \notag\\
&&\quad+\frac{1}{N_r-1}\sum^{N_r}_{j=1} [P^{(j)}_{\rm PPSE}(k)-\hat P_{\rm PPSE}(k)]
[P^{(j)}_{\rm PPSE}(k^\prime)-\hat P_{\rm PPSE}(k^\prime)], \label{eqC-18}
\end{eqnarray}
where $N_r$ is the number of realizations, and the subscript $(j)$ indicates a realization.
We execute the numerical simulations with $N_r=20$ different realizations of initial field configurations.
The results we present in the text is obtained from $\hat P_{\rm PPSE}(k)$ and 
$\hat C(k,k^\prime)$.
\section{\label{secD} Calculation of gravitational waves}
We use the method introduced in~\cite{Dufaux:2007pt} to calculate the spectrum of gravitational waves.
The evolution of gravitational waves is described by spatial metric perturbations $h_{ij}$ around the FRW background
\begin{equation}
ds^2 = -dt^2 + R^2(t)(\delta_{ij}+h_{ij})dx^idx^j. \label{eqD-1}
\end{equation}
Let us define the proper fluctuations as
\begin{equation}
\chi_{ij}({\bf x},\tau) \equiv R(\tau)h_{ij}({\bf x},\tau), \label{eqD-2}
\end{equation}
and denote their Fourier components as 
\begin{equation}
\tilde{\chi}_{ij}({\bf k},\tau) = \int d^3{\bf x}e^{i{\bf k\cdot x}}\chi_{ij}({\bf x},\tau). \label{eqD-3}
\end{equation}
In the Fourier space, the equations of motion for metric perturbations reduce to the form~\cite{Kawasaki:2011vv}
\begin{equation}
\left[\frac{\partial^2}{\partial y^2}+1\right]\tilde{\chi}_{ij} = \frac{S_{ij}}{k^2}, \label{eqD-4}
\end{equation}
where we defined the variable $y\equiv k\tau$ and the source
\begin{equation}
S_{ij}({\bf k},\tau) \equiv 16\pi GR(\tau)T^{\mathrm{TT}}_{ij}({\bf k},\tau). \label{eqD-5}
\end{equation}
The transverse traceless (TT) part of the stress-energy tensor is computed by applying the projection operator
\begin{align}
T^{\mathrm{TT}}_{ij}({\bf k},\tau) &= \Lambda_{ij,kl}(\hat{k})T_{ij}({\bf k},\tau) \notag\\
&= \Lambda_{ij,kl}(\hat{k})\{\partial_k\Phi^*\partial_l\Phi\}({\bf k},\tau), \label{eqD-6}\\
\Lambda_{ij,kl}(\hat{k}) &= P_{ik}(\hat{k})P_{jl}(\hat{k}) - \frac{1}{2}P_{ij}(\hat{k})P_{kl}(\hat{k}), \label{eqD-7}\\
P_{ij}(\hat{k}) &= \delta_{ij}-\hat{k}_i\hat{k}_j, \label{eqD-8}
\end{align}
where $\hat{k}={\bf k}/|{\bf k}|$, and $\{\partial_k\Phi^*\partial_l\Phi\}({\bf k},\tau)$ is the Fourier transform of $\partial_k\Phi^*({\bf x},\tau)\partial_l\Phi({\bf x},\tau)$.

There are two homogeneous solutions of Eq.~(\ref{eqD-4}), $\cos y$ and $\sin y$. The special solution is given by the time integral of the source term convoluted with
two homogeneous solutions
\begin{equation}
\tilde{\chi}_{ij}({\bf k},\tau) = C^{(1)}_{ij}({\bf k},\tau)\cos k\tau + C^{(2)}_{ij}({\bf k},\tau)\sin k\tau, \label{eqD-9}
\end{equation}
where
\begin{equation}
C^{(1)}_{ij} ({\bf k},\tau) = -\int^y_{y_i}dy'\sin k\tau'\frac{S_{ij}({\bf k},\tau')}{k^2} \equiv \frac{16\pi G}{k^2}\bar{C}^{(1)}_{ij}({\bf k},\tau), \label{eqD-10} 
\end{equation}
and
\begin{equation}
C^{(2)}_{ij} ({\bf k},\tau) = \int^y_{y_i}dy'\cos k\tau\frac{S_{ij}({\bf k},\tau')}{k^2} \equiv \frac{16\pi G}{k^2}\bar{C}^{(2)}_{ij}({\bf k},\tau). \label{eqD-11} 
\end{equation}

The energy density of the gravitational waves is given by
\begin{equation}
\rho_{\mathrm{gw}}(\tau) = \frac{1}{32\pi G R^4}\langle\chi'_{ij}({\bf x},\tau)\chi'_{ij}({\bf x},\tau)\rangle, \label{eqD-12}
\end{equation}
where a prime denotes a derivative with respect to conformal time $\tau$. We replace the ensemble average in Eq.~(\ref{eqD-12})
by an average over a volume $V$ of the comoving simulation box,
\begin{equation}
\rho_{\mathrm{gw}}(\tau) = \frac{1}{32\pi G R^4}\frac{1}{V}\int\frac{d^3{\bf k}}{(2\pi)^3}\chi'_{ij}({\bf k},\tau)\chi'^*_{ij}({\bf k},\tau). \label{eqD-13}
\end{equation}
Substituting the solution given by Eq.~(\ref{eqD-9}), and ignoring the terms with higher order in $RH$, we obtain
\begin{equation}
\rho_{\mathrm{gw}}(\tau) = \frac{4\pi G}{R^4V}\int\frac{d^3{\bf k}}{(2\pi)^3}\frac{1}{k^2}\sum_{ij}\left(\left|\bar{C}^{(1)}_{ij}\right|^2 + \left|\bar{C}^{(2)}_{ij}\right|^2 \right), \label{eqD-14}
\end{equation}
where we averaged over a period of the oscillation of $\chi_{ij}({\bf k},\tau)$ with time. The spectrum of gravitational waves is given by
\begin{equation}
\frac{d\rho_{\mathrm{gw}}}{d\ln k} = \frac{Gk}{2\pi^2VR^4}\int d\Omega_k\sum_{ij}\left(\left|\bar{C}^{(1)}_{ij}\right|^2 + \left|\bar{C}^{(2)}_{ij}\right|^2 \right). \label{eqD-15}
\end{equation}

We define the dimensionless energy spectrum of gravitational waves
\begin{equation}
\Omega_{\mathrm{gw}}(k,\tau) \equiv \frac{d\rho_{\mathrm{gw}}/d\ln k}{\rho_c(\tau)} = \frac{4}{3\pi V}\frac{G^2}{R^4H^2}S_k(\tau), \label{eqD-16}
\end{equation}
where
\begin{equation}
S_k(\tau) \equiv k\int d\Omega_k\sum_{ij}\left(\left|\bar{C}^{(1)}_{ij}\right|^2 + \left|\bar{C}^{(2)}_{ij}\right|^2 \right), \label{eqD-17}
\end{equation}
and $\rho_c(\tau)$ is the critical density of the universe at conformal time $\tau$
\begin{equation}
\rho_c(\tau) = \frac{3H^2}{8\pi G}. \label{eqD-18}
\end{equation}
In radiation dominated background, we expect that $\Omega_{\mathrm{gw}}(k,\tau)\propto S_k(\tau)$, since $R^4H^2$ remains constant.
Therefore, we calculate $S_k(\tau)$ in the numerical studies. Note that $\Omega_{\mathrm{gw}}(k,\tau)$ given by Eq.~(\ref{eqD-16})
does not correspond to the spectrum observed today.
The spectrum of gravitational waves observed today is obtained by multiplying $\Omega_{\mathrm{gw}}(k,\tau)$ with a damping factor
which arises from the following matter-dominated epoch.

As we discussed in Sec.~\ref{sec3-3}, we subtract the contribution of gravitational waves produced at the initial epoch
where the defects do not relax into the scaling regime. This subtraction can be executed as follows.
Since the energy density of gravitational waves shifts as $\propto R^{-4}$, the quantity $S_k(\tau)\propto R(\tau)^4d\rho_{\mathrm{gw}}(\tau)/d\ln k$
is proportional to the number of gravitons with the momentum $k$ per comoving volume.
Therefore, the difference of the spectral function
\begin{equation}
\Delta S_k(\tau) \equiv S_k(\tau) - S_k(\tau_g) \label{eqD-19}
\end{equation}
gives the spectrum of gravitational waves produced during the time interval [$\tau_g,\tau$].
Here $\tau_g$ is a reference time which we treat as a free parameter in the numerical simulations.
\section{\label{secE} Surface mass density of domain walls}
In this appendix, following Ref.~\cite{2000csot.book.....V}, we shortly comment on the surface mass density of domain walls in the models used in numerical simulations.
First, consider the model of scalar field $\phi$ with the potential given by
\begin{equation}
V(\phi) = \frac{\lambda_r}{4}(\phi^2-\eta_r^2)^2, \label{eqE-1}
\end{equation}
where $\lambda_r$ and $\eta_r$ are constants. This model has a discrete $Z_2$ symmetry in which the scalar field transforms as $\phi\to -\phi$.
In the flat Minkowski background, the classical field equation of this model has a static solution
\begin{equation}
\phi(z) = \eta_r\tanh\left[\sqrt{\frac{\lambda_r}{2}}\eta_r z\right], \label{eqE-2}
\end{equation}
which describes a domain wall lying perpendicular to the $z$-axis. By integrating the energy density of the field over the direction
perpendicular to the surface of the wall, we obtain the surface mass density
\begin{equation}
\sigma_{\mathrm{wall}} = \int^{\infty}_{-\infty}dz\left(\frac{d\phi}{dz}\right)^2 = \frac{4}{3}\sqrt{\frac{\lambda_r}{2}}\eta_r^3. \label{eqE-3}
\end{equation}

Next, let us consider the axionic model with the Lagrangian given by Eq.~(\ref{eq2-1-1}). If we restrict ourselves to low energy configurations at the QCD scale,
we can put $\Phi=\eta e^{i\theta}$, which gives the effective potential for $\theta$
\begin{equation}
{\cal L}_{\theta} = \frac{\eta^2}{2}(\partial_{\mu}\theta)^2 + \frac{m^2\eta^2}{N_{\mathrm{DW}}^2}(1-\cos N_{\mathrm{DW}}\theta). \label{eqE-4}
\end{equation}
The classical field equation in the Minkowski background again yields a planner solution perpendicular to the $z$-axis
\begin{equation}
\theta(z) = \frac{2\pi k}{N_{\mathrm{DW}}} + \frac{4}{N_{\mathrm{DW}}}\tan^{-1}\exp(mz), \qquad k=0,1,\dots,N_{\mathrm{DW}}-1. \label{eqE-5}
\end{equation}
Integrating out the energy density, we obtain the surface mass density of the domain wall
\begin{equation}
\sigma_{\mathrm{wall}} = \int^{\infty}_{-\infty}dz\eta^2\left(\frac{d\theta}{dz}\right)^2 = \frac{8m\eta^2}{N_{\mathrm{DW}}^2}. \label{eqE-6}
\end{equation}

Comparing Eqs.~(\ref{eqE-3}) and (\ref{eqE-6}), the values of $\sigma_{\mathrm{wall}}$ coincide between these two models if
we set $\lambda_r=4m^2/3\eta^2$, $\eta_r=(3/2)^{1/2}\eta$, and $N_{\mathrm{DW}}=2$.
Also, the width of domain walls becomes the same order $\delta_w\simeq (\lambda_r/2)^{-1/2}\eta_r^{-1} \simeq m^{-1}$.
Substituting these values of $\lambda_r$ and $\eta_r$ into Eq.~(\ref{eqE-1}), we obtain the potential given by (\ref{eq3-3}).

We note that the surface mass density of axionic domain wall, given by Eq.~(\ref{eqE-6}) should be modified
if we include the structure of the neutral pion field which varies inside the wall.
In this case, the surface mass density is given by~\cite{Huang:1985tt}
\begin{equation}
\sigma_{\mathrm{wall}} = 4.32f_{\pi}m_{\pi}\eta/N_{\mathrm{DW}} \simeq 9.23m\eta^2/N_{\mathrm{DW}}^2, \label{eqE-7}
\end{equation}
where $f_{\pi}$ is the pion decay constant, and $m_{\pi}$ is the mass of the pion.
In the second equality, we used $m_u/m_d\simeq 0.48$~\cite{Kim:2008hd}, where $m_u$ and $m_d$ are the mass of up quark and down quark, respectively.


\begin{thebibliography}{66}
\expandafter\ifx\csname natexlab\endcsname\relax\def\natexlab#1{#1}\fi
\expandafter\ifx\csname bibnamefont\endcsname\relax
  \def\bibnamefont#1{#1}\fi
\expandafter\ifx\csname bibfnamefont\endcsname\relax
  \def\bibfnamefont#1{#1}\fi
\expandafter\ifx\csname citenamefont\endcsname\relax
  \def\citenamefont#1{#1}\fi
\expandafter\ifx\csname url\endcsname\relax
  \def\url#1{\texttt{#1}}\fi
\expandafter\ifx\csname urlprefix\endcsname\relax\def\urlprefix{URL }\fi
\providecommand{\bibinfo}[2]{#2}
\providecommand{\eprint}[2][]{\url{#2}}

\bibitem[{\citenamefont{Weinberg}(1978)}]{Weinberg:1977ma}
\bibinfo{author}{\bibfnamefont{S.}~\bibnamefont{Weinberg}},
  \bibinfo{journal}{Phys.Rev.Lett.} \textbf{\bibinfo{volume}{40}},
  \bibinfo{pages}{223} (\bibinfo{year}{1978});
\bibinfo{author}{\bibfnamefont{F.}~\bibnamefont{Wilczek}},
  \bibinfo{journal}{Phys.Rev.Lett.} \textbf{\bibinfo{volume}{40}},
  \bibinfo{pages}{279} (\bibinfo{year}{1978}).

\bibitem[{\citenamefont{Peccei and Quinn}(1977{\natexlab{a}})}]{Peccei:1977hh}
\bibinfo{author}{\bibfnamefont{R.}~\bibnamefont{Peccei}} \bibnamefont{and}
  \bibinfo{author}{\bibfnamefont{H.~R.} \bibnamefont{Quinn}},
  \bibinfo{journal}{Phys.Rev.Lett.} \textbf{\bibinfo{volume}{38}},
  \bibinfo{pages}{1440} (\bibinfo{year}{1977}{\natexlab{a}});
\bibinfo{author}{\bibfnamefont{R.}~\bibnamefont{Peccei}} \bibnamefont{and}
  \bibinfo{author}{\bibfnamefont{H.~R.} \bibnamefont{Quinn}},
  \bibinfo{journal}{Phys.Rev.} \textbf{\bibinfo{volume}{D16}},
  \bibinfo{pages}{1791} (\bibinfo{year}{1977}{\natexlab{b}}).

\bibitem[{\citenamefont{Kim and Carosi}(2010)}]{Kim:2008hd}
\bibinfo{author}{\bibfnamefont{J.~E.} \bibnamefont{Kim}} \bibnamefont{and}
  \bibinfo{author}{\bibfnamefont{G.}~\bibnamefont{Carosi}},
  \bibinfo{journal}{Rev.Mod.Phys.} \textbf{\bibinfo{volume}{82}},
  \bibinfo{pages}{557} (\bibinfo{year}{2010}), \eprint{0807.3125}.

\bibitem[{\citenamefont{Kim}(1987)}]{Kim:1986ax}
\bibinfo{author}{\bibfnamefont{J.~E.} \bibnamefont{Kim}},
  \bibinfo{journal}{Phys.Rept.} \textbf{\bibinfo{volume}{150}},
  \bibinfo{pages}{1} (\bibinfo{year}{1987}).

\bibitem[{\citenamefont{Kim}(1979)}]{Kim:1979if}
\bibinfo{author}{\bibfnamefont{J.~E.} \bibnamefont{Kim}},
  \bibinfo{journal}{Phys.Rev.Lett.} \textbf{\bibinfo{volume}{43}},
  \bibinfo{pages}{103} (\bibinfo{year}{1979});
\bibinfo{author}{\bibfnamefont{M.~A.} \bibnamefont{Shifman}},
  \bibinfo{author}{\bibfnamefont{A.}~\bibnamefont{Vainshtein}},
  \bibnamefont{and} \bibinfo{author}{\bibfnamefont{V.~I.}
  \bibnamefont{Zakharov}}, \bibinfo{journal}{Nucl.Phys.}
  \textbf{\bibinfo{volume}{B166}}, \bibinfo{pages}{493} (\bibinfo{year}{1980}).

\bibitem[{\citenamefont{Dine et~al.}(1981)\citenamefont{Dine, Fischler, and
  Srednicki}}]{Dine:1981rt}
\bibinfo{author}{\bibfnamefont{M.}~\bibnamefont{Dine}},
  \bibinfo{author}{\bibfnamefont{W.}~\bibnamefont{Fischler}}, \bibnamefont{and}
  \bibinfo{author}{\bibfnamefont{M.}~\bibnamefont{Srednicki}},
  \bibinfo{journal}{Phys.Lett.} \textbf{\bibinfo{volume}{B104}},
  \bibinfo{pages}{199} (\bibinfo{year}{1981});
\bibinfo{author}{\bibfnamefont{A.}~\bibnamefont{Zhitnitsky}},
  \bibinfo{journal}{Sov.J.Nucl.Phys.} \textbf{\bibinfo{volume}{31}},
  \bibinfo{pages}{260} (\bibinfo{year}{1980}).

\bibitem[{\citenamefont{Preskill et~al.}(1983)\citenamefont{Preskill, Wise, and
  Wilczek}}]{Preskill:1982cy}
\bibinfo{author}{\bibfnamefont{J.}~\bibnamefont{Preskill}},
  \bibinfo{author}{\bibfnamefont{M.~B.} \bibnamefont{Wise}}, \bibnamefont{and}
  \bibinfo{author}{\bibfnamefont{F.}~\bibnamefont{Wilczek}},
  \bibinfo{journal}{Phys.Lett.} \textbf{\bibinfo{volume}{B120}},
  \bibinfo{pages}{127} (\bibinfo{year}{1983});
\bibinfo{author}{\bibfnamefont{L.}~\bibnamefont{Abbott}} \bibnamefont{and}
  \bibinfo{author}{\bibfnamefont{P.}~\bibnamefont{Sikivie}},
  \bibinfo{journal}{Phys.Lett.} \textbf{\bibinfo{volume}{B120}},
  \bibinfo{pages}{133} (\bibinfo{year}{1983});
\bibinfo{author}{\bibfnamefont{M.}~\bibnamefont{Dine}} \bibnamefont{and}
  \bibinfo{author}{\bibfnamefont{W.}~\bibnamefont{Fischler}},
  \bibinfo{journal}{Phys.Lett.} \textbf{\bibinfo{volume}{B120}},
  \bibinfo{pages}{137} (\bibinfo{year}{1983}).

\bibitem[{\citenamefont{Vilenkin and Shellard}(2000)}]{2000csot.book.....V}
\bibinfo{author}{\bibfnamefont{A.}~\bibnamefont{Vilenkin}} \bibnamefont{and}
  \bibinfo{author}{\bibfnamefont{E.~P.~S.} \bibnamefont{Shellard}},
  \emph{\bibinfo{title}{{Cosmic Strings and Other Topological Defects}}}
  (\bibinfo{publisher}{Cambridge University Press}, \bibinfo{year}{2000}).

\bibitem[{\citenamefont{Sikivie}(1982)}]{Sikivie:1982qv}
\bibinfo{author}{\bibfnamefont{P.}~\bibnamefont{Sikivie}},
  \bibinfo{journal}{Phys.Rev.Lett.} \textbf{\bibinfo{volume}{48}},
  \bibinfo{pages}{1156} (\bibinfo{year}{1982}).

\bibitem[{\citenamefont{Georgi and Wise}(1982)}]{Georgi:1982ph}
\bibinfo{author}{\bibfnamefont{H.}~\bibnamefont{Georgi}} \bibnamefont{and}
  \bibinfo{author}{\bibfnamefont{M.~B.} \bibnamefont{Wise}},
  \bibinfo{journal}{Phys.Lett.} \textbf{\bibinfo{volume}{B116}},
  \bibinfo{pages}{123} (\bibinfo{year}{1982}).

\bibitem[{\citenamefont{Hiramatsu et~al.}(2012)\citenamefont{Hiramatsu,
  Kawasaki, Saikawa, and Sekiguchi}}]{Hiramatsu:2012gg}
\bibinfo{author}{\bibfnamefont{T.}~\bibnamefont{Hiramatsu}},
  \bibinfo{author}{\bibfnamefont{M.}~\bibnamefont{Kawasaki}},
  \bibinfo{author}{\bibfnamefont{K.}~\bibnamefont{Saikawa}}, \bibnamefont{and}
  \bibinfo{author}{\bibfnamefont{T.}~\bibnamefont{Sekiguchi}},
  \bibinfo{journal}{Phys. Rev.} \textbf{\bibinfo{volume}{D85}},
  \bibinfo{pages}{105020} (\bibinfo{year}{2012}), \eprint{1202.5851}.

\bibitem[{\citenamefont{Zeldovich et~al.}(1974)\citenamefont{Zeldovich,
  Kobzarev, and Okun}}]{Zeldovich:1974uw}
\bibinfo{author}{\bibfnamefont{Y.}~\bibnamefont{Zeldovich}},
  \bibinfo{author}{\bibfnamefont{I.~Y.} \bibnamefont{Kobzarev}},
  \bibnamefont{and} \bibinfo{author}{\bibfnamefont{L.}~\bibnamefont{Okun}},
  \bibinfo{journal}{Zh.Eksp.Teor.Fiz.} \textbf{\bibinfo{volume}{67}},
  \bibinfo{pages}{3} (\bibinfo{year}{1974}).

\bibitem[{\citenamefont{Gelmini et~al.}(1989)\citenamefont{Gelmini, Gleiser,
  and Kolb}}]{Gelmini:1988sf}
\bibinfo{author}{\bibfnamefont{G.~B.} \bibnamefont{Gelmini}},
  \bibinfo{author}{\bibfnamefont{M.}~\bibnamefont{Gleiser}}, \bibnamefont{and}
  \bibinfo{author}{\bibfnamefont{E.~W.} \bibnamefont{Kolb}},
  \bibinfo{journal}{Phys.Rev.} \textbf{\bibinfo{volume}{D39}},
  \bibinfo{pages}{1558} (\bibinfo{year}{1989});
\bibinfo{author}{\bibfnamefont{S.~E.} \bibnamefont{Larsson}},
  \bibinfo{author}{\bibfnamefont{S.}~\bibnamefont{Sarkar}}, \bibnamefont{and}
  \bibinfo{author}{\bibfnamefont{P.~L.} \bibnamefont{White}},
  \bibinfo{journal}{Phys.Rev.} \textbf{\bibinfo{volume}{D55}},
  \bibinfo{pages}{5129} (\bibinfo{year}{1997}), \eprint{hep-ph/9608319}.

\bibitem[{\citenamefont{Lazarides and Shafi}(1982)}]{Lazarides:1982tw}
\bibinfo{author}{\bibfnamefont{G.}~\bibnamefont{Lazarides}} \bibnamefont{and}
  \bibinfo{author}{\bibfnamefont{Q.}~\bibnamefont{Shafi}},
  \bibinfo{journal}{Phys.Lett.} \textbf{\bibinfo{volume}{B115}},
  \bibinfo{pages}{21} (\bibinfo{year}{1982});
\bibinfo{author}{\bibfnamefont{S.~M.} \bibnamefont{Barr}},
  \bibinfo{author}{\bibfnamefont{X.}~\bibnamefont{Gao}}, \bibnamefont{and}
  \bibinfo{author}{\bibfnamefont{D.}~\bibnamefont{Reiss}},
  \bibinfo{journal}{Phys.Rev.} \textbf{\bibinfo{volume}{D26}},
  \bibinfo{pages}{2176} (\bibinfo{year}{1982});
\bibinfo{author}{\bibfnamefont{K.}~\bibnamefont{Choi}} \bibnamefont{and}
  \bibinfo{author}{\bibfnamefont{J.~E.} \bibnamefont{Kim}},
  \bibinfo{journal}{Phys.Rev.Lett.} \textbf{\bibinfo{volume}{55}},
  \bibinfo{pages}{2637} (\bibinfo{year}{1985}).
  
  \bibitem[{\citenamefont{Lyth}(1990)}]{Lyth:1989pb}
\bibinfo{author}{\bibfnamefont{D.~H.} \bibnamefont{Lyth}},
  \bibinfo{journal}{Phys.Lett.} \textbf{\bibinfo{volume}{B236}},
  \bibinfo{pages}{408} (\bibinfo{year}{1990});
\bibinfo{author}{\bibfnamefont{M.~S.} \bibnamefont{Turner}} \bibnamefont{and}
  \bibinfo{author}{\bibfnamefont{F.}~\bibnamefont{Wilczek}},
  \bibinfo{journal}{Phys.Rev.Lett.} \textbf{\bibinfo{volume}{66}},
  \bibinfo{pages}{5} (\bibinfo{year}{1991});
\bibinfo{author}{\bibfnamefont{A.~D.} \bibnamefont{Linde}},
  \bibinfo{journal}{Phys.Lett.} \textbf{\bibinfo{volume}{B259}},
  \bibinfo{pages}{38} (\bibinfo{year}{1991});
\bibinfo{author}{\bibfnamefont{M.}~\bibnamefont{Kawasaki}} \bibnamefont{and}
  \bibinfo{author}{\bibfnamefont{T.}~\bibnamefont{Sekiguchi}},
  \bibinfo{journal}{Prog.Theor.Phys.} \textbf{\bibinfo{volume}{120}},
  \bibinfo{pages}{995} (\bibinfo{year}{2008}), \eprint{0705.2853};
\bibinfo{author}{\bibfnamefont{M.~P.} \bibnamefont{Hertzberg}},
  \bibinfo{author}{\bibfnamefont{M.}~\bibnamefont{Tegmark}}, \bibnamefont{and}
  \bibinfo{author}{\bibfnamefont{F.}~\bibnamefont{Wilczek}},
  \bibinfo{journal}{Phys.Rev.} \textbf{\bibinfo{volume}{D78}},
  \bibinfo{pages}{083507} (\bibinfo{year}{2008}), \eprint{0807.1726};
\bibinfo{author}{\bibfnamefont{K.~J.} \bibnamefont{Mack}},
  \bibinfo{journal}{JCAP} \textbf{\bibinfo{volume}{1107}}, \bibinfo{pages}{021}
  (\bibinfo{year}{2011}), \eprint{0911.0421}.

\bibitem[{\citenamefont{Rai and Senjanovic}(1994)}]{Rai:1992xw}
\bibinfo{author}{\bibfnamefont{B.}~\bibnamefont{Rai}} \bibnamefont{and}
  \bibinfo{author}{\bibfnamefont{G.}~\bibnamefont{Senjanovic}},
  \bibinfo{journal}{Phys.Rev.} \textbf{\bibinfo{volume}{D49}},
  \bibinfo{pages}{2729} (\bibinfo{year}{1994}), \eprint{hep-ph/9301240}.

\bibitem[{\citenamefont{Kamionkowski and
  March-Russell}(1992)}]{Kamionkowski:1992mf}
\bibinfo{author}{\bibfnamefont{M.}~\bibnamefont{Kamionkowski}}
  \bibnamefont{and}
  \bibinfo{author}{\bibfnamefont{J.}~\bibnamefont{March-Russell}},
  \bibinfo{journal}{Phys.Lett.} \textbf{\bibinfo{volume}{B282}},
  \bibinfo{pages}{137} (\bibinfo{year}{1992}), \eprint{hep-th/9202003};
\bibinfo{author}{\bibfnamefont{R.}~\bibnamefont{Holman}},
  \bibinfo{author}{\bibfnamefont{S.~D.} \bibnamefont{Hsu}},
  \bibinfo{author}{\bibfnamefont{T.~W.} \bibnamefont{Kephart}},
  \bibinfo{author}{\bibfnamefont{E.~W.} \bibnamefont{Kolb}},
  \bibinfo{author}{\bibfnamefont{R.}~\bibnamefont{Watkins}},
  \bibnamefont{et~al.}, \bibinfo{journal}{Phys.Lett.}
  \textbf{\bibinfo{volume}{B282}}, \bibinfo{pages}{132} (\bibinfo{year}{1992}),
  \eprint{hep-ph/9203206};
\bibinfo{author}{\bibfnamefont{B.~A.} \bibnamefont{Dobrescu}},
  \bibinfo{journal}{Phys.Rev.} \textbf{\bibinfo{volume}{D55}},
  \bibinfo{pages}{5826} (\bibinfo{year}{1997}), \eprint{hep-ph/9609221}.
  
  \bibitem[{\citenamefont{Barr and Seckel}(1992)}]{Barr:1992qq}
\bibinfo{author}{\bibfnamefont{S.~M.} \bibnamefont{Barr}} \bibnamefont{and}
  \bibinfo{author}{\bibfnamefont{D.}~\bibnamefont{Seckel}},
  \bibinfo{journal}{Phys.Rev.} \textbf{\bibinfo{volume}{D46}},
  \bibinfo{pages}{539} (\bibinfo{year}{1992}).

\bibitem[{\citenamefont{Dine}(1992)}]{Dine:1992vx}
\bibinfo{author}{\bibfnamefont{M.}~\bibnamefont{Dine}} (\bibinfo{year}{1992}),
  \eprint{hep-th/9207045}.

\bibitem[{\citenamefont{Hiramatsu
  et~al.}(2011{\natexlab{a}})\citenamefont{Hiramatsu, Kawasaki, and
  Saikawa}}]{Hiramatsu:2010yn}
\bibinfo{author}{\bibfnamefont{T.}~\bibnamefont{Hiramatsu}},
  \bibinfo{author}{\bibfnamefont{M.}~\bibnamefont{Kawasaki}}, \bibnamefont{and}
  \bibinfo{author}{\bibfnamefont{K.}~\bibnamefont{Saikawa}},
  \bibinfo{journal}{JCAP} \textbf{\bibinfo{volume}{1108}}, \bibinfo{pages}{030}
  (\bibinfo{year}{2011}{\natexlab{a}}), \eprint{1012.4558}.

\bibitem[{\citenamefont{Dine et~al.}(2010)\citenamefont{Dine, Takahashi, and
  Yanagida}}]{Dine:2010eb}
\bibinfo{author}{\bibfnamefont{M.}~\bibnamefont{Dine}},
  \bibinfo{author}{\bibfnamefont{F.}~\bibnamefont{Takahashi}},
  \bibnamefont{and} \bibinfo{author}{\bibfnamefont{T.~T.}
  \bibnamefont{Yanagida}}, \bibinfo{journal}{JHEP}
  \textbf{\bibinfo{volume}{1007}}, \bibinfo{pages}{003} (\bibinfo{year}{2010}),
  \eprint{1005.3613}.

\bibitem[{\citenamefont{Moroi and Nakayama}(2011)}]{Moroi:2011be}
\bibinfo{author}{\bibfnamefont{T.}~\bibnamefont{Moroi}} \bibnamefont{and}
  \bibinfo{author}{\bibfnamefont{K.}~\bibnamefont{Nakayama}},
  \bibinfo{journal}{Phys.Lett.} \textbf{\bibinfo{volume}{B703}},
  \bibinfo{pages}{160} (\bibinfo{year}{2011}), \eprint{1105.6216}.

\bibitem[{\citenamefont{Hamaguchi et~al.}(2012)\citenamefont{Hamaguchi,
  Nakayama, and Yokozaki}}]{Hamaguchi:2011nm}
\bibinfo{author}{\bibfnamefont{K.}~\bibnamefont{Hamaguchi}},
  \bibinfo{author}{\bibfnamefont{K.}~\bibnamefont{Nakayama}}, \bibnamefont{and}
  \bibinfo{author}{\bibfnamefont{N.}~\bibnamefont{Yokozaki}},
  \bibinfo{journal}{Phys.Lett.} \textbf{\bibinfo{volume}{B708}},
  \bibinfo{pages}{100} (\bibinfo{year}{2012}), \eprint{1107.4760}.

\bibitem[{\citenamefont{Davis}(1986)}]{Davis:1986xc}
\bibinfo{author}{\bibfnamefont{R.~L.} \bibnamefont{Davis}},
  \bibinfo{journal}{Phys.Lett.} \textbf{\bibinfo{volume}{B180}},
  \bibinfo{pages}{225} (\bibinfo{year}{1986}).

\bibitem[{\citenamefont{Davis and Shellard}(1989)}]{Davis:1989nj}
\bibinfo{author}{\bibfnamefont{R.}~\bibnamefont{Davis}} \bibnamefont{and}
  \bibinfo{author}{\bibfnamefont{E.}~\bibnamefont{Shellard}},
  \bibinfo{journal}{Nucl.Phys.} \textbf{\bibinfo{volume}{B324}},
  \bibinfo{pages}{167} (\bibinfo{year}{1989});
\bibinfo{author}{\bibfnamefont{A.}~\bibnamefont{Dabholkar}} \bibnamefont{and}
  \bibinfo{author}{\bibfnamefont{J.~M.} \bibnamefont{Quashnock}},
  \bibinfo{journal}{Nucl.Phys.} \textbf{\bibinfo{volume}{B333}},
  \bibinfo{pages}{815} (\bibinfo{year}{1990});
\bibinfo{author}{\bibfnamefont{R.}~\bibnamefont{Battye}} \bibnamefont{and}
  \bibinfo{author}{\bibfnamefont{E.}~\bibnamefont{Shellard}},
  \bibinfo{journal}{Nucl.Phys.} \textbf{\bibinfo{volume}{B423}},
  \bibinfo{pages}{260} (\bibinfo{year}{1994}{\natexlab{a}}),
  \eprint{astro-ph/9311017};
\bibinfo{author}{\bibfnamefont{R.}~\bibnamefont{Battye}} \bibnamefont{and}
  \bibinfo{author}{\bibfnamefont{E.}~\bibnamefont{Shellard}},
  \bibinfo{journal}{Phys.Rev.Lett.} \textbf{\bibinfo{volume}{73}},
  \bibinfo{pages}{2954} (\bibinfo{year}{1994}{\natexlab{b}}),
  \eprint{astro-ph/9403018}.

\bibitem[{\citenamefont{Hagmann and Sikivie}(1991)}]{Hagmann:1990mj}
\bibinfo{author}{\bibfnamefont{C.}~\bibnamefont{Hagmann}} \bibnamefont{and}
  \bibinfo{author}{\bibfnamefont{P.}~\bibnamefont{Sikivie}},
  \bibinfo{journal}{Nucl.Phys.} \textbf{\bibinfo{volume}{B363}},
  \bibinfo{pages}{247} (\bibinfo{year}{1991}).

\bibitem[{\citenamefont{Hagmann et~al.}(2001)\citenamefont{Hagmann, Chang, and
  Sikivie}}]{Hagmann:2000ja}
\bibinfo{author}{\bibfnamefont{C.}~\bibnamefont{Hagmann}},
  \bibinfo{author}{\bibfnamefont{S.}~\bibnamefont{Chang}}, \bibnamefont{and}
  \bibinfo{author}{\bibfnamefont{P.}~\bibnamefont{Sikivie}},
  \bibinfo{journal}{Phys.Rev.} \textbf{\bibinfo{volume}{D63}},
  \bibinfo{pages}{125018} (\bibinfo{year}{2001}), \eprint{hep-ph/0012361}.

\bibitem[{\citenamefont{Yamaguchi et~al.}(1999)\citenamefont{Yamaguchi,
  Kawasaki, and Yokoyama}}]{Yamaguchi:1998gx}
\bibinfo{author}{\bibfnamefont{M.}~\bibnamefont{Yamaguchi}},
  \bibinfo{author}{\bibfnamefont{M.}~\bibnamefont{Kawasaki}}, \bibnamefont{and}
  \bibinfo{author}{\bibfnamefont{J.}~\bibnamefont{Yokoyama}},
  \bibinfo{journal}{Phys.Rev.Lett.} \textbf{\bibinfo{volume}{82}},
  \bibinfo{pages}{4578} (\bibinfo{year}{1999}), \eprint{hep-ph/9811311}.
  
  \bibitem[{\citenamefont{Hiramatsu
  et~al.}(2011{\natexlab{b}})\citenamefont{Hiramatsu, Kawasaki, Sekiguchi,
  Yamaguchi, and Yokoyama}}]{Hiramatsu:2010yu}
\bibinfo{author}{\bibfnamefont{T.}~\bibnamefont{Hiramatsu}},
  \bibinfo{author}{\bibfnamefont{M.}~\bibnamefont{Kawasaki}},
  \bibinfo{author}{\bibfnamefont{T.}~\bibnamefont{Sekiguchi}},
  \bibinfo{author}{\bibfnamefont{M.}~\bibnamefont{Yamaguchi}},
  \bibnamefont{and} \bibinfo{author}{\bibfnamefont{J.}~\bibnamefont{Yokoyama}},
  \bibinfo{journal}{Phys.Rev.} \textbf{\bibinfo{volume}{D83}},
  \bibinfo{pages}{123531} (\bibinfo{year}{2011}{\natexlab{b}}),
  \eprint{1012.5502}.  

\bibitem[{\citenamefont{Barr et~al.}(1987)\citenamefont{Barr, Choi, and
  Kim}}]{Barr:1986hs}
\bibinfo{author}{\bibfnamefont{S.~M.} \bibnamefont{Barr}},
  \bibinfo{author}{\bibfnamefont{K.}~\bibnamefont{Choi}}, \bibnamefont{and}
  \bibinfo{author}{\bibfnamefont{J.~E.} \bibnamefont{Kim}},
  \bibinfo{journal}{Nucl.Phys.} \textbf{\bibinfo{volume}{B283}},
  \bibinfo{pages}{591} (\bibinfo{year}{1987}).

\bibitem[{\citenamefont{Ryden et~al.}(1990)\citenamefont{Ryden, Press, and
  Spergel}}]{Ryden:1989vj}
\bibinfo{author}{\bibfnamefont{B.~S.} \bibnamefont{Ryden}},
  \bibinfo{author}{\bibfnamefont{W.~H.} \bibnamefont{Press}}, \bibnamefont{and}
  \bibinfo{author}{\bibfnamefont{D.~N.} \bibnamefont{Spergel}},
  \bibinfo{journal}{Astrophys. J.} \textbf{\bibinfo{volume}{357}},
  \bibinfo{pages}{293} (\bibinfo{year}{1990}).

\bibitem[{\citenamefont{Yamaguchi}(1999)}]{Yamaguchi:1999yp}
\bibinfo{author}{\bibfnamefont{M.}~\bibnamefont{Yamaguchi}},
  \bibinfo{journal}{Phys.Rev.} \textbf{\bibinfo{volume}{D60}},
  \bibinfo{pages}{103511} (\bibinfo{year}{1999}), \eprint{hep-ph/9907506};
\bibinfo{author}{\bibfnamefont{M.}~\bibnamefont{Yamaguchi}} \bibnamefont{and}
  \bibinfo{author}{\bibfnamefont{J.}~\bibnamefont{Yokoyama}},
  \bibinfo{journal}{Phys.Rev.} \textbf{\bibinfo{volume}{D66}},
  \bibinfo{pages}{121303} (\bibinfo{year}{2002}), \eprint{hep-ph/0205308};
\bibinfo{author}{\bibfnamefont{M.}~\bibnamefont{Yamaguchi}} \bibnamefont{and}
  \bibinfo{author}{\bibfnamefont{J.}~\bibnamefont{Yokoyama}},
  \bibinfo{journal}{Phys.Rev.} \textbf{\bibinfo{volume}{D67}},
  \bibinfo{pages}{103514} (\bibinfo{year}{2003}), \eprint{hep-ph/0210343}.

\bibitem[{\citenamefont{Harari and Sikivie}(1987)}]{Harari:1987ht}
\bibinfo{author}{\bibfnamefont{D.}~\bibnamefont{Harari}} \bibnamefont{and}
  \bibinfo{author}{\bibfnamefont{P.}~\bibnamefont{Sikivie}},
  \bibinfo{journal}{Phys.Lett.} \textbf{\bibinfo{volume}{B195}},
  \bibinfo{pages}{361} (\bibinfo{year}{1987}).

\bibitem[{\citenamefont{Chang et~al.}(1999)\citenamefont{Chang, Hagmann, and
  Sikivie}}]{Chang:1998tb}
\bibinfo{author}{\bibfnamefont{S.}~\bibnamefont{Chang}},
  \bibinfo{author}{\bibfnamefont{C.}~\bibnamefont{Hagmann}}, \bibnamefont{and}
  \bibinfo{author}{\bibfnamefont{P.}~\bibnamefont{Sikivie}},
  \bibinfo{journal}{Phys.Rev.} \textbf{\bibinfo{volume}{D59}},
  \bibinfo{pages}{023505} (\bibinfo{year}{1999}), \eprint{hep-ph/9807374}.

\bibitem[{\citenamefont{Kawasaki and Saikawa}(2011)}]{Kawasaki:2011vv}
\bibinfo{author}{\bibfnamefont{M.}~\bibnamefont{Kawasaki}} \bibnamefont{and}
  \bibinfo{author}{\bibfnamefont{K.}~\bibnamefont{Saikawa}},
  \bibinfo{journal}{JCAP} \textbf{\bibinfo{volume}{1109}}, \bibinfo{pages}{008}
  (\bibinfo{year}{2011}), \eprint{1102.5628}.

\bibitem[{\citenamefont{Hiramatsu
  et~al.}(2010{\natexlab{a}})\citenamefont{Hiramatsu, Kawasaki, and
  Saikawa}}]{Hiramatsu:2010yz}
\bibinfo{author}{\bibfnamefont{T.}~\bibnamefont{Hiramatsu}},
  \bibinfo{author}{\bibfnamefont{M.}~\bibnamefont{Kawasaki}}, \bibnamefont{and}
  \bibinfo{author}{\bibfnamefont{K.}~\bibnamefont{Saikawa}},
  \bibinfo{journal}{JCAP} \textbf{\bibinfo{volume}{1005}}, \bibinfo{pages}{032}
  (\bibinfo{year}{2010}{\natexlab{a}}), \eprint{1002.1555}.

\bibitem[{\citenamefont{Maggiore}(2007)}]{Maggiore:1900zz}
\bibinfo{author}{\bibfnamefont{M.}~\bibnamefont{Maggiore}},
  \emph{\bibinfo{title}{{Gravitational Waves. Vol. 1: Theory and Experiments}}}
  (\bibinfo{publisher}{Oxford University Press}, \bibinfo{year}{2007}).

\bibitem[{\citenamefont{Vilenkin}(1981)}]{Vilenkin:1981zs}
\bibinfo{author}{\bibfnamefont{A.}~\bibnamefont{Vilenkin}},
  \bibinfo{journal}{Phys.Rev.} \textbf{\bibinfo{volume}{D23}},
  \bibinfo{pages}{852} (\bibinfo{year}{1981}).

\bibitem[{\citenamefont{Komatsu et~al.}(2011)}]{Komatsu:2010fb}
\bibinfo{author}{\bibfnamefont{E.}~\bibnamefont{Komatsu}} \bibnamefont{et~al.}
  (\bibinfo{collaboration}{WMAP Collaboration}),
  \bibinfo{journal}{Astrophys.J.Suppl.} \textbf{\bibinfo{volume}{192}},
  \bibinfo{pages}{18} (\bibinfo{year}{2011}), \eprint{1001.4538}.

\bibitem[{\citenamefont{Baker et~al.}(2006)\citenamefont{Baker, Doyle,
  Geltenbort, Green, van~der Grinten et~al.}}]{Baker:2006ts}
\bibinfo{author}{\bibfnamefont{C.}~\bibnamefont{Baker}},
  \bibinfo{author}{\bibfnamefont{D.}~\bibnamefont{Doyle}},
  \bibinfo{author}{\bibfnamefont{P.}~\bibnamefont{Geltenbort}},
  \bibinfo{author}{\bibfnamefont{K.}~\bibnamefont{Green}},
  \bibinfo{author}{\bibfnamefont{M.}~\bibnamefont{van~der Grinten}},
  \bibnamefont{et~al.}, \bibinfo{journal}{Phys.Rev.Lett.}
  \textbf{\bibinfo{volume}{97}}, \bibinfo{pages}{131801}
  (\bibinfo{year}{2006}), \eprint{hep-ex/0602020}.

\bibitem[{\citenamefont{Raffelt}(2008)}]{Raffelt:2006cw}
\bibinfo{author}{\bibfnamefont{G.~G.} \bibnamefont{Raffelt}},
  \bibinfo{journal}{Lect.Notes Phys.} \textbf{\bibinfo{volume}{741}},
  \bibinfo{pages}{51} (\bibinfo{year}{2008}), \eprint{hep-ph/0611350}.

\bibitem[{\citenamefont{Raffelt}(1990)}]{Raffelt:1990yz}
\bibinfo{author}{\bibfnamefont{G.~G.} \bibnamefont{Raffelt}},
  \bibinfo{journal}{Phys.Rept.} \textbf{\bibinfo{volume}{198}},
  \bibinfo{pages}{1} (\bibinfo{year}{1990}).

\bibitem[{\citenamefont{Schlattl et~al.}(1999)\citenamefont{Schlattl, Weiss,
  and Raffelt}}]{Schlattl:1998fz}
\bibinfo{author}{\bibfnamefont{H.}~\bibnamefont{Schlattl}},
  \bibinfo{author}{\bibfnamefont{A.}~\bibnamefont{Weiss}}, \bibnamefont{and}
  \bibinfo{author}{\bibfnamefont{G.}~\bibnamefont{Raffelt}},
  \bibinfo{journal}{Astropart.Phys.} \textbf{\bibinfo{volume}{10}},
  \bibinfo{pages}{353} (\bibinfo{year}{1999}), \eprint{hep-ph/9807476}.

\bibitem[{\citenamefont{Inoue et~al.}(2008)\citenamefont{Inoue, Akimoto, Ohta,
  Mizumoto, Yamamoto et~al.}}]{Inoue:2008zp}
\bibinfo{author}{\bibfnamefont{Y.}~\bibnamefont{Inoue}},
  \bibinfo{author}{\bibfnamefont{Y.}~\bibnamefont{Akimoto}},
  \bibinfo{author}{\bibfnamefont{R.}~\bibnamefont{Ohta}},
  \bibinfo{author}{\bibfnamefont{T.}~\bibnamefont{Mizumoto}},
  \bibinfo{author}{\bibfnamefont{A.}~\bibnamefont{Yamamoto}},
  \bibnamefont{et~al.}, \bibinfo{journal}{Phys.Lett.}
  \textbf{\bibinfo{volume}{B668}}, \bibinfo{pages}{93} (\bibinfo{year}{2008}),
  \eprint{0806.2230}.

\bibitem[{\citenamefont{Andriamonje et~al.}(2007)}]{Andriamonje:2007ew}
\bibinfo{author}{\bibfnamefont{S.}~\bibnamefont{Andriamonje}}
  \bibnamefont{et~al.} (\bibinfo{collaboration}{CAST Collaboration}),
  \bibinfo{journal}{JCAP} \textbf{\bibinfo{volume}{0704}}, \bibinfo{pages}{010}
  (\bibinfo{year}{2007}), \eprint{hep-ex/0702006}.

\bibitem[{\citenamefont{Arik et~al.}(2009)}]{Arik:2008mq}
\bibinfo{author}{\bibfnamefont{E.}~\bibnamefont{Arik}} \bibnamefont{et~al.}
  (\bibinfo{collaboration}{CAST Collaboration}), \bibinfo{journal}{JCAP}
  \textbf{\bibinfo{volume}{0902}}, \bibinfo{pages}{008} (\bibinfo{year}{2009}),
  \eprint{0810.4482}.

\bibitem[{\citenamefont{Irastorza et~al.}(2011)\citenamefont{Irastorza,
  Avignone, Caspi, Carmona, Dafni et~al.}}]{Irastorza:2011gs}
\bibinfo{author}{\bibfnamefont{I.}~\bibnamefont{Irastorza}},
  \bibinfo{author}{\bibfnamefont{F.}~\bibnamefont{Avignone}},
  \bibinfo{author}{\bibfnamefont{S.}~\bibnamefont{Caspi}},
  \bibinfo{author}{\bibfnamefont{J.}~\bibnamefont{Carmona}},
  \bibinfo{author}{\bibfnamefont{T.}~\bibnamefont{Dafni}},
  \bibnamefont{et~al.}, \bibinfo{journal}{JCAP}
  \textbf{\bibinfo{volume}{1106}}, \bibinfo{pages}{013} (\bibinfo{year}{2011}),
  \eprint{1103.5334};
\bibinfo{author}{\bibfnamefont{I.}~\bibnamefont{Irastorza}}
  \bibnamefont{et~al.} (\bibinfo{collaboration}{IAXO Collaboration})
  (\bibinfo{year}{2012}), \eprint{1201.3849}.

\bibitem[{\citenamefont{Raffelt}(1999)}]{Raffelt:1999tx}
\bibinfo{author}{\bibfnamefont{G.~G.} \bibnamefont{Raffelt}},
  \bibinfo{journal}{Ann.Rev.Nucl.Part.Sci.} \textbf{\bibinfo{volume}{49}},
  \bibinfo{pages}{163} (\bibinfo{year}{1999}), \eprint{hep-ph/9903472}.

\bibitem[{\citenamefont{Raffelt and Weiss}(1995)}]{Raffelt:1994ry}
\bibinfo{author}{\bibfnamefont{G.}~\bibnamefont{Raffelt}} \bibnamefont{and}
  \bibinfo{author}{\bibfnamefont{A.}~\bibnamefont{Weiss}},
  \bibinfo{journal}{Phys.Rev.} \textbf{\bibinfo{volume}{D51}},
  \bibinfo{pages}{1495} (\bibinfo{year}{1995}), \eprint{hep-ph/9410205}.

\bibitem[{\citenamefont{Raffelt}(1986)}]{Raffelt:1985nj}
\bibinfo{author}{\bibfnamefont{G.~G.} \bibnamefont{Raffelt}},
  \bibinfo{journal}{Phys.Lett.} \textbf{\bibinfo{volume}{B166}},
  \bibinfo{pages}{402} (\bibinfo{year}{1986}).

\bibitem[{\citenamefont{Isern et~al.}(2008)\citenamefont{Isern, Garcia-Berro,
  Torres, and Catalan}}]{Isern:2008nt}
\bibinfo{author}{\bibfnamefont{J.}~\bibnamefont{Isern}},
  \bibinfo{author}{\bibfnamefont{E.}~\bibnamefont{Garcia-Berro}},
  \bibinfo{author}{\bibfnamefont{S.}~\bibnamefont{Torres}}, \bibnamefont{and}
  \bibinfo{author}{\bibfnamefont{S.}~\bibnamefont{Catalan}}
  (\bibinfo{year}{2008}), \eprint{0806.2807};
\bibinfo{author}{\bibfnamefont{J.}~\bibnamefont{Isern}},
  \bibinfo{author}{\bibfnamefont{S.}~\bibnamefont{Catalan}},
  \bibinfo{author}{\bibfnamefont{E.}~\bibnamefont{Garcia-Berro}},
  \bibnamefont{and} \bibinfo{author}{\bibfnamefont{S.}~\bibnamefont{Torres}},
  \bibinfo{journal}{J.Phys.Conf.Ser.} \textbf{\bibinfo{volume}{172}},
  \bibinfo{pages}{012005} (\bibinfo{year}{2009}), \eprint{0812.3043}.

\bibitem[{\citenamefont{Isern et~al.}(1992)\citenamefont{Isern, Hernanz, and
  Garcia-Berro}}]{1992ApJ...392L..23I}
\bibinfo{author}{\bibfnamefont{J.}~\bibnamefont{Isern}},
  \bibinfo{author}{\bibfnamefont{M.}~\bibnamefont{Hernanz}}, \bibnamefont{and}
  \bibinfo{author}{\bibfnamefont{E.}~\bibnamefont{Garcia-Berro}},
  \bibinfo{journal}{Astrophys. J.} \textbf{\bibinfo{volume}{392}},
  \bibinfo{pages}{L23} (\bibinfo{year}{1992});
\bibinfo{author}{\bibfnamefont{J.}~\bibnamefont{Isern}},
  \bibinfo{author}{\bibfnamefont{E.}~\bibnamefont{Garcia-Berro}},
  \bibinfo{author}{\bibfnamefont{L.}~\bibnamefont{Althaus}}, \bibnamefont{and}
  \bibinfo{author}{\bibfnamefont{A.}~\bibnamefont{Corsico}}
  (\bibinfo{year}{2010}), \eprint{1001.5248}.

\bibitem[{\citenamefont{Seto et~al.}(2001)\citenamefont{Seto, Kawamura, and
  Nakamura}}]{Seto:2001qf}
\bibinfo{author}{\bibfnamefont{N.}~\bibnamefont{Seto}},
  \bibinfo{author}{\bibfnamefont{S.}~\bibnamefont{Kawamura}}, \bibnamefont{and}
  \bibinfo{author}{\bibfnamefont{T.}~\bibnamefont{Nakamura}},
  \bibinfo{journal}{Phys.Rev.Lett.} \textbf{\bibinfo{volume}{87}},
  \bibinfo{pages}{221103} (\bibinfo{year}{2001}), \eprint{astro-ph/0108011}.

\bibitem[{\citenamefont{Kudoh et~al.}(2006)\citenamefont{Kudoh, Taruya,
  Hiramatsu, and Himemoto}}]{Kudoh:2005as}
\bibinfo{author}{\bibfnamefont{H.}~\bibnamefont{Kudoh}},
  \bibinfo{author}{\bibfnamefont{A.}~\bibnamefont{Taruya}},
  \bibinfo{author}{\bibfnamefont{T.}~\bibnamefont{Hiramatsu}},
  \bibnamefont{and} \bibinfo{author}{\bibfnamefont{Y.}~\bibnamefont{Himemoto}},
  \bibinfo{journal}{Phys.Rev.} \textbf{\bibinfo{volume}{D73}},
  \bibinfo{pages}{064006} (\bibinfo{year}{2006}), \eprint{gr-qc/0511145}.
  
  \bibitem[{\citenamefont{Nagasawa and Kawasaki}(1994)}]{Nagasawa:1994qu}
\bibinfo{author}{\bibfnamefont{M.}~\bibnamefont{Nagasawa}} \bibnamefont{and}
  \bibinfo{author}{\bibfnamefont{M.}~\bibnamefont{Kawasaki}},
  \bibinfo{journal}{Phys.Rev.} \textbf{\bibinfo{volume}{D50}},
  \bibinfo{pages}{4821} (\bibinfo{year}{1994}), \eprint{astro-ph/9402066};
\bibinfo{author}{\bibfnamefont{M.}~\bibnamefont{Nagasawa}} \bibnamefont{and}
  \bibinfo{author}{\bibfnamefont{M.}~\bibnamefont{Kawasaki}},
  \bibinfo{journal}{Nucl.Phys.Proc.Suppl.} \textbf{\bibinfo{volume}{43}},
  \bibinfo{pages}{82} (\bibinfo{year}{1995}).

\bibitem[{\citenamefont{Chang and Choi}(1993)}]{Chang:1993gm}
\bibinfo{author}{\bibfnamefont{S.}~\bibnamefont{Chang}} \bibnamefont{and}
  \bibinfo{author}{\bibfnamefont{K.}~\bibnamefont{Choi}},
  \bibinfo{journal}{Phys.Lett.} \textbf{\bibinfo{volume}{B316}},
  \bibinfo{pages}{51} (\bibinfo{year}{1993}), \eprint{hep-ph/9306216}.

\bibitem[{\citenamefont{Moroi and Murayama}(1998)}]{Moroi:1998qs}
\bibinfo{author}{\bibfnamefont{T.}~\bibnamefont{Moroi}} \bibnamefont{and}
  \bibinfo{author}{\bibfnamefont{H.}~\bibnamefont{Murayama}},
  \bibinfo{journal}{Phys.Lett.} \textbf{\bibinfo{volume}{B440}},
  \bibinfo{pages}{69} (\bibinfo{year}{1998}), \eprint{hep-ph/9804291}.

\bibitem[{\citenamefont{Buckley and Murayama}(2007)}]{Buckley:2007tm}
\bibinfo{author}{\bibfnamefont{M.~R.} \bibnamefont{Buckley}} \bibnamefont{and}
  \bibinfo{author}{\bibfnamefont{H.}~\bibnamefont{Murayama}},
  \bibinfo{journal}{JCAP} \textbf{\bibinfo{volume}{0707}}, \bibinfo{pages}{012}
  (\bibinfo{year}{2007}), \eprint{0705.0542}.

\bibitem[{\citenamefont{Hannestad et~al.}(2005)\citenamefont{Hannestad,
  Mirizzi, and Raffelt}}]{Hannestad:2005df}
\bibinfo{author}{\bibfnamefont{S.}~\bibnamefont{Hannestad}},
  \bibinfo{author}{\bibfnamefont{A.}~\bibnamefont{Mirizzi}}, \bibnamefont{and}
  \bibinfo{author}{\bibfnamefont{G.}~\bibnamefont{Raffelt}},
  \bibinfo{journal}{JCAP} \textbf{\bibinfo{volume}{0507}}, \bibinfo{pages}{002}
  (\bibinfo{year}{2005}), \eprint{hep-ph/0504059};
\bibinfo{author}{\bibfnamefont{S.}~\bibnamefont{Hannestad}},
  \bibinfo{author}{\bibfnamefont{A.}~\bibnamefont{Mirizzi}},
  \bibinfo{author}{\bibfnamefont{G.~G.} \bibnamefont{Raffelt}},
  \bibnamefont{and} \bibinfo{author}{\bibfnamefont{Y.~Y.} \bibnamefont{Wong}},
  \bibinfo{journal}{JCAP} \textbf{\bibinfo{volume}{1008}}, \bibinfo{pages}{001}
  (\bibinfo{year}{2010}), \eprint{1004.0695}.

\bibitem[{\citenamefont{Masso et~al.}(2002)\citenamefont{Masso, Rota, and
  Zsembinszki}}]{Masso:2002np}
\bibinfo{author}{\bibfnamefont{E.}~\bibnamefont{Masso}},
  \bibinfo{author}{\bibfnamefont{F.}~\bibnamefont{Rota}}, \bibnamefont{and}
  \bibinfo{author}{\bibfnamefont{G.}~\bibnamefont{Zsembinszki}},
  \bibinfo{journal}{Phys.Rev.} \textbf{\bibinfo{volume}{D66}},
  \bibinfo{pages}{023004} (\bibinfo{year}{2002}), \eprint{hep-ph/0203221}.

\bibitem[{\citenamefont{Cadamuro and Redondo}(2012)}]{Cadamuro:2011fd}
\bibinfo{author}{\bibfnamefont{D.}~\bibnamefont{Cadamuro}} \bibnamefont{and}
  \bibinfo{author}{\bibfnamefont{J.}~\bibnamefont{Redondo}},
  \bibinfo{journal}{JCAP} \textbf{\bibinfo{volume}{1202}}, \bibinfo{pages}{032}
  (\bibinfo{year}{2012}), \eprint{1110.2895}.

\bibitem[{\citenamefont{Hiramatsu
  et~al.}(2010{\natexlab{b}})\citenamefont{Hiramatsu, Kawasaki, and
  Takahashi}}]{Hiramatsu:2010dx}
\bibinfo{author}{\bibfnamefont{T.}~\bibnamefont{Hiramatsu}},
  \bibinfo{author}{\bibfnamefont{M.}~\bibnamefont{Kawasaki}}, \bibnamefont{and}
  \bibinfo{author}{\bibfnamefont{F.}~\bibnamefont{Takahashi}},
  \bibinfo{journal}{JCAP} \textbf{\bibinfo{volume}{1006}}, \bibinfo{pages}{008}
  (\bibinfo{year}{2010}{\natexlab{b}}), \eprint{1003.1779}.

\bibitem[{\citenamefont{Yoshida}(1990)}]{Yoshida:1990zz}
\bibinfo{author}{\bibfnamefont{H.}~\bibnamefont{Yoshida}},
  \bibinfo{journal}{Phys.Lett.} \textbf{\bibinfo{volume}{A150}},
  \bibinfo{pages}{262} (\bibinfo{year}{1990}).

\bibitem[{\citenamefont{Press et~al.}(1989)\citenamefont{Press, Ryden, and
  Spergel}}]{Press:1989yh}
\bibinfo{author}{\bibfnamefont{W.~H.} \bibnamefont{Press}},
  \bibinfo{author}{\bibfnamefont{B.~S.} \bibnamefont{Ryden}}, \bibnamefont{and}
  \bibinfo{author}{\bibfnamefont{D.~N.} \bibnamefont{Spergel}},
  \bibinfo{journal}{Astrophys.J.} \textbf{\bibinfo{volume}{347}},
  \bibinfo{pages}{590} (\bibinfo{year}{1989}).

\bibitem[{\citenamefont{Dufaux et~al.}(2007)\citenamefont{Dufaux, Bergman,
  Felder, Kofman, and Uzan}}]{Dufaux:2007pt}
\bibinfo{author}{\bibfnamefont{J.~F.} \bibnamefont{Dufaux}},
  \bibinfo{author}{\bibfnamefont{A.}~\bibnamefont{Bergman}},
  \bibinfo{author}{\bibfnamefont{G.~N.} \bibnamefont{Felder}},
  \bibinfo{author}{\bibfnamefont{L.}~\bibnamefont{Kofman}}, \bibnamefont{and}
  \bibinfo{author}{\bibfnamefont{J.-P.} \bibnamefont{Uzan}},
  \bibinfo{journal}{Phys.Rev.} \textbf{\bibinfo{volume}{D76}},
  \bibinfo{pages}{123517} (\bibinfo{year}{2007}), \eprint{0707.0875}.

\bibitem[{\citenamefont{Huang and Sikivie}(1985)}]{Huang:1985tt}
\bibinfo{author}{\bibfnamefont{M.}~\bibnamefont{Huang}} \bibnamefont{and}
  \bibinfo{author}{\bibfnamefont{P.}~\bibnamefont{Sikivie}},
  \bibinfo{journal}{Phys.Rev.} \textbf{\bibinfo{volume}{D32}},
  \bibinfo{pages}{1560} (\bibinfo{year}{1985}).

\end{thebibliography}
\end{document}